\documentclass[openany, a4papper,twoside, 12pt]{book}
\usepackage{amsmath}
\usepackage{epsfig}
\usepackage{graphicx}
\usepackage{hyperref}
\usepackage{amsbsy}
\usepackage{here}
\usepackage{SIunits}
\newlength{\bildtitel}
\newcommand{\epsbild}[5]
{
\begin{figure}[#5]
\begin{center}
\epsfig{file=#1,width=#3\linewidth,angle=#4}
\parbox{\bildtitel}{\scriptsize\caption{\label{#1}#2}}
\end{center}
\end{figure}
}
\begin{document}
\pagestyle{empty}
\begin{center}
\vspace{0.5cm}
{\Huge\    JAGIELLONIAN UNIVERSITY         \\
            INSTITUTE OF PHYSICS          \\
}
\vspace{3.0cm} \vspace{3.0cm} {\bf{\huge Thermal up-scattering
of very cold and ultra-cold neutrons in solid deuterium}}
\vspace{2.0cm}

{\Large                 Ma\l{}gorzata Kasprzak}

\vspace{2.0cm}
\end{center}
\vspace{4.0cm}
\begin{flushright}
{\large Master Thesis prepared at the Nuclear Physics Department\\
        academic supervisor: Prof. Dr hab. Kazimierz Bodek  \\
}
\end{flushright}
\vspace{1.0cm}
\begin{center}
{\large Krak\'ow 2004}
\end{center}

\frontmatter
\chapter*{Abstract}
The work presented in this thesis forms part of a program at the
Paul Scherrer Institute (PSI) to construct a high intensity
superthermal ultra-cold neutron (UCN) source based on solid
deuterium (sD$_2$) as UCN production medium. We carried out a set
of experiments to gain a better understanding of the properties
and the behaviour of solid deuterium as a cold neutron moderator
and ultra-cold neutron converter.

We present the measurements of the total neutron cross section as obtained
by transmission studies with very cold neutrons and ultra-cold neutrons in solid deuterium.
The experimental set-up and the methods of data analysis are described and also the
procedure of preparing the solid deuterium samples is given. The neutron
transmission studies are supported by optical investigation of the crystal
and by Raman spectroscopy. We have thus characterised the temperature
dependence of the neutron transmission through solid deuterium and we have
been able to identify the role that coherent neutron scattering plays for
the investigated deuterium samples.

\chapter*{Acknowledgements}

First of all, I would like to thank my supervisor, Prof. Kazimierz
Bodek for supporting this work with ideas, helpful comments and
suggestions and for providing advice on drafts.
\\\\
I would especially like to thank Dr Klaus Kirch for all his
support, guidance and for providing stimulating conversations.
Many of his great ideas allowed much of this work to be done.
\\\\
I would like to acknowledge Prof. Reinhard Kulessa for allowing me
preparing this thesis at the Nuclear Physics Department of the
Jagiellonian University and Prof. Bogus\l{}aw Kamys for a great
support.
\\\\
I also thank Tomek Bry\'s for his suggestions, fruitful discussions and for being a great friend.
\\\\
I thank fellow students at the Nuclear Physics Department: Joanna Przerwa,  Micha\l{} Janusz, Marcin Ku\'zniak and Tytus Smoli\'nski for providing a great atmosphere.
\\\\
Finally I would like to say a big 'thank you' to my family and all my friends for supporting me
totally and unconditionally throughout my studies. I would especially like
to thank my grandparents: Miros\l{}awa and Stanis\l{}aw Skrok for unremitting support and encouragement.

\pagestyle{headings}
\tableofcontents
\mainmatter
\chapter{Introduction}
\section{Physics of ultra-cold neutrons}
Ultra-cold neutrons (UCN) are neutrons of very low energy: the kinetic energy of a UCN is smaller than the effective Fermi potential $V_F$ of a given material:
\begin{eqnarray}
E \le V_F = \frac{2\pi\hbar^2 Nb}{M_n} 
\end{eqnarray}
where $M_n$ = 939~\mega\electronvolt~is the neutron mass, $N$ is the number density of atoms and $b$ is the bound coherent scattering length of the material. 
Materials which are used in UCN physics are normally chosen to have a high Fermi potential, typically of the order of \power{10}{-7}~\electronvolt~. For example 
Be ($V_F$ = 252~\nano\electronvolt) and $^{58}$Ni ($V_F$ = 335~\nano\electronvolt) are often incorporated as wall coatings in UCN equipment. 
UCN undergo total reflection at any angle of incidence from the surface of a material. As a consequence UCN can be trapped in bottles for times comparable
with the free neutron lifetime of about 887~\second. Alternatively, UCN can be stored in magnetic traps: the UCN are confined to a trap region defined by a 
strong magnetic field that acts as a barrier through the interaction of the neutron magnetic moment and the magnetic field~\cite{Huf,Doy}. 

The fact that one can store and observe UCN for such long periods makes them an excellent tool to study fundamental properties of the neutron.
Stored UCN offer the possibility of a precise measurement of the neutron lifetime that plays a key role for Big Bang nucleosynthesis and also is one
of the main input parameters in tests of the Standard Model. Moreover, stored UCN are ideal for electric dipole moment (EDM) experiments.
The existence of a permanent non-zero neutron EDM violates both parity (P) and time reversal (T) invariance and, through the conservation of the
combined symmetry CPT, also violates CP symmetry (C refers to charge conjugation). CP violation, which is believed to be responsible for the baryon asymmetry of 
the Universe, has been measured in the K$^0$ and the B$^0$ mesonic systems. These results are consistent with the description of CP violation in the Standard Model
using the Cabibo-Kobayashi-Maskawa mass mixing matrix (CKM) ansatz. 
CP violation as described in the Standard Model is however at a level which is orders of magnitudes too low to generate the observed
baryon asymmetry which suggests an existence of a mechanism for CP violation which is beyond the Standard Model. UCN can thus, through a measurement of their EDM,
form a very sensitive probe for physics beyond the Standard Model. Other fields where UCN can be used are also emerging: surface physics~\cite{Gol2} and the observation
of quantum states in the Earth's gravitational field~\cite{Abe, Nes}.

Many of the experiments with UCN are limited by the statistics that is achieved in present sources. Next generation experiments that aim to improve these sensitivities
rely on the development of new high intensity UCN sources.
\section{Methods of the UCN production}
Neutrons produced in a nuclear reaction in a spallation source or in a fission reaction in a nuclear reactor have energies of the order of~\mega\electronvolt. 
These fast neutrons are moderated to thermal velocities by a moderator (e.g. heavy water) and can subsequently be slowed down into the
cold regime upon interaction with a cold source (e.g. at the Institute Laue-Langevin (ILL) liquid deuterium at a temperature of 25~\kelvin~acts as a cold source).
To obtain neutrons with energy in the UCN region two methods can be used: 
\begin{itemize}
\item
the conventional method which relies on extracting slow neutrons from the tail of the Maxwellian distribution of neutrons from the cold source,
\item
a method in which cold neutrons are converted to UCN through a superthermal UCN production mechanism.
\end{itemize}
\subsection{Conventional UCN extraction}
The velocity spectrum of the neutrons in the cold source follows a Maxwellian velocity distribution.   
There will thus be a fraction, albeit a small fraction, of neutrons of UCN energy present
in the low energy tail of the spectrum. The thermal spectrum of the moderator of  
temperature $T_0$ is given by:
\begin{equation}
\rho(v)dv = \frac{2\Phi_0}{\alpha}\frac{v^2}{\alpha^2}\exp(-\frac{v^2}{\alpha^2})\frac{dv}{\alpha} \label{eq:density}
\end{equation}
where $\Phi_0$ is the total thermal neutron flux and $\alpha=\sqrt{2k_BT_0/M_n}$ is the mean thermal velocity.
Integration of \ref{eq:density} for the energies ranging from zero to the potential $V_F$ that corresponds to the Fermi potential
of the wall material gives the maximum UCN density:
\begin{equation}
\rho_{UCN} = \int_0^{V_F}{\rho(v)dv} = \frac{2}{3}\frac{\Phi_0}{\alpha}\left(\frac{V_F}{k_BT_0}\right)^{\frac{3}{2}} 
\end{equation}
Ultra-cold neutrons can be extracted from the cold source through a vertical guide. Gravity is 
acting on the neutrons traveling upwards and decelerate faster neutrons to the lower energies.
The mechanical deceleration can be used as well: 
a moving scatterer interacts with the neutrons and will carry away some of the neutron momentum. The number of UCN 
generated however is limited by Liouville's theorem:  the phase space density of the neutrons must be conserved if 
the particle move under the action of forces which are derivable from a potential. 
The final UCN density cannot be higher than the density at the cold neutron source because the compression 
in the neutron velocity space dilutes the neutron number density. 
    
The method described here is used at the UCN source operating at the Institute Laue-Langevin in Grenoble, France. 
It is in fact the strongest source of UCNs in the world. Extraction of the UCN from a liquid deuterium cold source is done through a vertical guide. 
Neutrons that travel the 17 meters upward through the guide and pass through a curved guide part at the end have
kinetic energies in the very-cold neutron region. These VCN are guided into a turbine where they interact with moving
mirror blades that slow them further down, by mechanical deceleration, to UCN energies. The details of the turbine can be found in Ref.~\cite{Gol2}.   
The UCN density obtained in this set-up reaches 40~\centi\rpcubic\metre~in a storage bottle placed immediately outside the turbine. 
Further improvement in this density is not expected because of the limitation imposed by Liouville's theorem.
\subsection{Superthermal UCN production}
\subsubsection{The principle of superthermal UCN production}     
UCN can also be created from cold neutrons in a superthermal process. In this process neutrons interact with a medium that acts as a moderator to the cold neutrons 
and converts these neutrons into UCN by inelastic scattering. The medium must posses specific properties such as well defined energy levels and excitations that enable this down-scattering to take place. Superthermal UCN production was first proposed by Golub and Pendlenbury in a paper in which the UCN production in superfluid $^4$He through this mechanism was studied~\cite{Gol}. The understand the principle, consider a moderator with two energy levels, a ground state and an excited state, separated by an energy $\Delta$.
The moderator in the ground state can be excited by absorbing an energy $\Delta$ through inelastic interactions with neutrons of energy $E_{UCN} + \Delta$.
The neutrons will thus be down-scattered to UCN energies.
The inverse is also possible: the moderator falls back to its ground state by releasing an energy $\Delta$ and UCN can absorb this energy and thus 
be up-scattered out of the UCN energy region. 
The principle of detailed balance for this system leads to a relationship between the down-scattering rate (UCN production) and
the up-scattering rate (UCN losses):
\begin{equation}
\sigma_{up} = \sigma_{down}\exp{\left(-\frac{\Delta}{k_BT}\right)}
\end{equation}
where $\sigma_{up}$ is the up-scattering rate of a neutron with energy $E_{UCN}$ and $\sigma_{down}$ is the down-scattering rate of a neutron with 
energy $E_{UCN}+\Delta$.  
At low enough temperatures the up-scattering rate will be suppressed and the down-scattering will dominate
resulting in a net production rate of UCN. It is noteworthy that this process is not limited by the Liouville's theorem for the neutrons alone.
   
The UCN density that will be obtained with a system in equilibrium depends on the ratio of the production 
rate $P$ and the loss rate 1/$\tau$:
\begin{equation}
\rho = P\cdot\tau 
\end{equation} 
The loss factor is determined by a combination of loss mechanisms such as the neutron lifetime and losses
specific to the design of the superthermal source such as nuclear absorption in the moderator, wall losses etc.
\subsubsection{Superthermal UCN sources}
There is a range of materials that could, in principle, be suitable as superthermal converters. 
Superthermal UCN sources based on superfluid helium and on solid D$_2$ have already reached an advanced 
stage and have shown to be generating high densities of UCN. Furthermore converters such as CD$_4$~\cite{Pok1} and O$_2$~\cite{Liu1,Liu2} are currently being studied and appear to be promising candidates for further progress in this field.

The first superthermal converter that was proposed is superfluid helium~\cite{Gol}. 
One of the major advantages of superfluid helium is that the storage lifetime of UCN in such a medium is very long: $^4$He has got
zero neutron absorption cross section and the up-scattering modes can be strongly suppressed by keeping  helium at temperature
of 0.5~\kelvin. UCN are produced in superfluid helium by down-scattering cold neutrons through creation of phonons. The dispersion curves of a free neutron
and a phonon intersect, apart from at the origin, only at one energy of 1~\milli\electronvolt. 
This means that, due to conservation of both energy and momentum, neutrons of only
one wavelength, 9~\angstrom, can be down-scattered to UCN energies. The fact that the production of UCN can proceed through the interaction 
of cold neutron of one single energy only results in a helium UCN production rate about two orders of magnitude lower as compared with 
solid deuterium where there is no such restriction on the incident neutron energy in the down-scattering process. Furthermore,
efficient extraction of UCN from the superfluid has proved to be extremely difficult and thus most superthermal UCN sources
are incorporated in single dedicated experiments (a neutron lifetime experiment at NIST~\cite{Doy1}, 
neutron EDM experiments at LANL~\cite{Lanl} and ILL~\cite{Suss}). A superthermal UCN source based on superfluid helium that will feed several 
independent experiments is being studied only at Osaka~\cite{Mas}.

After the first work on $^4$He UCN sources, other materials emerged as a possible medium to create UCN. Most noticeable is the development 
of the solid D$_2$ based sources (that have an absorption lifetime of 150~\milli\second, which is about four orders of magnitude lower than that for He) 
together with the idea of the ``thin film superthermal sources''~\cite{Gol1,Yu}. The thin film
superthermal source is based on the notion that the UCN density in a storage volume that contains  
converter material is independent of the actual volume of the converter
material. The use of the pulsed neutron source in combination with the converter with a short neutron lifetime was first pointed
out by Pokotilovski~\cite{Pok2}. In this scheme the moderator is connected to a UCN storage volume only for a short time. 
After the neutron pulse has occurred a valve closes separating the conversion and storage regions, this enables the UCN to survive for long times
as they are no longer interacting with the absorbing converter material. 

Superthermal UCN sources based on solid deuterium  are currently under intense investigation at PSI~\cite{Psi} and at Los Alamos~\cite{Hill,Kir,Sau,Mor}.
The production of UCN using sD$_2$ has first been studied at PNPI Gatchina where a gain factor of 10 in
UCN production was obtained compared to the UCN production using liquid deuterium source~\cite{Ser,Alt}.  
At Los Alamos a prototype sD$_2$ based UCN source was operated and
UCN densities in excess of 100~\centi\rpcubic\metre~have been reported. The UCN source being built at PSI, also based on sD$_2$, 
aims at achieving densities of 3000~\centi\reciprocal\metre~in a 2~\cubic\metre~volume. 
A detailed discussion about sD$_2$ as a UCN moderator is presented in the next chapter.

\chapter{sD$_2$ as a super-thermal source} \label{sd2mod}
A good superthermal converter needs to possess a number of properties that are directly related to the creation and losses of UCN.
In this chapter the properties of deuterium are described with the focus on its use as superthermal UCN converter material. Some of the properties of deuterium are
already quite favourable: deuterium has a relatively low neutron absorption cross section (the absorption cross section for
thermal neutrons is 519~\micro\barn) and it is a light element.
Also, deuterium has a large down-scattering cross section because the phonon density of states has a large overlap with the
cold neutron spectrum.
\section{Neutron scattering in sD$_2$}
\subsection{Coherent and incoherent scattering}
To investigate the suitability of using solid deuterium as a superthermal UCN converter it is important to understand the
neutron scattering from this material.
The interaction potential between a neutron and a rigid array of N nuclei in the first Born approximation
is~\cite{Lov}:
\begin{equation}
V(\mathbf{r})=\frac{2\pi\hbar^2}{M_n}\sum_{l}b_l\delta(\mathbf{r}-\mathbf{R}_l) \label{eq:fpot}
\end{equation}
where $M_n$ is the neutron mass, $b_l$ is the bound nucleus scattering length for the $l$th nucleus
and $\mathbf{R}_l$ is the position vector of the \textsl{l}th nucleus.
The final state $\psi_{\mathbf{k'}}$ and the initial state $\psi_{\mathbf{k}}$  of a neutron,
defined by the wave vectors  $\mathbf{k'}$ and  $\mathbf{k}$, respectively, are given by plane waves.
For a large box of volume $L^3$ the states $\psi_{\mathbf{k}}$ and $\psi_{\mathbf{k'}}$ are:
\begin{eqnarray}
\psi_{\mathbf{k}} = \frac{\exp(i\mathbf{k\cdot{}r})}{L^{\frac{3}{2}}} ~~\textrm{with energy $E = \frac{\hbar^2k^2}{2M_n}$} \label{eq:inst}
\end{eqnarray}
and
\begin{equation}
\psi_{\mathbf{k'}} = \frac{\exp(i\mathbf{k'\cdot{}r})}{L^{\frac{3}{2}}} ~~\textrm{with energy $E' = \frac{\hbar^2k'^2}{2M_n}$} \label{eq:outst}
\end{equation}
The initial state of the system composed of the incident neutron and the scatterer is described by:
\begin{equation}
|k\rangle~|\lambda\rangle~\equiv~|k\lambda\rangle \label{eq:syst}
\end{equation}
where $|\lambda\rangle$ denotes the state of the scatterer with energy $E_{\lambda}$.\\
The transition rate from the initial state $\psi_{\mathbf{k}}$
to the final state $\psi_{\mathbf{k'}}$  is given by the Fermi's golden rule~\cite{Sch}:
\begin{equation}
dW_{k\to{}k'}=\frac{2\pi}{\hbar}\bigg|\int{}d\mathbf{r}\psi_{\mathbf{k}'}^{*}V\psi_{\mathbf{k}}^{}\bigg|^2\rho_{\mathbf{k}'}^{}(E')~\delta(E+E_{\lambda}-E'-E_{\lambda'}) \label{eq:gold}
\end{equation}
where $V$ refers to the interaction potential that causes the transition.
$\rho_{\mathbf{k}'}^{}(E')dE'$ is the density of the neutron final states per unit energy interval
\begin{equation}
\rho_{\mathbf{k}'}^{}(E') = \frac{L^3M_n~k'}{8\pi^3\hbar^2}d\Omega. \label{eq:dens}
\end{equation}
The cross section is obtained by dividing the transition rate $dW$ by the incident flux of neutrons $\left(\hbar{}k/(M_nL^3)\right)$.
Using the Dirac notation we can write the double differential cross section as:
\begin{equation}
\frac{d^2\sigma}{d\Omega{}dE'} = \frac{k'}{k}|\langle\mathbf{k}'\lambda'|~V~|\mathbf{k}\lambda\rangle|^2~\delta(\hbar\omega + E_{\lambda} - E_{\lambda'}) \label{eq:pcross_1}
\end{equation}
where $\hbar\omega$ is the defined by:
\begin{equation}
\hbar\omega = E - E' = \frac{\hbar^2}{2M_n}(k^2 - k'^2)
\end{equation}
Inserting expression~\ref{eq:fpot} for the scattering potential one obtain the differential cross section per solid angle per atom:
\begin{equation}
\frac{d^2\sigma}{d\Omega{}dE'} = \frac{1}{N}\frac{k'}{k}\sum_{\lambda{}'\lambda{}}p_{\lambda}\big|\langle\lambda'|\sum_lb_l\exp(i\mathbf{Q}\cdot\mathbf{R}_l^{})|\lambda\rangle\big|^2\delta(\hbar\omega+E_{\lambda}-E_{\lambda'}) \label{eq:pcross_2}
\end{equation}
where $\mathbf{Q}\hbar = (\mathbf{k} - \mathbf{k'})\hbar$ is the momentum transfer at the scattering.
$p_{\lambda}$ represents the weight for the state $|\lambda\rangle$, which is proportional to the Boltzmann factor
$\exp(-E_{\lambda}/k_BT)$.\\
We can write equation~\ref{eq:pcross_2} as:
\begin{eqnarray}
\frac{d^2\sigma}{d\Omega{}dE'} = \frac{1}{N}\frac{k'}{k}\sum_{\lambda{}'\lambda{}}p_{\lambda}\sum_{l,l'}\overline{b_{l}^{*}b_{l'}^{}}
\langle\lambda|\exp(-i\mathbf{Q}\cdot\mathbf{R}_l^{})|\lambda'\rangle\nonumber\\
\times\langle\lambda'|\exp(i\mathbf{Q}\cdot\mathbf{R}_l^{})|\lambda\rangle~
\delta(\hbar\omega+E_{\lambda}-E_{\lambda'}) \label{eq:pcross_3}
\end{eqnarray}
The quantity $\overline{b_{l'}^{*}b_{l}^{}}$ is the average of $b_{l'}^{*}b_{l}^{}$ over a random nuclear
spin orientation and a random isotope distribution and is described by:
\begin{equation}
\overline{b_{l'}^{*}b_{l}^{}} = |\overline{b}|^2 + \delta_{l,l'}(\overline{|b^2|} - |\overline{b}|^2) \label{eq:slen}
\end{equation}
Substituting \ref{eq:slen} into \ref{eq:pcross_3}, the cross section can be split in two parts:
\begin{equation}
\frac{d^2\sigma}{d\Omega{}dE'} = \left(\frac{d^2\sigma}{d\Omega{}dE'}\right)_{coh} + \left(\frac{d^2\sigma}{d\Omega{}dE'}\right)_{inc}
\end{equation}
where the coherent cross section is:
\begin{eqnarray}
\left(\frac{d^2\sigma}{d\Omega{}dE'}\right)_{coh} = \frac{1}{N}\frac{k'}{k}\sum_{\lambda{}'\lambda{}}p_{\lambda}b_{coh}^2\big|\langle\lambda'|\sum_l\exp(i\mathbf{Q}\cdot\mathbf{R}_l^{})|\lambda\rangle\big|^2~\delta(\hbar\omega+E_{\lambda}-E_{\lambda'})\nonumber \\
\equiv \frac{k'}{k}b_{coh}^2~S_{coh}(\mathbf{Q},\omega) \label{eq:cohcross}
\end{eqnarray}
and the incoherent cross section
\begin{eqnarray}
\left(\frac{d^2\sigma}{d\Omega{}dE'}\right)_{inc} = \frac{1}{N}\frac{k'}{k}b_{inc}^2\sum_{\lambda{}'\lambda{}}p_{\lambda}\sum_l\big|\langle\lambda'|\exp(i\mathbf{Q}\cdot\mathbf{R}_l^{})|\lambda\rangle\big|^2~\delta(\hbar\omega+E_{\lambda}-E_{\lambda'})\nonumber \\
\equiv \frac{k'}{k}b_{inc}^2~S_{inc}(\mathbf{Q},\omega) \label{eq:inccross}
\end{eqnarray}
$S(Q,\omega)$ represents the response function or dynamic
structure factor and $\hbar\omega$ is the energy lost by the
neutron in the scattering process. The response function is given
by the Fourier transform of the density correlation function
$G(r,t)$  which represents the probability that, given a particle
located at the origin at time $t=0$, any particle is found at
position $r$ at time $t$.
\subsubsection{Elastic scattering}
If the scattering system has no internal structure (no excited states) then the energy of the scattered neutron is
identical to that of the incident neutron. Such scattering is referred to as elastic scattering.
For elastic scattering $E_{\lambda}=E_{\lambda'}$ and the total coherent and incoherent cross section reduces to:
\begin{equation}
\sigma_{coh} = 4\pi{}b_{coh}^2
\end{equation}
\begin{equation}
\sigma_{inc} = 4\pi{}b_{inc}^2
\end{equation}
In the coherent scattering there is a strong interference between the waves scattered from neighbouring nuclei.
Thus the coherent scattering is present only if strict geometrical condition are satisfied.
In the incoherent scattering there is no interference at all and the cross section is isotropic.

When a neutron with spin $I = 1/2$ interacts with a nucleus of a deuterium atom of spin $I = 1$ it can
do so in states of total spin $S = 3/2$ or $S = 1/2$. Therefore the scattering lengths
$b^{(+)}$ and  $b^{(-)}$ should be associated with these two states. We have 4 substates of spin 3/2 and 2 substates of spin 1/2,
so the statistical weight of the $S=3/2$ state is $2/3$ and that of $S=1/2$ state is $1/3$.
The scattering lengths $b_{coh}$ and $b_{inc}$ are thus averaged over these two scattering modes:
\begin{equation}
b_{coh} = \frac{2}{3}b^{(+)} + \frac{1}{3}b^{(-)} \label{cohsl}
\end{equation}
and
\begin{equation}
b_{inc} = \sqrt{\frac{2}{3}|b^{(+)}-b_{coh}|^2 + \frac{1}{3}|b^{(-)}-b_{coh}|^2}
\end{equation}
Using the measured values $b^{(+)}$ = 9.5~\femto\meter~and $b^{(-)}$ = 1.0~\femto\meter~we find $\sigma_{coh} = 4\pi{}b_{coh}^2$ = 5.59~\barn~and
$\sigma_{inc} = 4\pi{}b_{inc}^2 $= 2.04~\barn.
\subsubsection{Scattering from static inhomogeneities} \label{inhom}
If the neutron scattering takes place on nuclei that move we are dealing with an energy dependent function $S(Q,\omega)$.
For elastic scattering with a static target, where $\hbar\omega$ is zero, the response function $S(Q,\omega) = S(Q)$, where $S(Q)$ is called ``structure factor''.
All the information about the static distribution of atoms in the scattering system is then represented by the structure factor.
The scattering from static inhomogeneities with radius $R$ and uniform density difference to the surrounding medium $\delta\rho = \rho - \rho_0$ leads to~\cite{Gol2}:
\begin{equation}
S(Q) = (4\pi{}R^3\delta{}\rho{})^2\left(\frac{j_1(QR)}{QR}\right)^2
\end{equation}
where $j_1(x)$ is the spherical Bessel function of order 1.
Thus the total coherent elastic cross section is
\begin{eqnarray}
\sigma_{tot} = \int{}d\Omega\left(\frac{d\sigma}{d\Omega}\right) = \frac{2\pi{}b^2}{k^2}\int{}S(Q)QdQ \nonumber\\ \nonumber \\
= \frac{2\pi{}b_{coh}^2}{k^2}(4\pi{}R^2\delta{}\rho{})^2\int_{k\theta_1}^{2k}\frac{[j_1(QR)]^2}{Q}dQ \label{eq:elastic}
\end{eqnarray}
where $\theta_1$ is the detector aperture.
The integral in \ref{eq:elastic} is~\cite{Eng}:
\begin{equation}
\int_{k\theta_1}^{2k}\frac{[j_1(QR)]^2}{Q}dQ\equiv{}g(kR\theta_1) = \frac{1}{4}[j_0^2(kR\theta_1)+j_1^2(kR\theta_1)-j_0^2(2kR)-j_1^2(2kR)] \label{eq:factor}
\end{equation}
In section \ref{vcnresults} this model will be used to interpret the transmission data from the VCN neutron experiment.
\subsection{Inelastic neutron scattering from deuterium molecules}
\subsubsection{Deuterium molecule} \label{deutmol}
The deuterium molecule consists of two deuterons of spin 1 and thus can form states of total nuclear
spin S = 0, 1 and 2. State of S = 0 and 2 refer to ortho-deuterium and state of S = 1 refers to para-deuterium.\footnote{At room temperature the deuterium is a mixture of 66.7$\%$ ortho-D$_2$ and 33.3$\%$ para-D$_2$. Higher ortho-D$_2$ concentration up to 98.5$\%$ can be achieved by using the ortho-para converters~\cite{Bod}.}
The molecular dynamics of deuterium is determined by the motions which are represented as the translation of the
center of mass, the inter-molecular oscillations and rotations. To excite the first vibrational state (vibrational quantum number $n=1$) of deuterium molecule an energy
of 386~\milli\electronvolt~is required and thus for the low energy neutrons\footnote{Cold neutrons have energies below 10~\milli\electronvolt}
the vibrational excitations do not contribute to the
scattering cross section. The rotational energy spectrum for diatomic molecules with the moment of interia
$I = mR^2$ is given by:
\begin{equation}
E_J = \frac{\hbar^2}{2I}J(J+1)
\end{equation}
Using for the deuterium molecule the values $m$ = 3.75~\giga\electronvolt~and $R$ =
0.37~\angstrom~\footnote{The separation between the two nuclei in D$_2$ is 0.74~\angstrom} the energy becomes $E_J$ = 3.75~$J(J+1)$~\milli\electronvolt.

The state with total nuclear spin $S = 0, 2$ can only have even values of rotational quantum number $J$, and for the para state
($S=1$) only odd values of $J$ are allowed.
The population of a state, $N_J$, with the rotation quantum number $J$ follows the Boltzmann distribution:
\begin{equation}
\frac{N_J}{N_0} = (2J+1)\exp{\left(-\frac{E_J}{k_BT}\right)}
\end{equation}
where $T$ is the temperature of the system.
At temperatures where deuterium is in the solid phase
$T \le$ 18~\kelvin~the molecules are in their lowest rotational state, which is $J = 0$ for ortho-deuterium and $J = 1$ for para-deuterium.
The transitions between even and odd $J$ states are prohibited by the spin selection rules in the case of a non-interacting gas.
For the solid phase such a transition is possible due to a spontaneous spin flipping, however, such a conversion is extremely
slow compared with the time scale of experiment~\cite{Liu1,Liu3}.
\subsubsection{Properties of the sD$_2$ lattice}
To evaluate the cross section for solid deuterium one has to consider the dynamics of the lattice. Many properties of solids can
be successfully described by a simple continuum Debye model. The dynamical properties of the solid
lattice are characterised by the Debye temperature $\theta_D$, which defines an effective maximum phonon frequency $\omega_D$ in the solid:
\begin{equation}
\theta_D = \frac{\hbar\omega_D}{k_B}
\end{equation}
For solid deuterium the Debye temperature is $\theta_D$ = 110~\kelvin~\cite{Sil}.\\
The normalized phonon density of states according to the Debye model is:
\begin{equation}
Z(\omega) = \frac{3~\omega^2}{\omega_D^3}
\end{equation}
The position vector of the $l$th atom in the crystal of N atoms can be written as:
\begin{equation}
\pmb{R}_l = \pmb{\rho}_l + \pmb{u}_l(t) \label{eq:posvec}
\end{equation}
where $\pmb{\rho}_l$ represents the equilibrium position of the $l$th atom.
The displacement $\mathbf{u}_{l}(t)$ of the atoms from their equilibrium configuration
for the harmonic lattice can be written as~\cite{Gol2}:
\begin{equation}
\mathbf{u}_l(t) = \sum_{s,\mathbf{q}}\left[\pmb{\xi}_i^{}(s,\mathbf{q})a_{s,\mathbf{q}}^{}+\boldsymbol{\xi}_i^{*}(s,\mathbf{q})a_{s,\mathbf{q}}^{+}\right] \label{eq:displ}
\end{equation}
where  $a_{s,\mathbf{q}}^{}$ and $a_{s,\mathbf{q}}^{+}$ are the phonon annihilation and creation operators
\begin{equation}
a_{s,\mathbf{q}}^{}|n_{s,\mathbf{q}}\rangle=~\sqrt{n_{s,\mathbf{q}}}~|n_{s,\mathbf{q}}-1\rangle \label{eq:phonon_1}
\end{equation}
\begin{equation}
a_{s,\mathbf{q}}^{+}|n_{s,\mathbf{q}}\rangle=~\sqrt{n_{s,\mathbf{q}}+1}~|n_{s,\mathbf{q}}+1\rangle \label{eq:phonon_2}
\end{equation}
and
\begin{equation}
\pmb{\xi}_i^{}(s,\mathbf{q})=\sqrt{\frac{\hbar}{2NM\omega_s(\mathbf{q})}}~\pmb{\gamma}_s(\mathbf{q})~\exp\left(i\mathbf{q}\cdot\pmb{\rho}\right) \label{eq:phonon_3}
\end{equation}
$\omega_s(\mathbf{q})$ represents the normal-mode frequencies, the index s=1,2,3 denotes the three solution for $\omega$ for each $\mathbf{q}$
~\cite{Lov} and $\pmb{\gamma}_s(\mathbf{q})$ is related to the polarization vector.
We assume that all the atoms have the same mass $M$.
Substituting \ref{eq:posvec} and \ref{eq:displ} into the matrix element of \ref{eq:inccross} gives
\begin{eqnarray}
\lefteqn{\langle\lambda'|\exp(i\mathbf{Q}\cdot\mathbf{R}_i)|\lambda\rangle = \exp(i\mathbf{Q}\cdot\pmb{\rho}_i)}\nonumber\\\nonumber\\
& & \times\prod_{s,\mathbf{q}}\langle\lambda'|\exp\left\{i\mathbf{Q}\cdot\left[\pmb{\xi}_i^{}(s,\mathbf{q})a_{s,\mathbf{q}}^{}+
\boldsymbol{\xi}_i^{*}(s,\mathbf{q})a_{s,\mathbf{q}}^{+}\right]\right\}|\lambda\rangle \label{eq:phonon_4}
\end{eqnarray}
Applying a Taylor expansion of the exponential function in the operator we get
\begin{eqnarray}
\exp\left\{i\mathbf{Q}\cdot\left[\pmb{\xi}_i^{}(s,\mathbf{q})a_{s,\mathbf{q}}^{}+
\boldsymbol{\xi}_i^{*}(s,\mathbf{q})a_{s,\mathbf{q}}^{+}\right]\right\} = 1+i\mathbf{Q}\cdot\left[\pmb{\xi}_i^{}(s,\mathbf{q})a_{s,\mathbf{q}}^{}+
\boldsymbol{\xi}_i^{*}(s,\mathbf{q})a_{s,\mathbf{q}}^{+}\right]+\nonumber\\\nonumber\\
-\frac{1}{2}\left[(\mathbf{Q}\cdot\pmb{\xi}_i)^2a_{s,\mathbf{q}}^{2}+(\mathbf{Q}\cdot\pmb{\xi}_i)^2(a_{s,\mathbf{q}}^{+})^2+|\mathbf{Q}\cdot\pmb{\xi}_i|^2
(a_{s,\mathbf{q}}^{}a_{s,\mathbf{q}}^{+}+a_{s,\mathbf{q}}^{+}a_{s,\mathbf{q}}^{}\right]+\ldots\nonumber\\
\end{eqnarray}
Since $a_{s,\mathbf{q}}^{}$ and $a_{s,\mathbf{q}}^{+}$ annihilate and create, respectively, phonons, we see that the second term changes the phonon number by one
while the third term changes the phonon number by two or zero.\\
To calculate the one-phonon contribution to the neutron cross section we can use \ref{eq:phonon_1}, \ref{eq:phonon_2} and \ref{eq:phonon_3} and write \ref{eq:phonon_4} as:
\begin{eqnarray}
\langle\lambda'|\exp(i\mathbf{Q}\cdot\mathbf{R}_i)|\lambda\rangle = \exp{[-W(\mathbf{Q})]}
\sqrt{\frac{\hbar}{2NM\omega_s(\mathbf{q})}}~i\mathbf{Q}\cdot\pmb{\gamma}(s,\mathbf{q})\nonumber\\\nonumber\\
\times\left\{\begin{array}{ll} \exp{[i\pmb{\rho}_i\cdot(\mathbf{Q}-\mathbf{q})]}\sqrt{n_{s,\mathbf{q}}+1} & \textrm{phonon creation}
\\\\ \exp{[i\pmb{\rho}_i\cdot(\mathbf{Q}+\mathbf{q})]}\sqrt{n_{s,\mathbf{q}}}&\textrm{phonon annihilation}\end{array}\right. \label{eq:phonon_5}
\end{eqnarray}
where the term $\exp[-W(\mathbf{Q})]$, the Debye-Waller factor, describes the zero phonon expansion.
For a cubic lattice $W(\mathbf{Q})$ is~\cite{Lov}:
\begin{equation}
W(\mathbf{Q})=\frac{1}{6}Q^2\langle\mathbf{u}^2\rangle
\end{equation}
where $\langle\mathbf{u}^2\rangle$ is the mean squared displacement of a nucleus.\\
Using \ref{eq:phonon_5} the incoherent double differential cross section of \ref{eq:inccross} becomes:
\begin{eqnarray}
\left(\frac{d^2\sigma}{d\Omega{}dE'}\right)_{inc} = \frac{k'}{k}b_{inc}^2\sum_{s,\mathbf{q}}\frac{\hbar}{2NM\omega_s(\mathbf{q})}
\exp[-2W(\mathbf{Q})]|\mathbf{Q}\cdot\pmb{\gamma}_{s}(\mathbf{q})|^2\nonumber\\\nonumber\\
\times\left\{\begin{array}{ll} [n_{s}(\mathbf{q})+1]~\delta(\hbar\omega_{s,\mathbf{q}}-\hbar\omega) & \textrm{phonon creation}
\\\\ n_{s}(\mathbf{q})~\delta(\hbar\omega_{s,\mathbf{q}}+\hbar\omega)&\textrm{phonon annihilation}\end{array}\right.
\end{eqnarray}
Finally, after replacing the summation over s,q by the normalized density of states~\cite{Lov} one obtains:
\begin{eqnarray}
\left(\frac{d^2\sigma}{d\Omega{}dE'}\right)_{inc} = \frac{3}{2M}b_{inc}^2\frac{k'}{k}\exp[-2W(\mathbf{Q})]
\left\{|\mathbf{Q}\cdot\pmb{\gamma}_{s}(\mathbf{q})|^2\right\}_{av}\nonumber\\\nonumber\\
\times\frac{Z(\omega)}{\omega}\left\{\begin{array}{ll} n(\omega)+1 & \textrm{if $\omega\geq 0$,  phonon creation}
\\\\ n(\omega)&\textrm{if $\omega<0$,  phonon annihilation}\end{array}\right.  \label{eq:phonon_6}
\end{eqnarray}
where $\hbar\omega$ is the energy lost by the neutron.
The subscript $av$ in~\ref{eq:phonon_6} stands for the  average over a surface in $\mathbf{Q}$ with constant $\omega$ and in cubic symmetry this average
is equal to $\frac{1}{3}Q^2$.
The coherent scattering is usually small because it is restricted by the energy and momentum conservation and very often it is sufficient to
use the ``incoherent approximation'', which consists of using~\ref{eq:phonon_6} with $b_{inc}^2$ replaced by $b_{coh}^2+b_{inc}^2$.
The validity of ``incoherent approximation'' is discussed in~\cite{Liu1}.

We can now apply the resulting formula~\ref{eq:phonon_6} to the solid ortho-deuterium system and find an expression for the cross section.
Deuterium crystallises in the hcp structure with an intermolecular separation
of 3.7~\angstrom. At low temperatures ( $T \le$ 18~\kelvin~) the vibrational $n$ and rotational $J$ quantum numbers
are zero. Thus the ortho-deuterium molecule is spherically symmetric and may be considered as a
single particle with a neutron scattering length $2b_{coh}j_0(\frac{Qa}{2})$~\cite{Nie2} where
$b_{coh}$ is the coherent scattering length of the deuteron (Eq.~\ref{cohsl}~). $j_0$ is a spherical
Bessel function, $a$ denotes the separation between the deuterons in the molecule ($a$ = 0.74~\angstrom) and $Q$ is the momentum transfer:
\begin{equation}
Q = \sqrt{k^2+(k')^2-2kk'\cos{\theta}}
\end{equation}
where $\theta$ is the angle between $\mathbf{k}$ and $\mathbf{k'}$.
Following Liu et al.~\cite{Liu3} we can use a cubic lattice to describe the solid deuterium structure so that
\begin{eqnarray}
\left(\frac{d^2\sigma}{d\Omega{}dE'}\right)_{inc}^{1~phonon} = \frac{1}{2M_{D_2}}~\left[2b_{coh}j_0\left(\frac{Qa}{2}\right)\right]^2~\frac{k'}{k}\exp[-2W(\mathbf{Q})]
~Q^2\nonumber\\\nonumber\\
\times\frac{Z(\omega)}{\omega}\left\{\begin{array}{ll} n(\omega)+1 & \textrm{if $\omega\geq 0$,  phonon creation}
\\\\ n(\omega)&\textrm{if $\omega<0$,  phonon annihilation}\end{array}\right. \label{eq:upscatt}
\end{eqnarray}
The population number $n(\omega)$ of phonons with the frequency
\begin{equation}
\omega = \frac{\hbar}{2M_n}Q^2
\end{equation}
is given by Bose-Einstein statistics
\begin{equation}
n(\omega) = \frac{1}{\exp\left(\frac{\hbar\omega}{k_BT}\right)-1}.
\end{equation}
Figure \ref{upscatt} shows the up-scattering cross section for VCN
calculated from \ref{eq:upscatt}. \epsbild{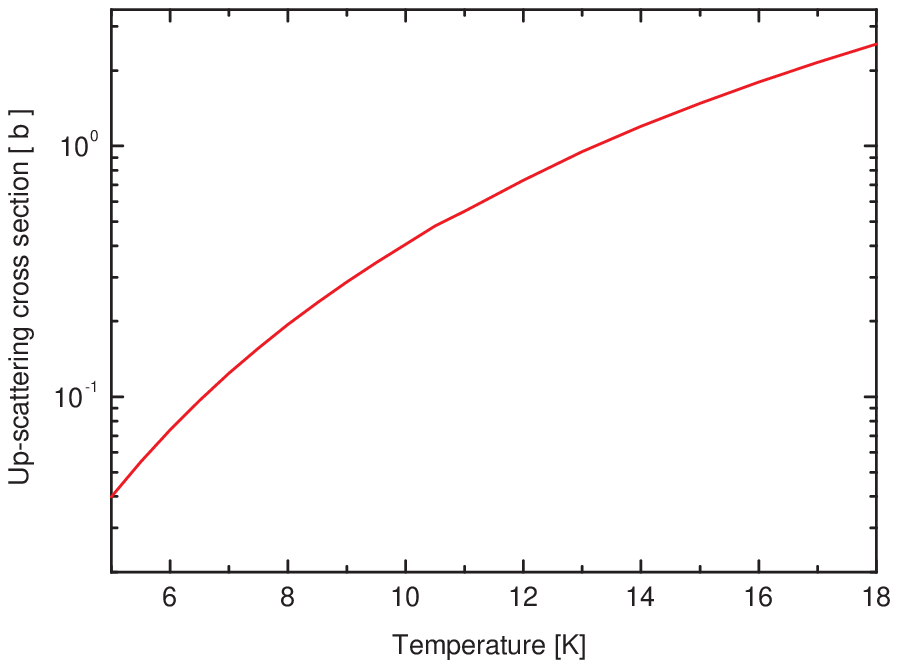}{The
calculated VCN up-scattering cross section vs. temperature of
solid D$_2$ with an initial neutron energy of
$7.9\cdot{}10^{-2}$~\milli\electronvolt, corresponding to a
velocity of 66~\metre\per\second~or a wavelength of 60~\angstrom.
\label{upscatt}}{1.0}{0}{!htb}

\chapter{Experiment}
\section{Experimental setup}
The experiment was carried out at the Intitute Laue-Langevin (Grenoble, France) on the VCN and UCN
beam-line during 2003. A total of 2 reactor cycles (reactor cycle 135 and 136,~50 days each) had been
allocated for these experiments. Reactor cycle 135 was devoted to the VCN measurements and
reactor cycle 136 was devoted to the measurement on the UCN test position.
\subsection{General overview of the VCN and UCN setup} \label{setup}
Figure \ref{setupvcn} shows a schematic view of the experimental
setup at the VCN beam position. Neutrons with velocities of
20~\metre\per\second~to 200~\metre\per\second~pass the setup as
discrete packets created by a neutron chopper and are detected in
a position sensitive micro-strip gas counter. The sD$_2$ crystal
was grown in the target cell with a volume of
40~\centi\cubic\metre. The cryostat with the target cell was
placed at the end of the neutron flight path between the chopper
and detector. During the experiment the detector was located at
two positions: ``front'' and ``extended''. For the ``front''
detector position the total length of the flight
path\footnote{Distance between the chopper and the middle of the
neutron detector} is 1434~\milli\metre. The collimator assembly of
a length of 1260~\milli\metre~consists of: the
20~\milli\metre~diameter boron collimator placed directly after
the chopper, the 20~\milli\metre~diameter cadmium collimator
placed before the target cell and four Cd/Cu collimators
(Fig.~\ref{coll}) of total length 150~\milli\metre~placed
600~\milli\metre~after the chopper. The reasons for having the
detector at two position are the calibration of time-of-flight
spectrum and the scattering measurements at small scattering
angles.

A similar setup was used for the UCN experiment (Fig.~\ref{setupucn}). For these measurements the cryostat with the sD$_2$ crystal
was placed in the flight path of length 1247~\milli\metre~directly after the chopper.
For the neutron scattering measurements a collimator with a length of 30~\milli\metre~and a diameter of 20~\milli\metre~was placed upstream of the chopper.
After passing through the cell neutrons enter
a neutron guide that consist of a movable tube. With this tube setup, scattering measurements at various acceptance solid angles could be performed.\\
\epsbild{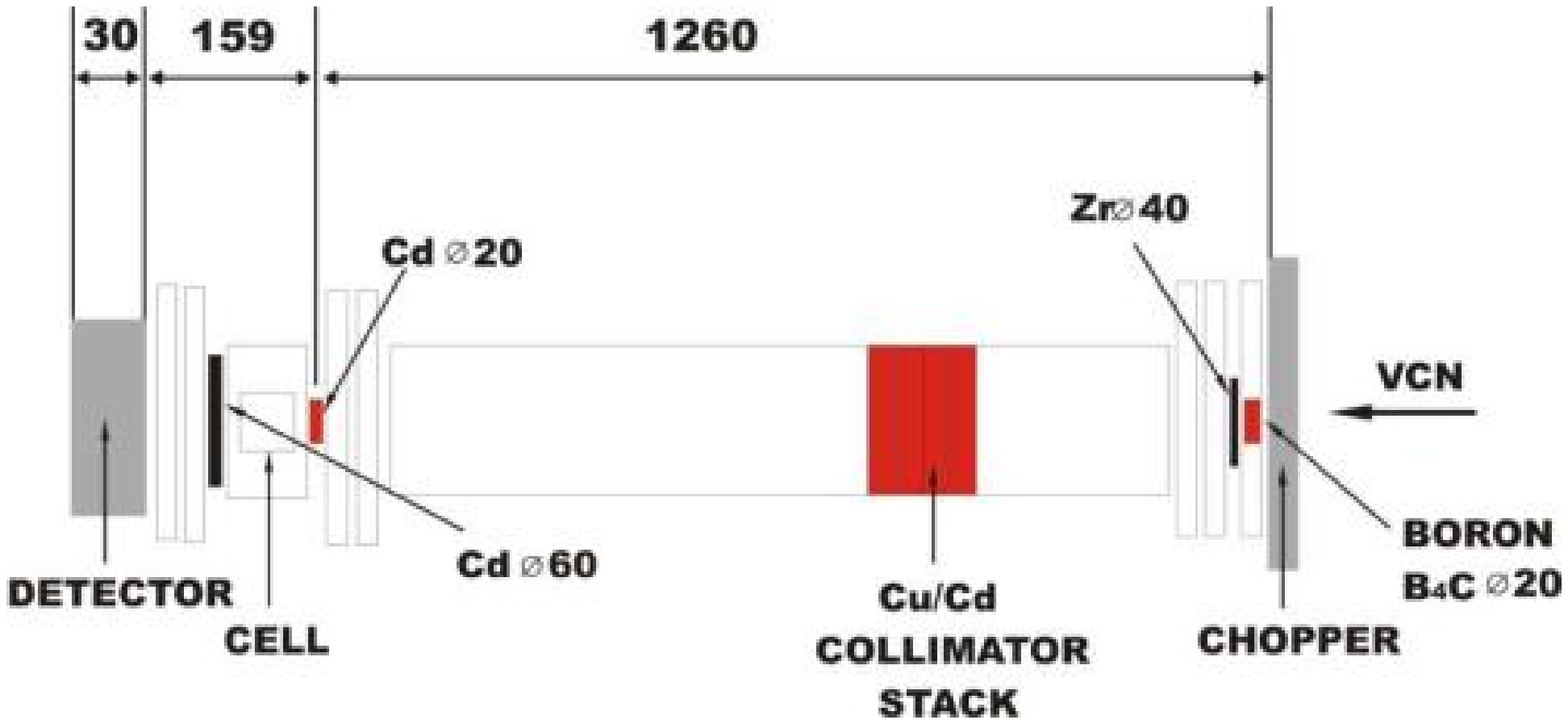}{The scheme of the VCN setup, all dimensions are given in~\milli\metre. The parts of collimation assembly are marked with red. The distance between the middle of the cell and the detector is 110~\milli\metre. \label{setupvcn}}{1.2}{90}{!htb}
\vspace{10cm}
\epsbild{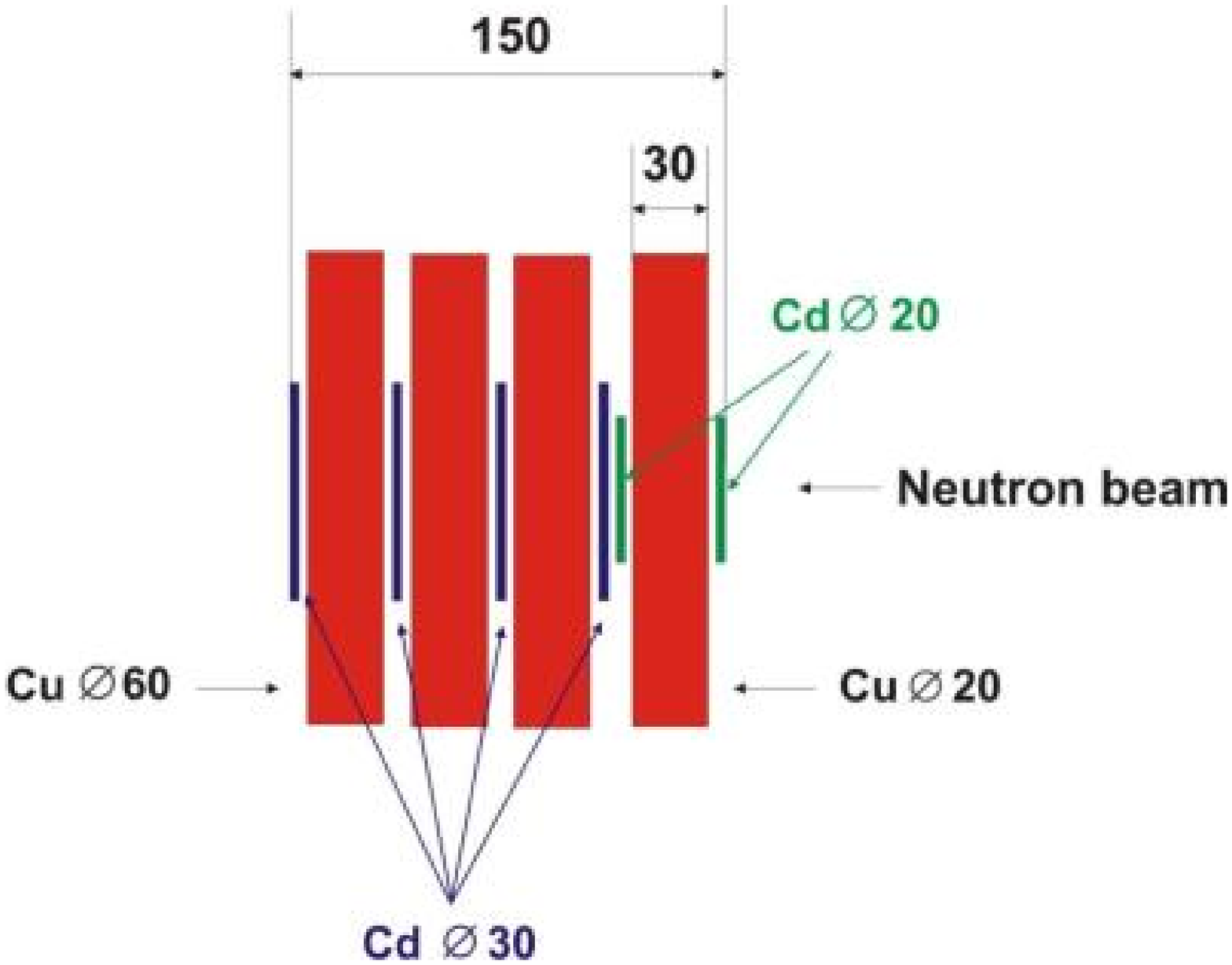}{The collimator stack which consists of four Cd/Cu collimators, all dimensions are given in~\milli\metre. The given diameters are
the inner diameters.\label{coll}}{1.0}{0}{tb}
\epsbild{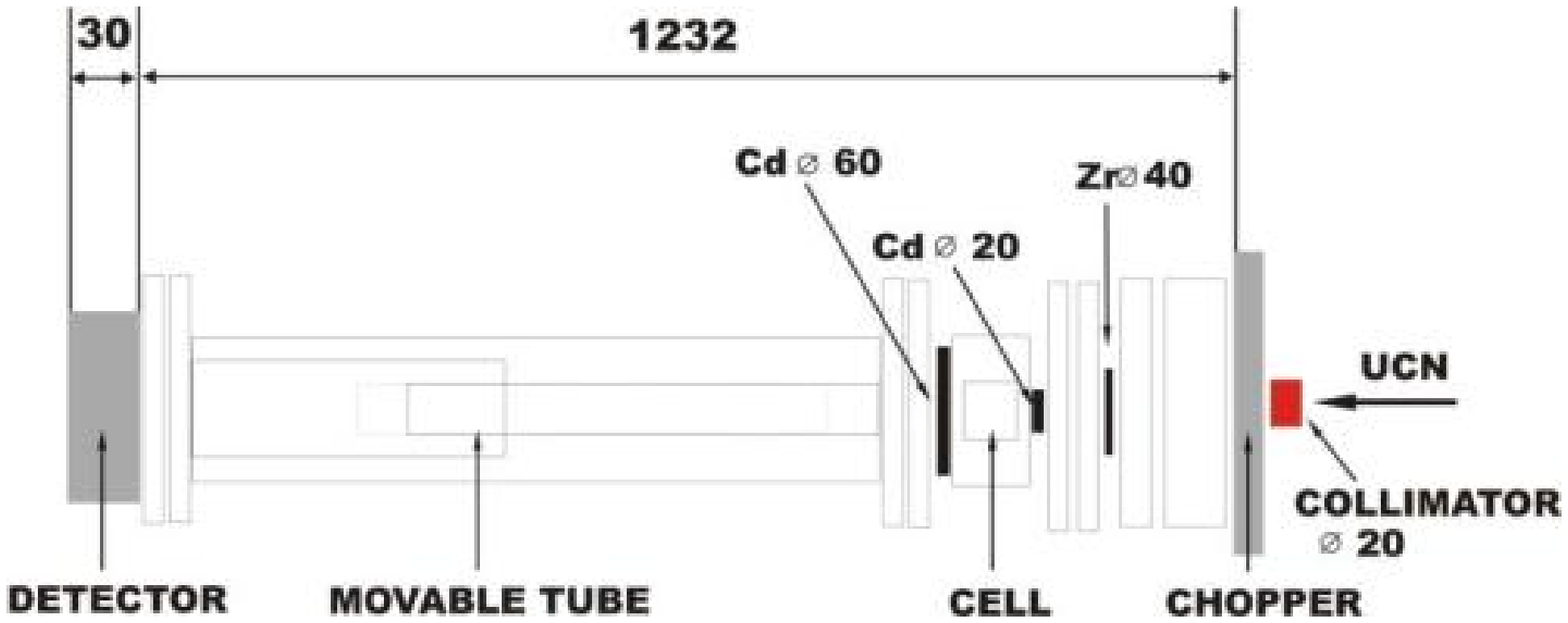}{The scheme of the UCN setup, all dimensions are given in~\milli\metre. \label{setupucn}}{1.2}{90}{!hbt}
\clearpage
\subsection{The cryostat and the target cell}
In this section the general concept of the target cell and the cryostat will be described together with the method
of preparing the crystal. Further details can be found elsewhere~\cite{Bod}.
Figure~\ref{cryostat} shows the cryostat designed to obtain
temperatures down to 5~\kelvin~in the sD$_2$ cell. These temperatures are required to study solid deuterium transmission
properties for the superthermal UCN source.

\epsbild{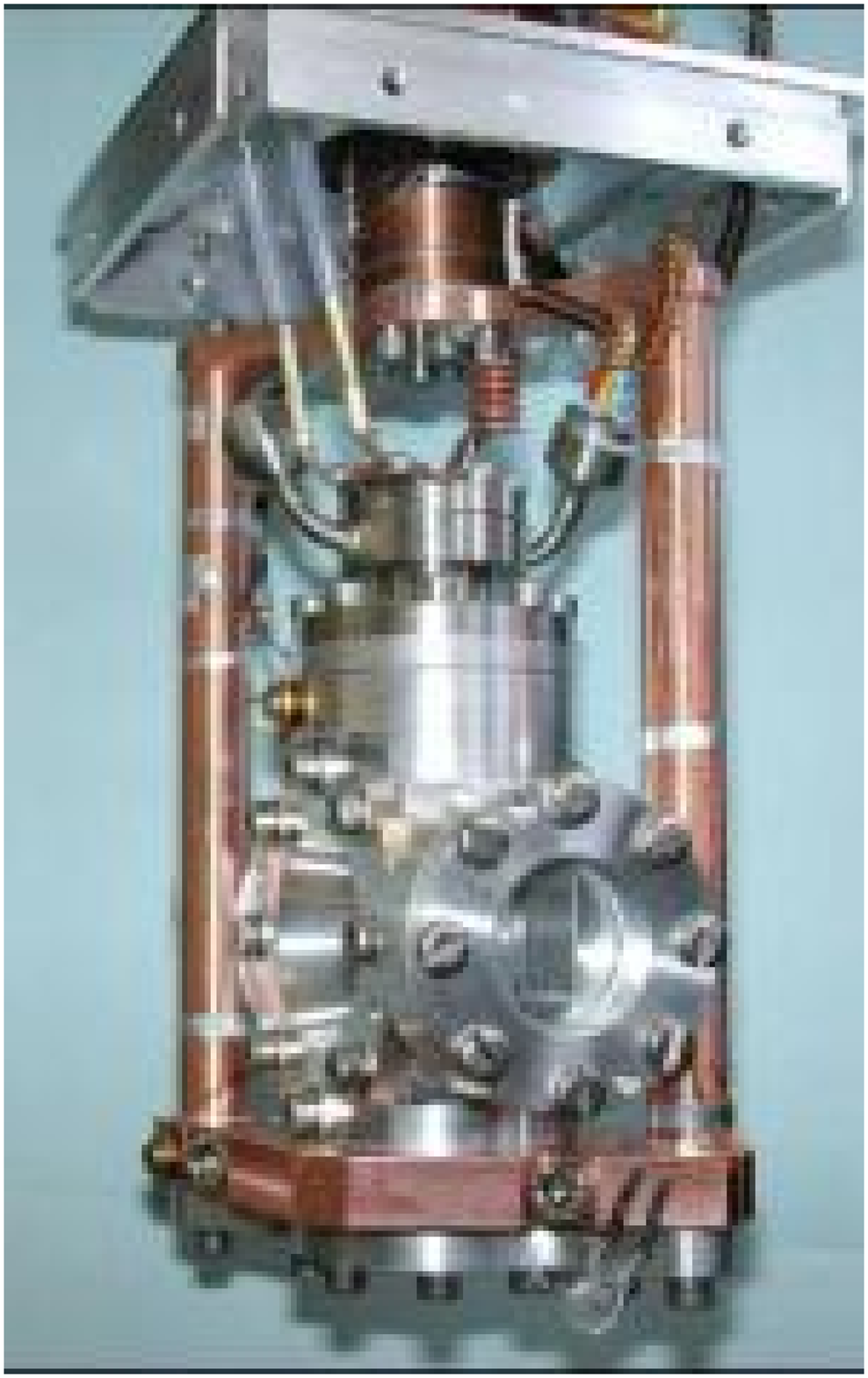}{The sD$_2$ cell mounted on the cryostat \label{cryostat}}{0.6}{0}{!htb}
The cell is made of an aluminum alloy AlMg$_3$ and has four windows: two optical windows and two windows for neutrons.
The optical windows, made from sapphire, serve for Raman spectroscopy as well as for
optical investigation of the  crystal.
The neutron windows, made from aluminium, are 0.15~\milli\metre~thick and have 26~\milli\metre~diameter. The distance between the entrance
window and the exit window is 10~\milli\metre.
\subsubsection{Method of freezing} \label{freezing}
The cell is designed in such a way that the crystal is frozen from the liquid phase. A high-resolution digital camera permanently monitors this process.
Gaseous D$_2$ kept in the aluminium storage container
is passed through the gas system into the cryostat and is subsequently liquefied in the target cell. The process of freezing is carried
out at a temperature close to the triple point (18.7~\kelvin), this procedure takes about 12~\hour.
The crystal that is obtained under these conditions is fully transparent and clear.
In Fig.~\ref{crystal} a series of pictures shows how the crystal is being grown from the melt.\\
\epsbild{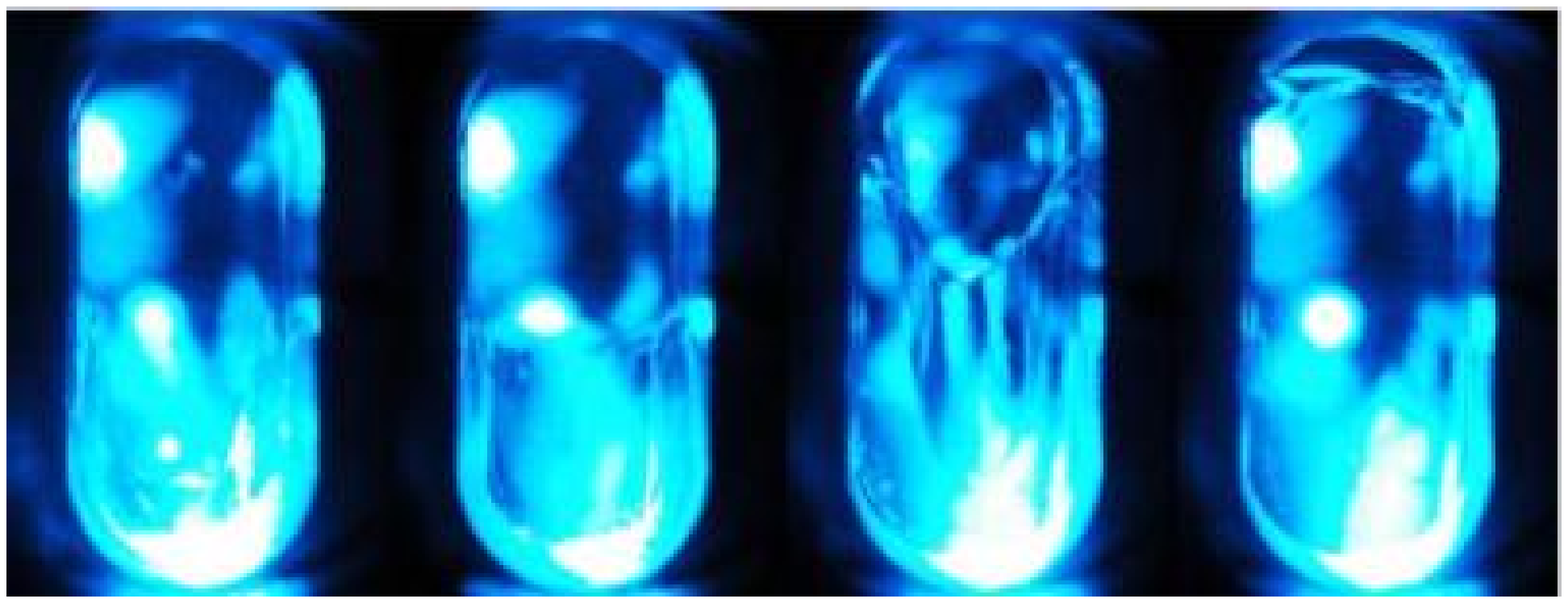}{A series of pictures showing the crystal being grown from the melt\label{crystal}}{1.}{0}{!hbt}

The speed at which the deuterium is frozen is a very important parameter in this process.
Figure \ref{slowfast} shows crystals grown at two cooling speeds and demonstrates the dependency of the crystal quality on the speed of cooling.
Because of a temperature gradient that arises during the filling of the cell the deuterium in the upper part of the cell will
be frozen before the complete crystal is formed. This will affect the flow of the rest of the deuterium as it moves into the cell
and can cause problems to the formation of the remaining part of the sample. However, we have developed an annealing technique in which the deuterium
is kept at a constant temperature for several hours. This annealing procedure has greatly improved the crystal homogeneity.
Figure \ref{anneal} shows the annealing process of a crystal as it is grown in the cell.
Once the crystal reached a temperature of 18~\kelvin~it was cooled down further to a temperature of 5~\kelvin. This last step in the process of
creating the final form of the solid deuterium was performed at a reduced cooling speed: in most cases with
a typical cooling speed of 2~\kelvin\per\hour.

The interpretation of the data is based on the measured neutron
transmission through the sample. It is thus important to know
accurately the amount of deuterium of the neutron beam passage.
The deuterium concentration as a function of position in the cell
is obtained by analysing the pictures as recorded by the digital
camera (section \ref{camera}). \epsbild{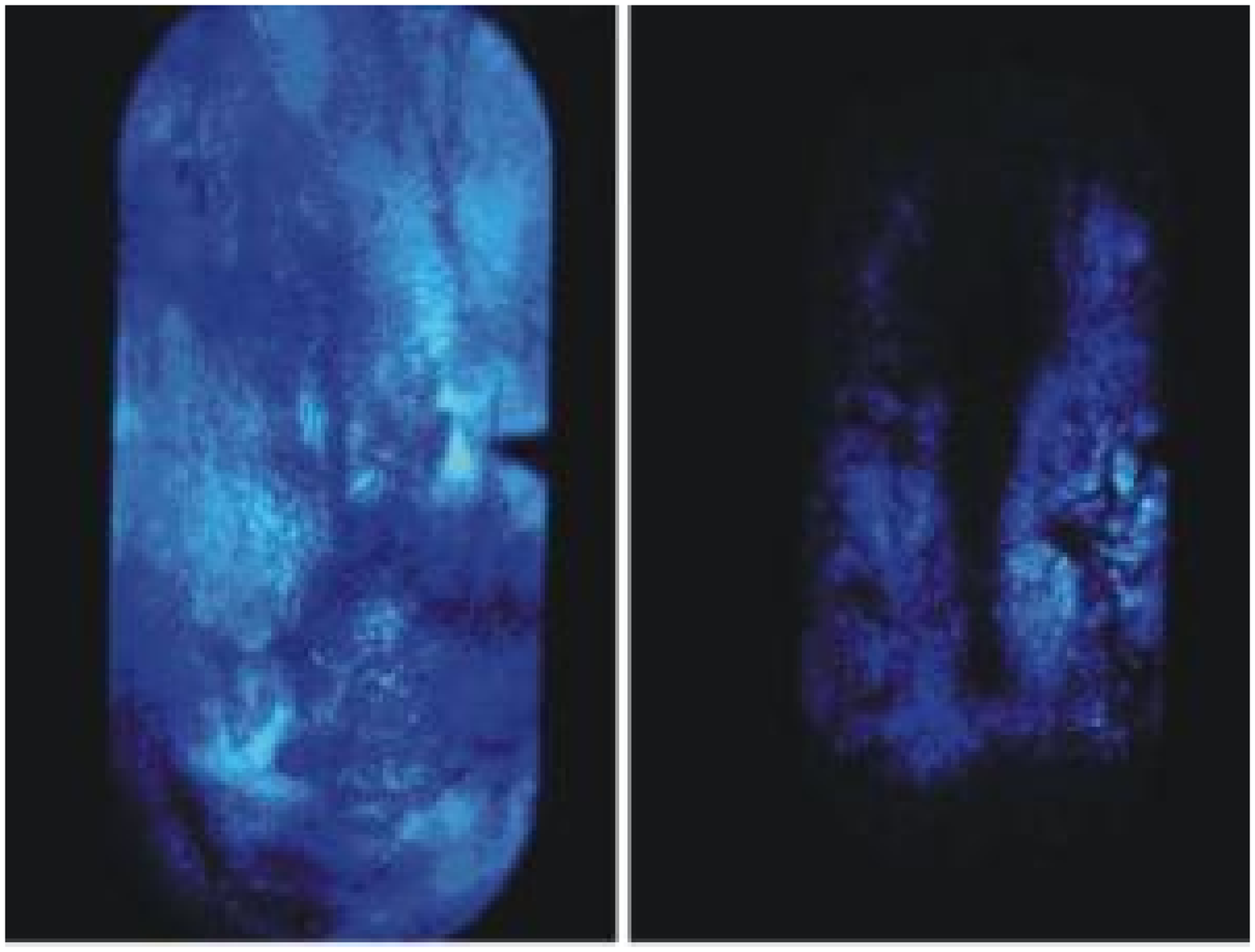}{Two
crystals at 5~\kelvin~that have been grown at two cooling speeds.
The crystal pictured on the right was grown at a speed of
2~\kelvin\per\hour~and the crystal pictured on the left was grown
at a speed of 14~\kelvin\per\hour.\label{slowfast}}{0.5}{0}{!htb}
\epsbild{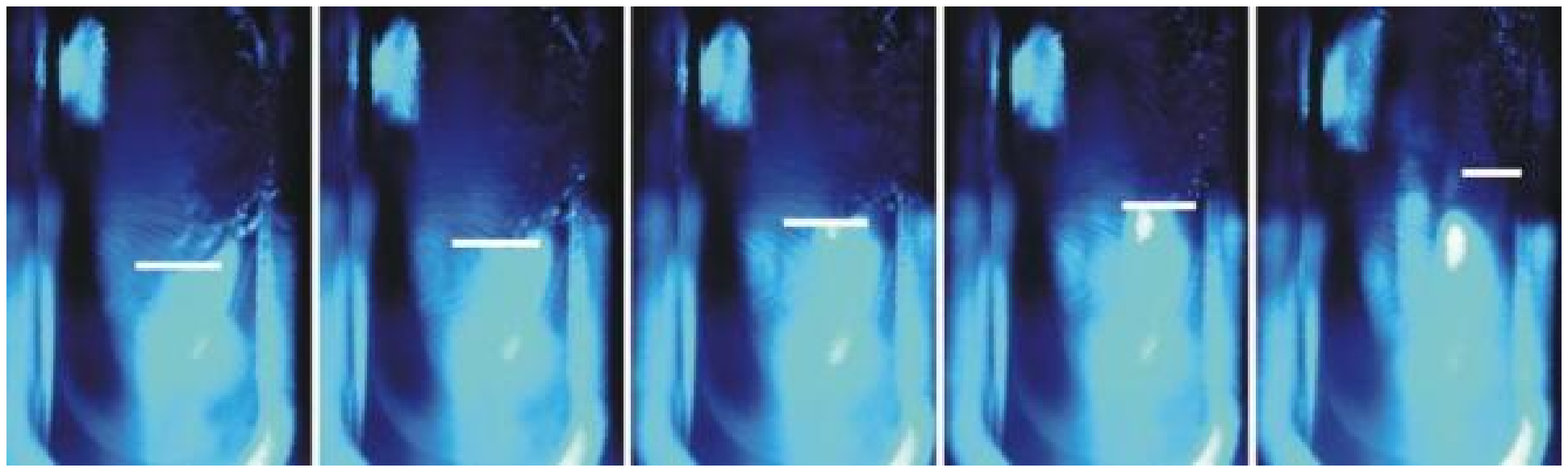}{The annealing process of a crystal, the
white line marks how the 'empty part' of the growing crystal is
filled with time. Shown is,  from left to right,  1 - start of
annealing, 2 - after 5~\hour, 3 - after 15~\hour, 4- after
30~\hour, 5 - stop  of annealing (40~\hour)
\label{anneal}}{1.0}{0}{!htb}
\subsection{The Raman spectroscopy setup}
The system which is used for Raman spectroscopy consists of three main parts: the argon-ion-laser, the Raman head, and the Raman spectrometer.
A detailed description of the setup and method of the data analysis can be found in Ref.~\cite{Bod}.

Raman spectroscopy has been used to investigate the ratio of ortho-D$_2$ to para-D$_2$ as well as the crystal structure.
Hydrogen contaminations in form of HD or H$_2$ can also be measured.
The ortho-para concentration can be calculated from the intensity ratio of the S$_0$(0) and S$_0$(1) lines of the rotational Raman spectrum (section \ref{craman})
while the lattice structure of the solid deuterium can be determined from the S$_0$(0) signal which produces a multiplet unique to the crystal structure.
From the relative line intensity within the multiplet one can deduce the crystal (crystallites) orientation with respect to the direction of the Raman collector (section \ref{craman}).
Figure \ref{raman} shows the Raman head mounted on the cell.
\epsbild{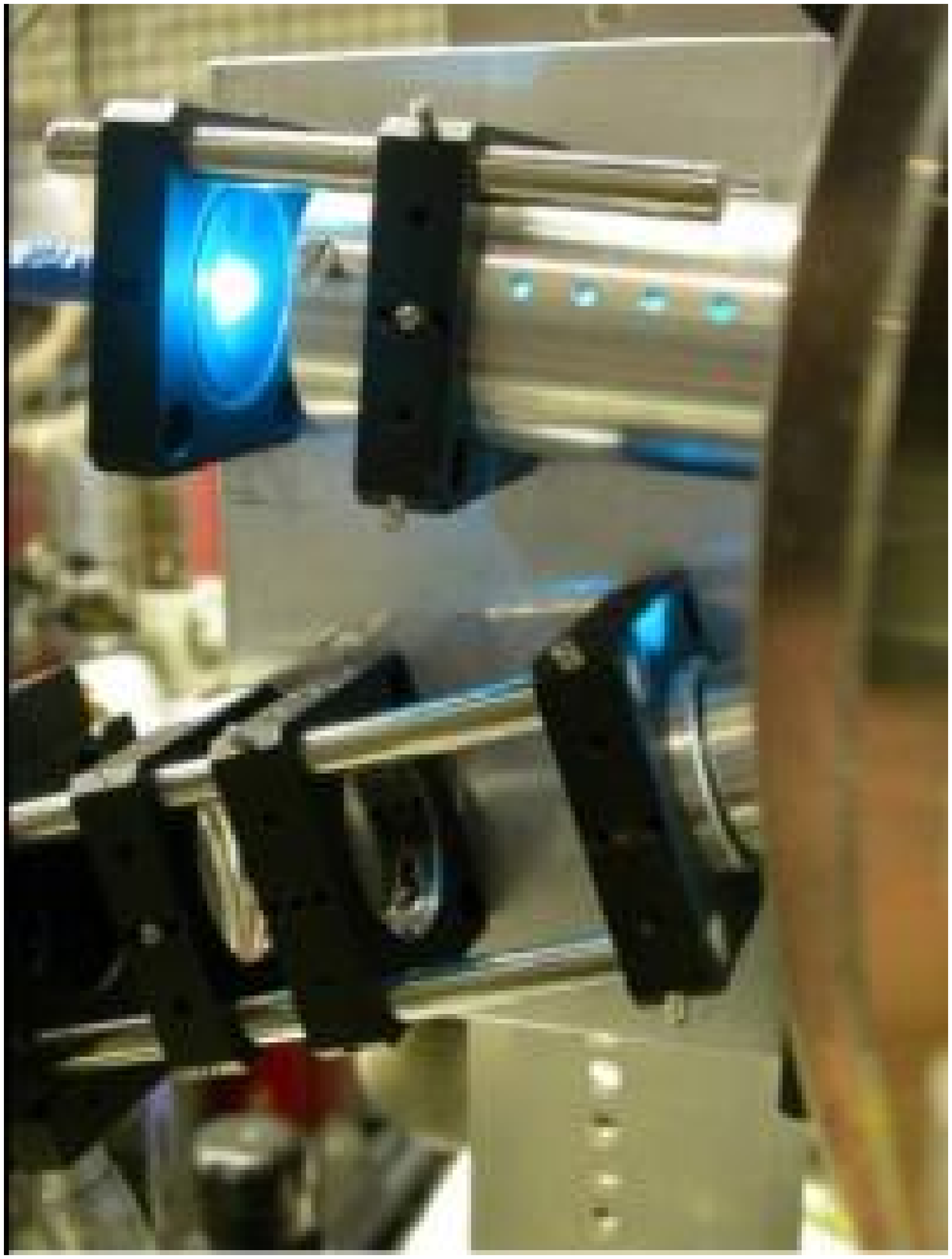}{The Raman head consists of the laser head (upper) for the light input and Raman collector (lower) for the scattered light. The Ar laser light is coupled to the laser head via optical fiber. \label{raman}}{.5}{0}{!htb}
\subsection{The Camera} \label{camera}
Throughout the measurements the digital camera is permanently mounted on the cell (Fig.~\ref{nikon}). The camera allowed to monitor
both the  optical properties and the total amount of deuterium that is formed. More information can be found in Ref.~\cite{Bod}.

The pictures generated by the camera are used to study the optical difference for a specific crystal at different temperatures
and prepared in different ways.
\epsbild{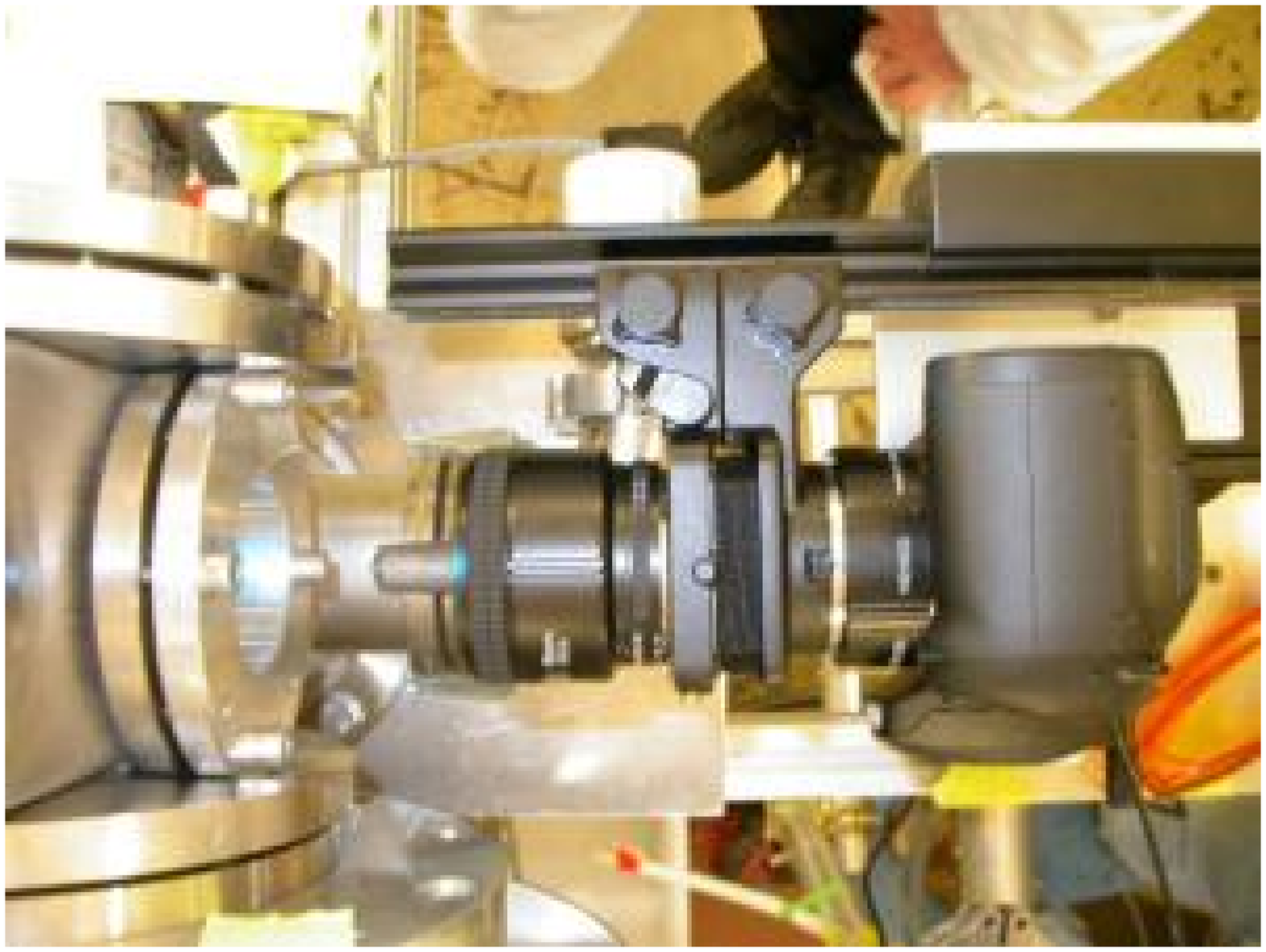}{Top view of the camera mounted close to the one of the optical windows; it allows to monitor the optical properties of the crystal. \label{nikon}}{.8}{0}{!htb}
\subsection{VCN/UCN detector and DAQ}
The same neutron detector and DAQ system has been used for the VCN transmission and the UCN transmission experiments.
Although the efficiency of the detector varies with the neutron energy, relative measurements are obtained directly by comparing
the neutron transmission through the different samples and measurements taken with the cell being empty.
\subsubsection{Detector}
The neutron detector, a so-called ``Bidim80'' developed at the ILL, is a 2-dimensional micro-strip detector that consist of an MSGC plate
placed in a sealed cylindrical metal container filled with a mixture of 50~\milli\bbar~of $^3$He and 1~\bbar~of CF$_4$.
The entrance window to the detector is a 100~\micro\metre~thick Al foil.
Because of the 50~\milli\bbar~overpressure the detector window was slightly bulged but this has no noticeable effect on the final results.
A part of the detector is shown in Fig.~\ref{bidim}: the MSGS plate and an aluminium diaphragm are mounted on a stainless steel flange
that is equipped with a valve through which the gas can be filled. The two
electrical feed-throughs mounted on a flange are used to apply the voltage onto the electrodes and to extract the neutron signal.

The MSGC is made of a semiconductive glass plate with a chromium pattern of strips, the total sensitive area
is  80x80 mm$^2$. The 10~\micro\metre~wide anodes are engraved on the front side of the plate, the  980~\micro\metre~cathodes are placed
on the rear side of the plate orientated perpendicular to the anodes. The spatial resolution for neutron detection is about 1.5~\milli\metre.
The X and Y position of the detected neutron is determined from pulse height asymmetries of signals from the electrodes in the following way:
\begin{equation}
X = \frac{A_0-A_1}{A_0+A_1}\cdot4~\centi\metre
\end{equation}
\begin{equation}
Y = \frac{A_2-A_3}{A_2+A_3}\cdot4~\centi\metre
\end{equation}
The position of the signal on the detector was obtained by
performing measurements with a cadmium mask (Fig.~\ref{mask})
placed on the front flange of the detector. Figure~\ref{maskview}
shows a two dimensional picture from the detector with the Cd mask
mounted. The distance between the window and the MSGC plate is
30~\milli\metre~(for a flat window).

The high voltage applied between the entrance window and
the anodes and cathodes is 1200~\volt. Neutron are detected through the reaction
$^3$He(n,p)$^3$H. The energy release in this reaction is 764~\kilo\electronvolt~and the cross section is 5330~\barn~for thermal neutrons.
Scaling the cross section to VCN and UCN energies gives a reaction cross section of the order of mega-barns and forms a very
efficient basis for neutron detection. The proton and triton emerging from the reaction will create electrons by ionising
the gas mixture. The electrons drift in the electric field toward the MSGC plate. On approaching the glass plate these electrons
are accelerated in very strong electric field in the vicinity of the anodes leading to secondary ionisation and creation of
avalanche. The electrons penetrate the glass plate and produce electron-hole pairs along its track, the number being proportional
to the energy loss. The electric field applied between the anodes and cathodes separates the pairs before they recombine; electrons drift towards the anode,
holes to the cathodes on the rear side. Due to the semiconductive properties of the glass, the signal from
the rear side is equal to the one on the front side.
The electric pulses thus generated are passed through a pre-amplifier before being converted by ADCs\footnote{Port Modules
and the MPA-3 BASE module}. A typical pulse height spectrum is shown in Fig.~\ref{pulses}.
\epsbild{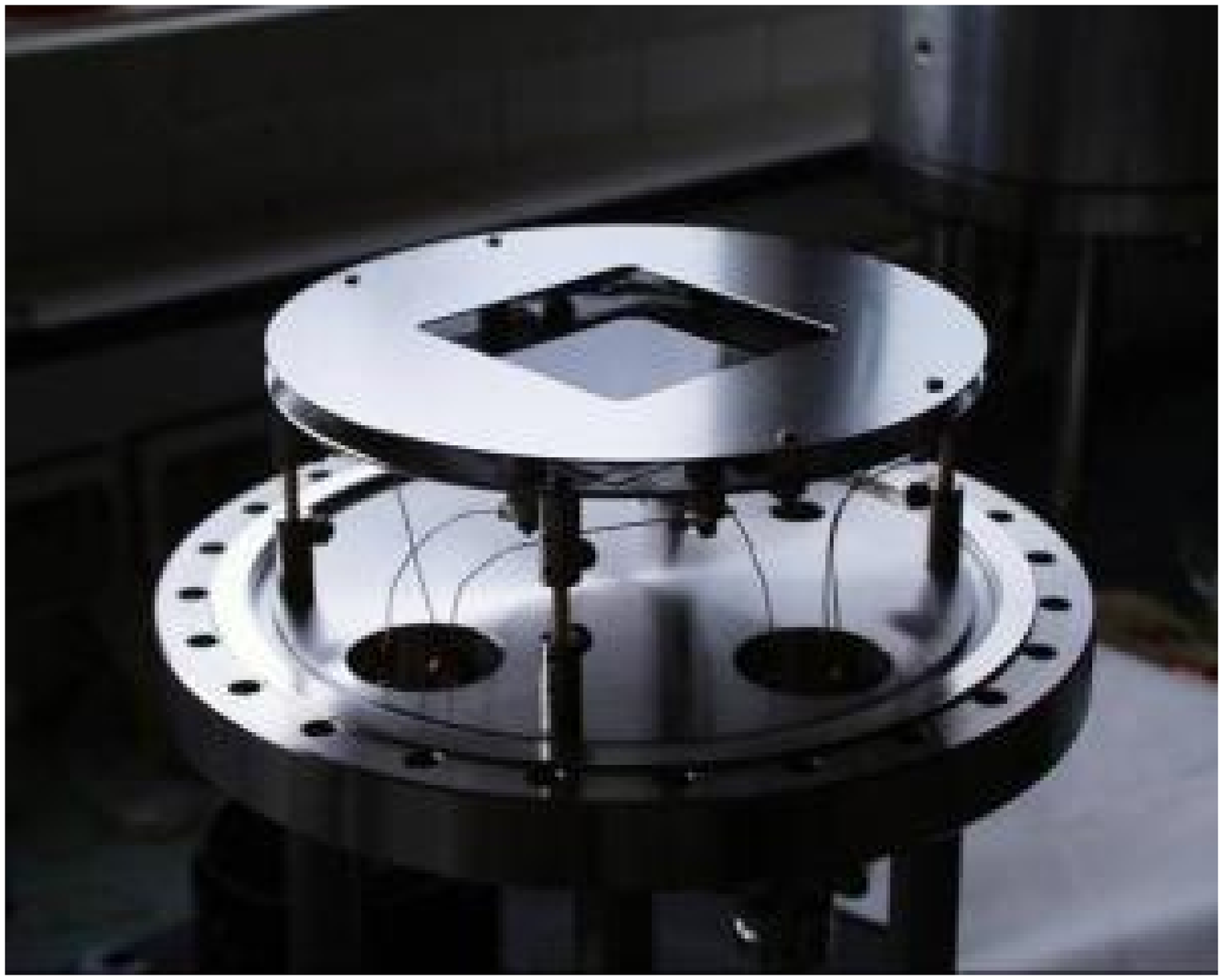}{The MSGC plate and an aluminium diaphragm of the ``Bidim80'' detector. \label{bidim}}{.8}{0}{htb}
\epsbild{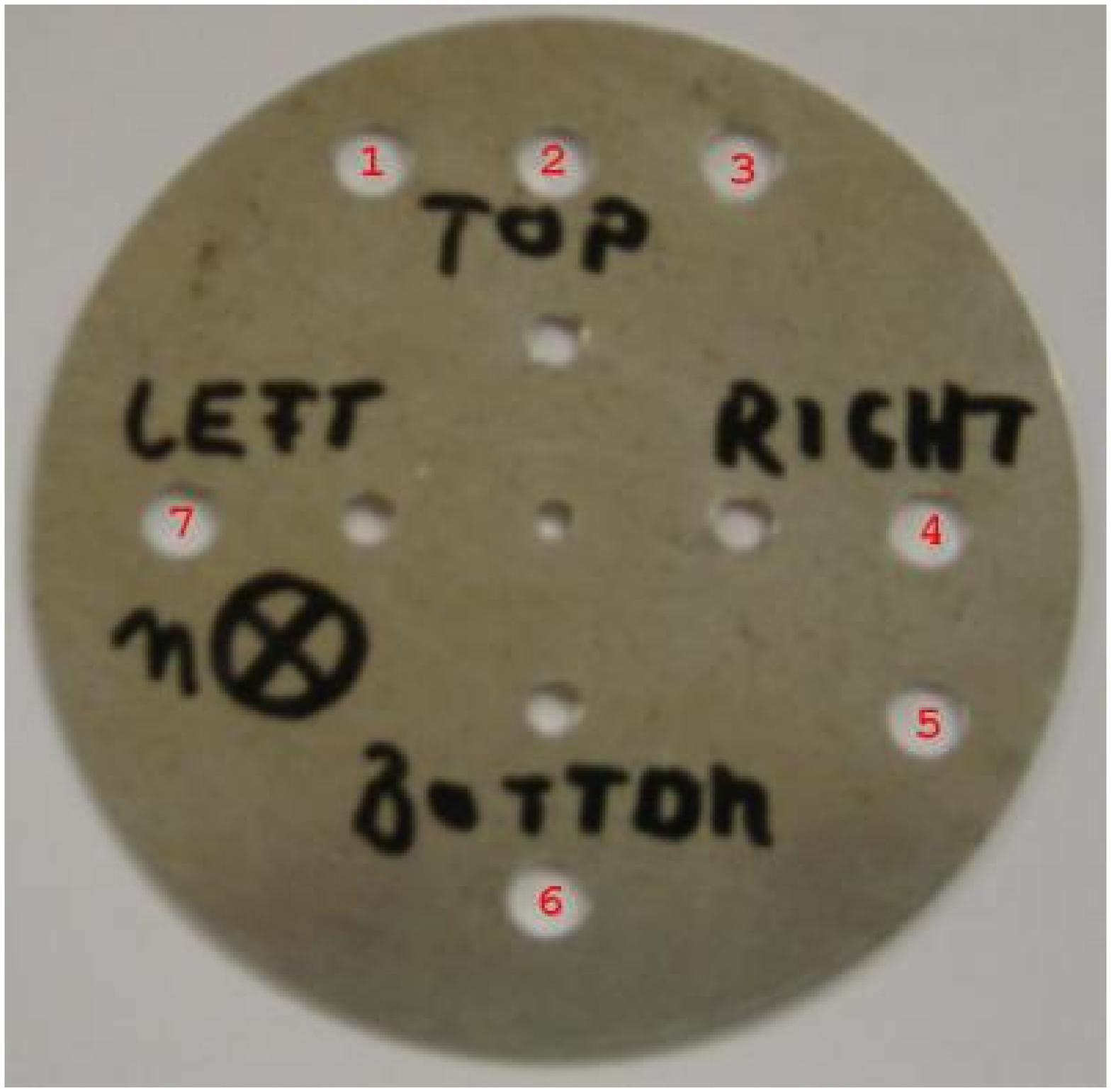}{The cadmium mask that is used to determine the position of the signal on the detector. \label{mask}}{.5}{0}{!hbt}
\epsbild{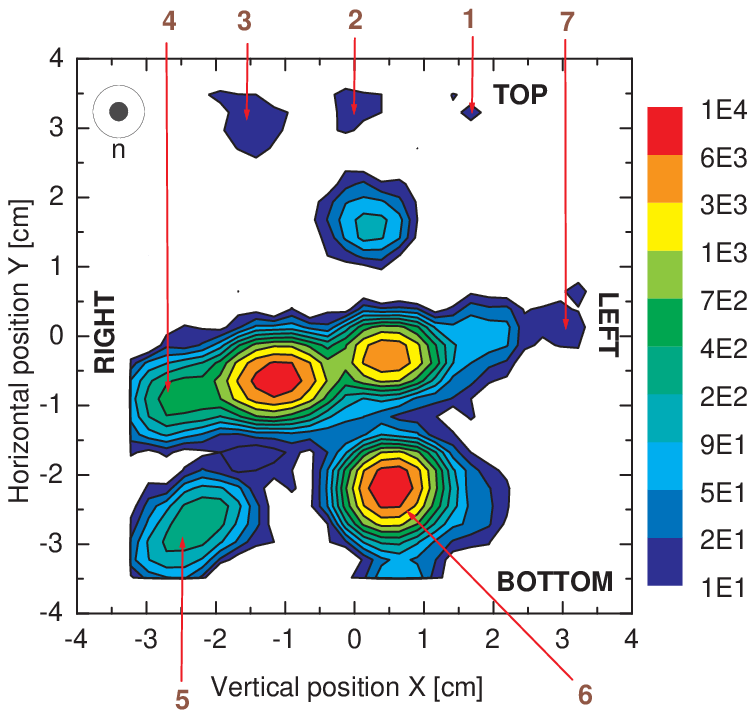}{A two dimensional picture from the detector with the Cd mask mounted, the numbering is the same as for Fig.~\ref{mask}. \label{maskview}}{.8}{0}{!htb}
\epsbild{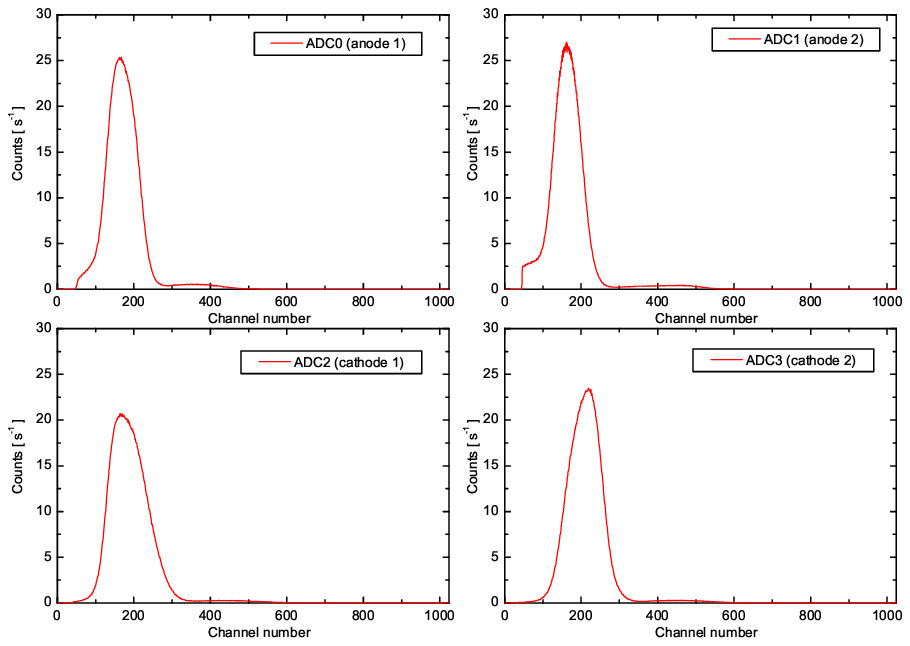}{Pulse height spectra as obtained from the measurement with the empty cell at the ``front'' detector position.\label{pulses}}{1.0}{0}{htb}
\clearpage
\subsubsection{DAQ and data format}
An MPA-3 Multiparameter System was used for data acquisition. The complete
description of this system can be found in Ref.~\cite{mpa}.
The main hardware modules of MPA-3 are:
\begin{itemize}
\item
An MPA-3 PCI Card with an internal 20~\mega\hertz~clock reset by a TTL signal from the chopper.
\item
An MPA-3 BASE module with four ADC ports for the connection of
external ADCs and two BNC type auxiliary connectors AUX 1 and 2. One of this is
used to accept a TTL signals from the chopper.
\item
ADC Port Modules: an interface for the 4 ADCs.
\end{itemize}
Data is written in a list mode file format and stored on the RAID disk system.
A detailed description can be found in Ref.~\cite{mpa}.
Each event contains the information on the ADCs channel, 16 bit ADC data and real time clock RTC value in
three 16 bit words rtc0, rtc1 and rtc2. The RTC value is calculated according to the formula (rtc2*65536+rtc1)*65536+rtc0. It starts from a preset value
which was rtc0=65528, rtc1=65536, rtc2=65536 and counts down in 50~\nano\second~intervals.
\subsection{The neutron chopper}
All the measurements of the neutron transmission through the crystal have been performed at specific neutron wavelengths selected using a neutron chopper
and time-of-flight techniques. For the UCN and VCN measurements different
choppers have been used.
\subsubsection{The VCN chopper} \label{vcnchopper}
The VCN chopper uses a rotating disk made of aluminium coated with a neutron absorbing Gd and $^6$LiF layer and it has two slits through which neutrons can pass.
A 20~\milli\metre~diameter boron collimator is placed directly after the chopper and is
centered on the beam axis. The chopper was operated at a rotation frequency of $\nu$ = 7~\hertz.  A 5~\micro\second~long TTL signal is generated
before the slit reaches the neutron beam center. This signal triggers the data acquisition system.
The cross section area of the beam that is cut by the chopper slit and the collimator at a given time defines the chopper opening function.
This can be calculated by integrating the overlapping area of the slit and collimator as the slit moves through the beam area.
The width of the TOF channel (80~\micro\second) defines the time step for this calculation.
The chopper opening functions have been calculated for the nominal
chopper frequency of $\nu$ = 7~\hertz~and also for a chopper frequency of $\nu$ = 3~\hertz~at which some measurement were carried out.
These chopper functions are plotted in Fig.~\ref{vcnchopper_1}. The transmission profiles obtained by the opening functions
can play an important role in the interpretation of the TOF spectra that are measured: the transmission profiles
have been used to deconvolve the raw data. The details of the deconvolution are given in Appendix~\ref{vcndeconv}.
\epsbild{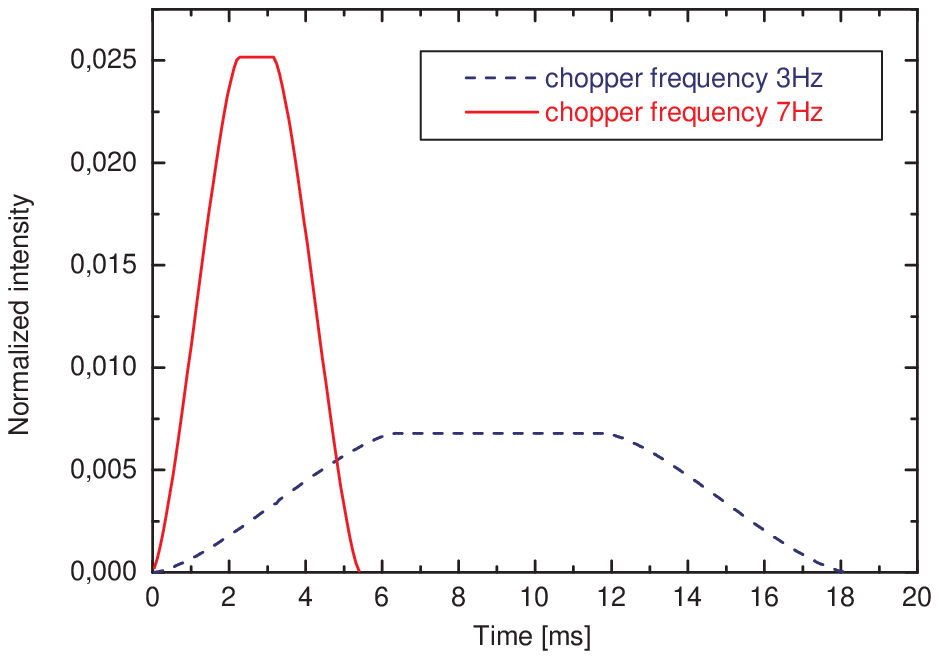}{The VCN chopper opening profiles obtained for two chopper frequencies: 3~\hertz~and 7~\hertz. \label{vcnchopper_1}}{1.0}{0}{!htb}
\subsubsection{The UCN chopper}
The UCN chopper consists of three rotating disks made from
polycarbonate: one disk rotates slowly while two other disks
rotate at the higher speed. The ratio of the speeds of the disks
is fixed at 1:6. The 'slow' disk opens a time window of
22~\milli\second~with a typical operation frequency of about
3~\hertz~to 4~\hertz~and produces a neutron pulse. The two fast
disks sharpen the rising and falling slope of the pulse. The disks
in the chopper are driven by a stepping motor, the stepping motor
is computer-controlled through a ``PCI 7324'' card and LabView
software. The trigger signal needed for the TOF measurements is
provided by the light being reflected by one of the reflection
sensors and being detected  by infrared diodes. The reflector is a
7~\milli\metre~long piece of aluminium mounted on the wheel of the
slow disk, which together with the rotation speed determines the
length of the pulse. To correct for this effect, a shaping stage
was applied. The length of the TTL signal, after passing this
stage is 5~\micro\second. A full chopper turn takes 864 steps. The
chopper was routinely operated at $\nu$ = 3~\hertz. The stepping
motor can be run at a slow enough speed that allows for an
experimental determination of the chopper opening function,
Fig.~\ref{ucnchopper_1} shows the neutron yield (and thus the
chopper opening) as a function of time (blue points). The results
with the best statistic are obtained at a chopper speed of 2
steps/\second~which corresponds to 2~\milli\hertz. The curve thus
measured has been fitted with a polynomial function,
Fig.~\ref{ucnchopper_1} shows the resulting curves. In order to
get the chopper opening function for the typically used speed of
2500 steps/\second, we used the obtained polynomial parameters to
evaluate appropriate values and the result of this procedure is
presented in the Fig.~\ref{ucnchopper_2}. The effect of
convolution of the obtained opening function with the raw UCN TOF
spectrum was calculated to be negligible in the analysis of the
data. The stepping motor can also be operated in a way so as to
obtain a precise calibration of the TOF spectra. This technique is
described in Appendix~\ref{ucncal}. \epsbild{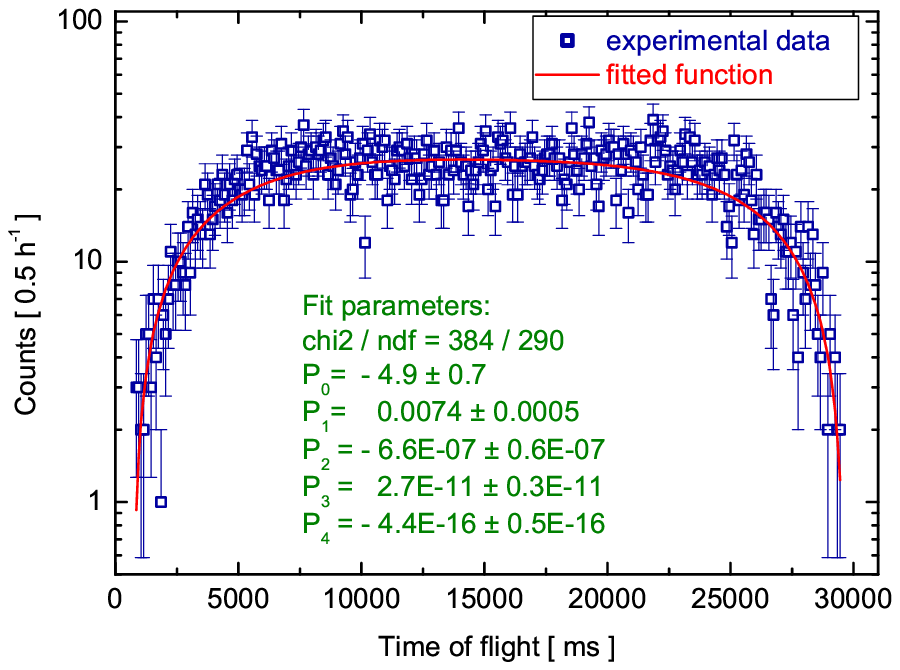}{The
chopper opening function for the chopper speed 2
steps/\second~(2~\milli\hertz) plotted together with a fitted
polynomial function (red line). The bin width is
100~\milli\second. \label{ucnchopper_1}}{1.0}{0}{htb}
\epsbild{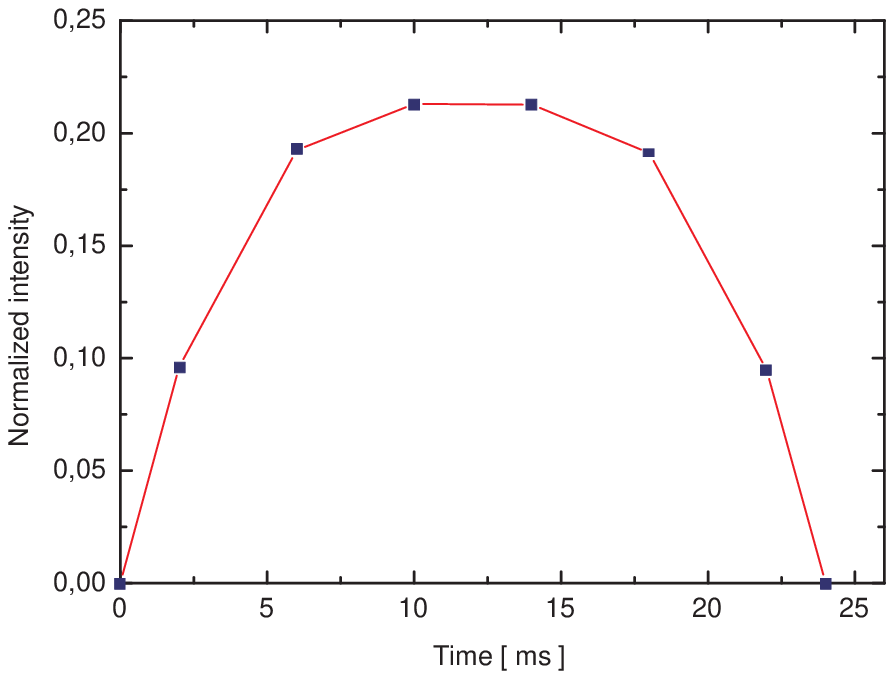}{The chopper opening profile for a
chopper speed of 2500 steps/\second~(3~\hertz).
\label{ucnchopper_2}}{1.0}{0}{!htb} \clearpage

\chapter{Results and data analysis} 
A typical set of data consist of neutron transmission measurements through liquid and solid  deuterium.
It took from 4 to 7 days. Measurements with the deuterium in the gaseous phase
were also carried out in some cases. Once the transmission measurements of the deuterium in the liquid phase
had been completed, the crystal was grown from 
the melt and was subsequently annealed (section~\ref{freezing}). The transmission through solid deuterium is measured as
a function of the crystal temperature range from 5~\kelvin to 18~\kelvin~in 2~\kelvin~steps in most cases.
from 18~\kelvin. Simultaneously  Raman Spectroscopy was performed as well as an optical 
investigation with the digital camera. For the ``front'' detector position, two sets of measurements
with deuterium containing a high ortho-D$_2$ concentration were carried out and one set of measurement with
normal deuterium was done.  
We also performed a few additional measurements testing 
in what way the method of freezing influences the neutron transmission. 
This section focuses on regular measurements. Others are only shown to emphasise some important effects that have been observed.

The analysis of one deuterium crystal consists of the following:
\begin{itemize}
\item
Determination of ortho-D$_2$ concentration from the Raman spectra (usually from the liquid measurements).
\item
Optical inspection of the measured crystal using pictures recorded by the digital camera.
\item
Analysis of the S$_0$(0) line splitting in the ortho solid which gives information about the relative orientation
of the crystal (or crystallites).
\item
Analysis of the neutron data including measurements with liquid, solid at various temperatures and an empty cell.
\end{itemize}
\section{Deuterium properties}\label{craman}
To investigate the ratio c$_0$ of ortho-D$_2$ to para-D$_2$ deuterium in the measured sample the Raman spectroscopy method has been used.
Figure~\ref{lraman} shows spectrum of rotational excitations for liquid ortho-deuterium with c$_0$ = (98.7 $\pm$ 0.2)$\%$. 
The ortho-deuterium  concentration is determined by comparing the Raman line intensities for the J=0$\to$2 and J=1$\to$3 
transition (lines S$_0$(0) and S$_0$(1) respectively) and taking into account the proper transition matrix element.
The Ar II (493.3209~\nano\metre) line which is seen in Fig.~\ref{lraman} is produced by the laser. 
\epsbild{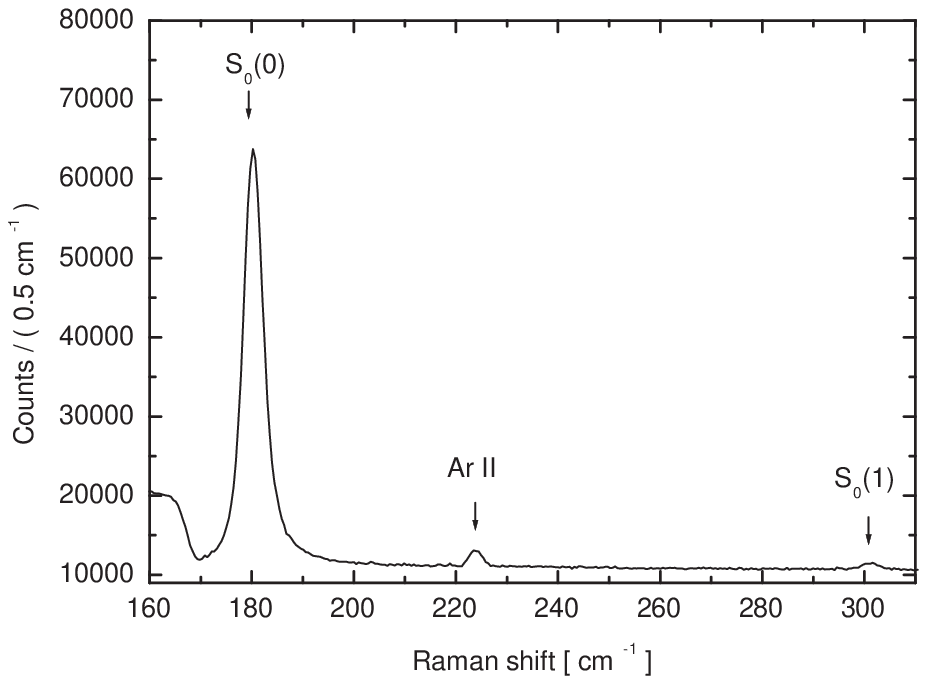}{A Raman spectrum of liquid o-D$_2$ ((98.7 $\pm$ 0.2)$\%$). The spectrum contains data from 14 single files with 200 s exposure time each.\label{lraman}}{.8}{0}{!htb}  

Figure~\ref{s18raman} shows the Raman signal for solid ortho-deuterium at temperature of 18~\kelvin. The S$_0$(0) line is split into a triplet which is a signature of the
hcp structure of the crystal~\cite{Kra}. The hcp symmetry of deuterium crystal makes the J = 2 level form three energy bands belonging to 
m = $\pm$ 1 ($\alpha$) m = $\pm$ 2 ($\beta$) and m = 0 ($\gamma$). Change of the  S$_0$ line shape with crystal temperature is shown in Fig.~\ref{merged} 

The pictures which were taken with the digital camera allow us to study the optical transparency of the deuterium crystal at various temperatures. 
It appears that only carefully prepared crystal at temperature of 18~\kelvin~is fully transparent (Fig.~\ref{temp}).   
\epsbild{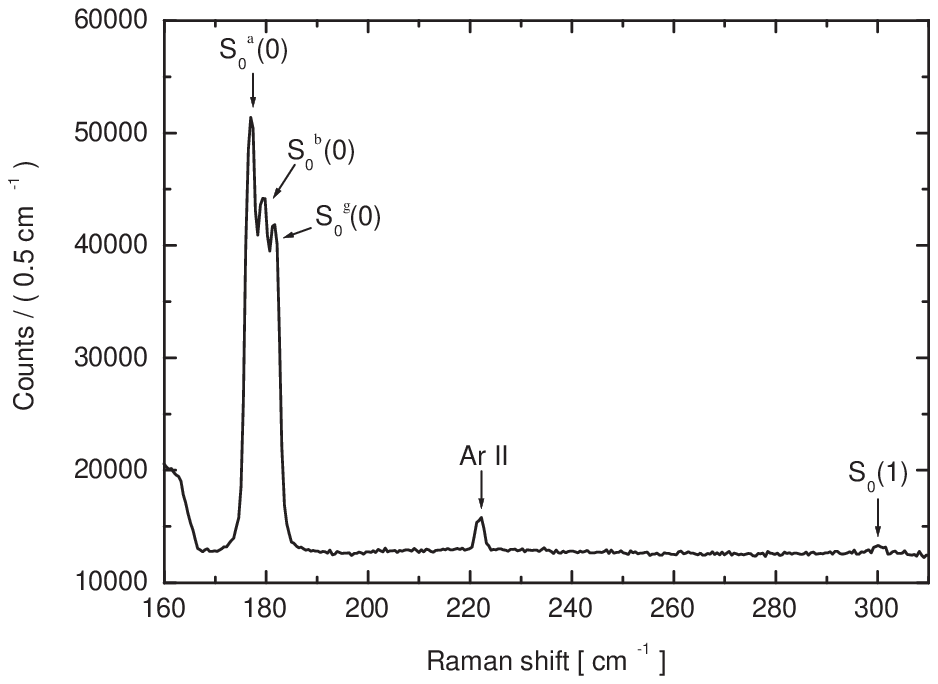}{A Raman spectrum of solid ortho-deuterium at 18K. The spectrum contains data from 14 single files with 200s exposure time each. \label{s18raman}}{.8}{0}{!htb}
\epsbild{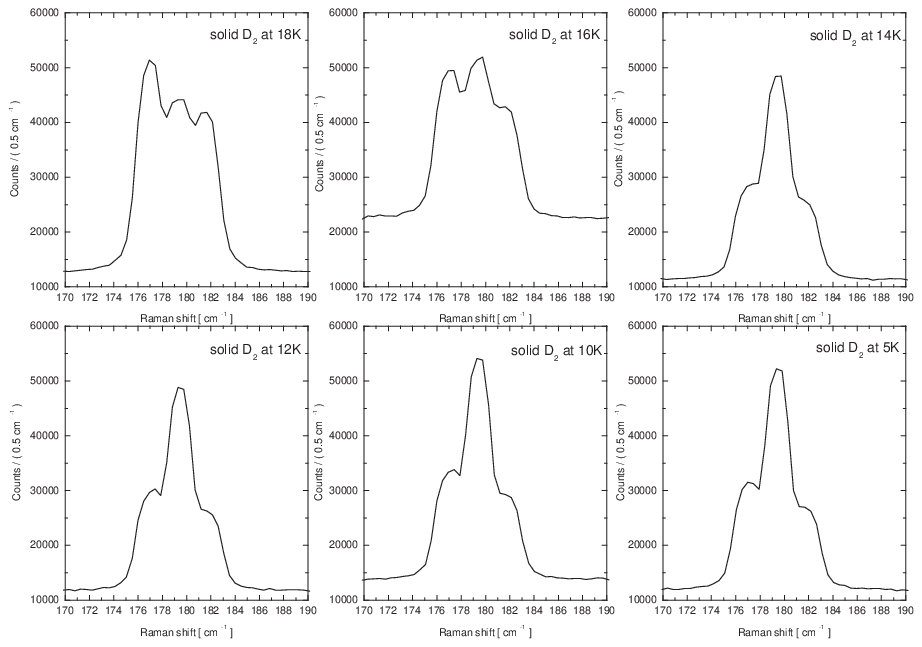}{The J=0$\to$2 transition for the deuterium crystal at various temperatures. The relative m$_j$ = 0, 1, 2 intensity ratio is
changing with the temperature which indicates that the crystal orientation is changing as well~\cite{Kra}.\label{merged}}{0.8}{0}{!htb}  
\epsbild{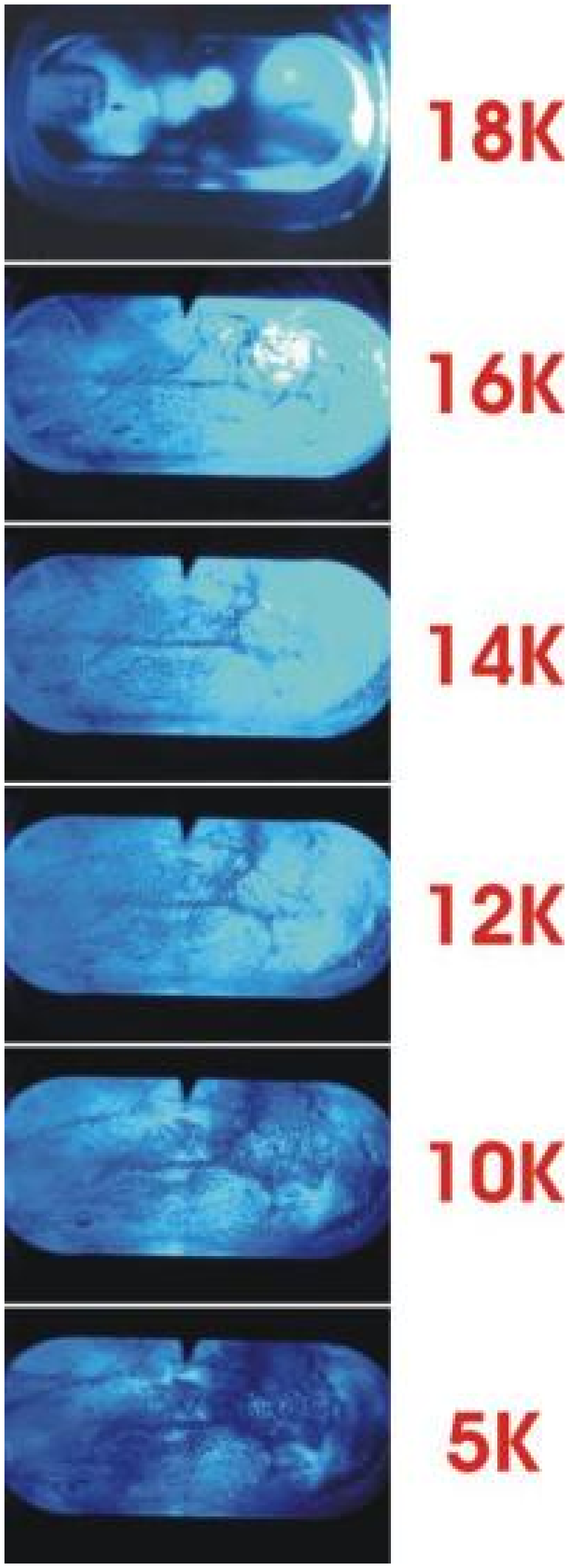}{Images of the crystal taken at various temperatures. \label{temp}}{0.5}{0}{!htb}
\clearpage
\section{Neutron data analysis}
The neutron total cross section was evaluated from the neutron beam attenuation 
in the deuterium sample. The neutron count-rate $N_0$ has been measured for the beam
passing through an empty target cell and the neutron count-rate $N_1$ for the beam passing 
through the target cell filled with deuterium has been measured.
These count-rates are related through:
\begin{equation}
N_1 = N_0\exp{(-\rho\sigma_{tot}l)}
\end{equation}
where $\rho$ is the density of the sample, $l$ is the length of the cell
and $\sigma_{tot}$ is the total cross section.
The density of a deuterium sample was calculated from the molar volume $V_L$, taken from Ref.~\cite{Sou}, in the following way:
\begin{equation}
\rho = \frac{N_A}{V_L}
\end{equation} 
where $N_A$ is Avogadro's number. The molar volume $V_L$ is equal to 23.21~\centi\cubic\metre\per\mole~for liquid deuterium at 19~\kelvin~
and  20.34~\centi\cubic\metre\per\mole~for solid deuterium.

A single run takes normally 1800~\second. In most cases data were taken for 2 hours at one temperature.
The empty cell transmission measurements have been also performed in the range from 5~\kelvin~to 20~\kelvin. No noticeable temperature dependency has been found.
\subsection{VCN data} \label{vcnresults}
Figure~\ref{liqsol} shows neutron total cross section 
versus neutron wave vector $k$ that have been measured with liquid deuterium and with solid deuterium at 18~\kelvin~and 5~\kelvin. The solid deuterium data were obtained with the same crystal with a high ortho-deuterium concentration of (98.7 $\pm$ 0.2)$\%$.

The fitting was done with the MINUIT package following two  $\chi^2$ minimisation methods:
Simplex~\cite{Ned} and Migrad~\cite{Fle}. Both lead to the same results.

For the liquid deuterium we fitted the total cross section vs. $k$ using:
\begin{equation}
\sigma_{tot} = \sigma_0 + \frac{A}{k}  \label{eq:linear}
\end{equation}
where $k = 2\pi/\lambda$ is the neutron wave vector in the material and $\sigma_0$ can be interpreted as an energy independent 
incoherent elastic cross section.
The total cross section $\sigma_{tot}$ is the sum of energy independent incoherent elastic cross section
$\sigma_0$ (4.1~\barn) and the capture plus inelastic cross section (up-scattering)
which both are inversely proportional to the neutron wave vector. In Fig.~\ref{l19}, the experimental data is plotted together with the fitted function. 
One notices that there is no visible deviation from the 1/$v$ law. The parameter $A$ which is a measure of inelastic cross section  
is $A$ = (0.742 $\pm$ 0.009)~\barn\usk\reciprocal\angstrom. 
The incoherent cross section value obtained from the fit is $\sigma_0$ = (4.19 $\pm$ 0.11)~\barn~and agrees with the theoretical value to within one standard deviation.
\epsbild{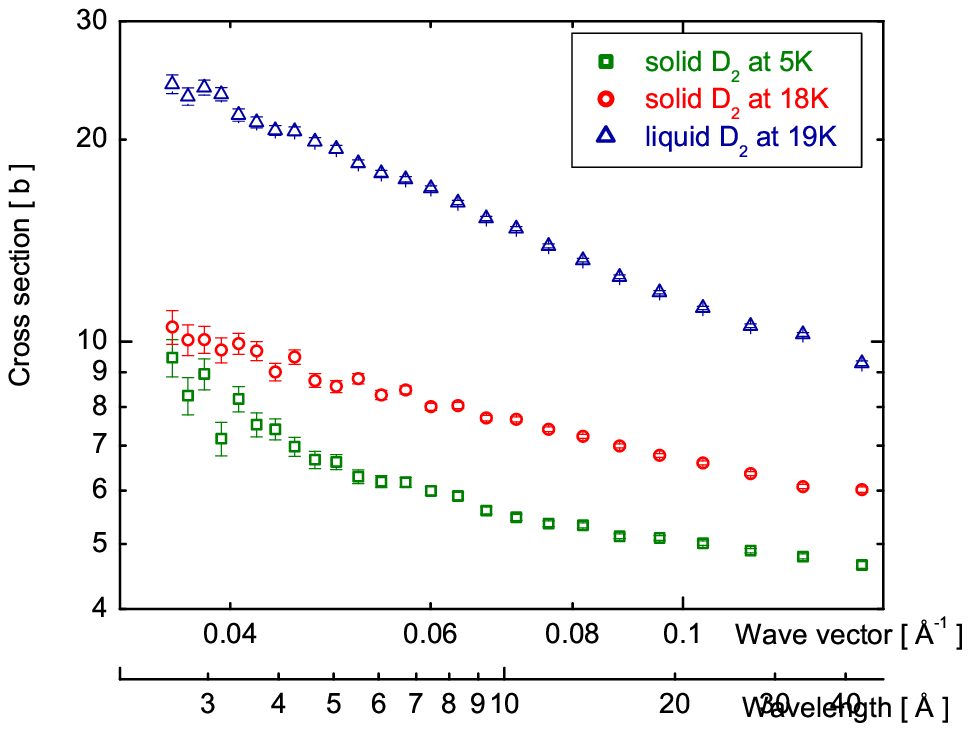}{The total cross section per deuterium molecule for liquid deuterium and solid deuterium plotted versus the neutron wave vector $k$. \label{liqsol}}{1.0}{0}{!htb}
\epsbild{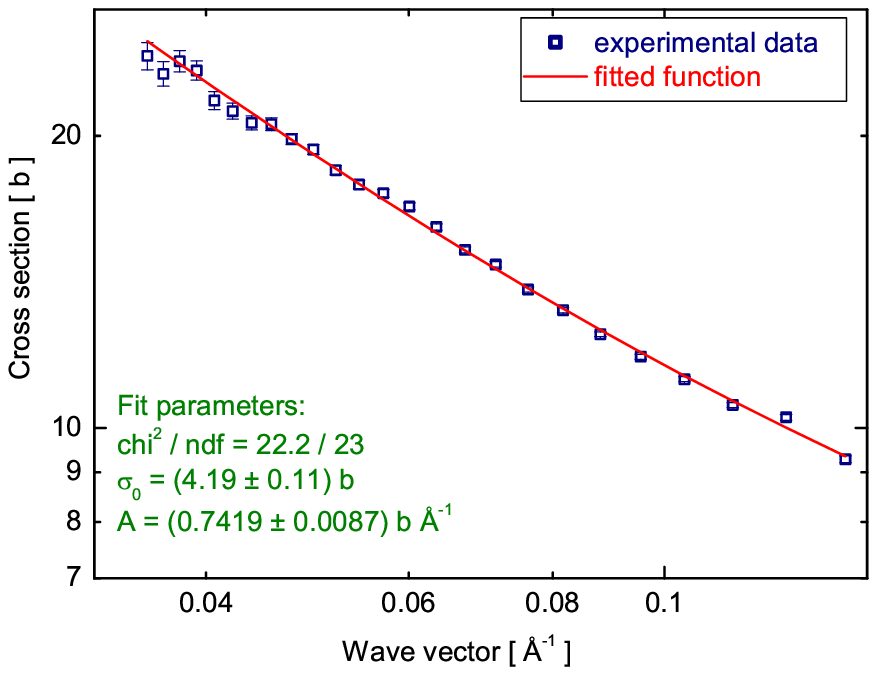}{The total cross section per deuterium molecule for liquid deuterium at 19~\kelvin~plotted versus the neutron wave vector $k$. The solid curve shows the result 
of a fit of the form $\sigma_{tot} = \sigma_0 + A/k$. \label{l19}}{1.0}{0}{!htb}
\clearpage
The fitted function used for liquid deuterium is not suitable for the transmission data taken with the solid deuterium.
Different dynamics of neutron scattering from the solid phase requires more advanced models, 
e.g. a model that includes the elastic coherent scattering due to inhomogeneities in the crystal.  
The fitted function which was applied to the data from the measurements of solid deuterium at 18~\kelvin~follows the idea of
Engelmann and Steyerl. In their paper~\cite{Eng} they show that very cold neutron transmission can be used to
investigate inhomogeneities in the matter (section~\ref{inhom}). The developed technique is alternative to the more conventional small angle scattering. 
In this model the elastic scattering is described by the Eq.~\ref{eq:elastic} of chapter 2. 
The total cross section may be written as: 
\begin{equation}
\sigma_{tot} = \sigma_0 + \frac{A}{k} + B\frac{g(kR\theta_1)}{k^2} \label{eq:model1}
\end{equation}     
where $g(kR\theta_1)$ is given by Eq.~\ref{eq:factor}, $\theta_1$ = 20~\degree~is the detector aperture and $R$ refers to the radius of
submicroscopic structures.
\epsbild{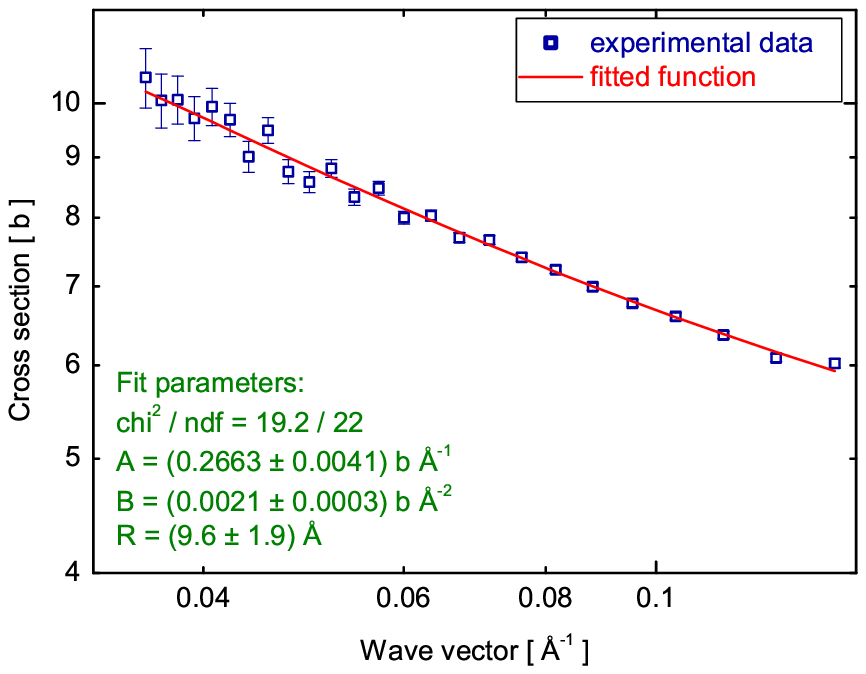}{The total cross section per deuterium molecule for solid deuterium at 18~\kelvin~plotted versus the neutron wave vector. The curve through the experimental
points is a fit according to Eq.~\ref{eq:model1}.\label{s18}}{1.0}{0}{!htb}
\\
Figure \ref{s18} shows the experimental data and the obtained fit. The incoherent elastic contribution $\sigma_0$ was taken from the measurement
with liquid deuterium and $A, B$ and $R$ were varied.

The parameter A = (0.266 $\pm$ 0.004)~\barn\usk\reciprocal\angstrom~obtained from the fit is 2.8 times lower than that of liquid deuterium at a temperature of 19~\kelvin. 
This indicates strongly suppressed up-scattering for the solid deuterium as compared with liquid deuterium and
agrees with a calculation presented by Keinert~\cite{Kei}: for the neutron energy of 200~\nano\electronvolt~Keinert calculated the up-scattering rate 
for solid ortho-deuterium at temperature of 17~\kelvin~to be 3.1 times lower than that of liquid deuterium at a temperature of 19~\kelvin.

The interpretation of parameter $B$ is the following~\cite{Gol2}:
\begin{equation}
B = 8\pi^3b^2R^4\delta\rho^2
\end{equation}
where $b$ is the bound coherent scattering length, $\delta\rho = \rho - \rho_0$ is the density difference to the surrounding medium.
 
Parameter $R$ refers to the radius of spherical inhomogeneities and the value obtained from the fit is $R$ = (9.6 $\pm$ 1.9)~\angstrom.

The same fitting procedure was applied to the data from the transmission measurements of the crystal at temperatures of 16~\kelvin, 14~\kelvin~and 12~\kelvin. The experimental
data and obtained fits are shown in Figs.~\ref{s16},~\ref{s14} and~\ref{s12}, respectively.
\epsbild{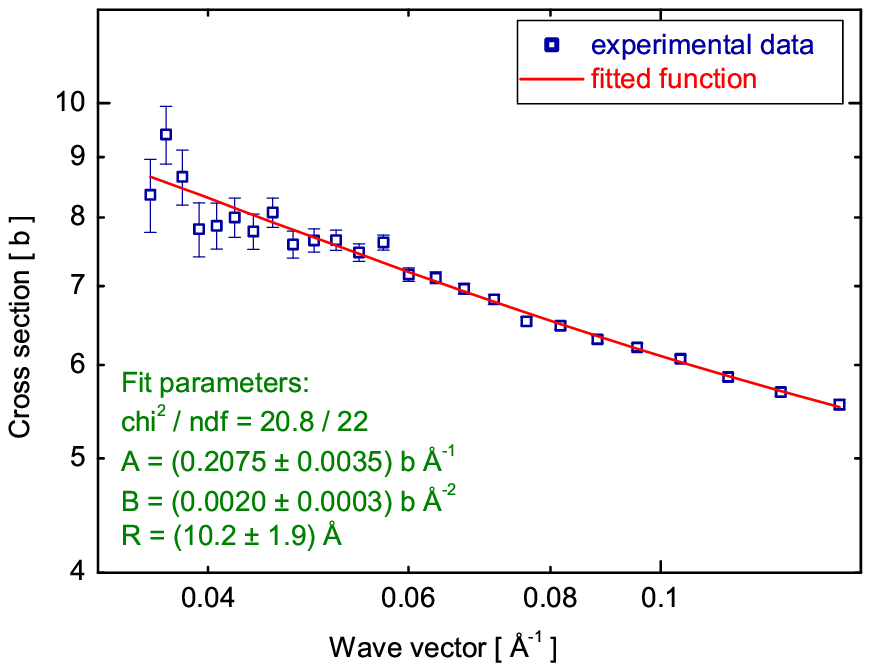}{The total cross section per deuterium molecule for sD$_2$ at 16~\kelvin. The curve through the experimental
points is a fit according to Eq.~\ref{eq:model1}. \label{s16}}{1.0}{0}{!htb}
\epsbild{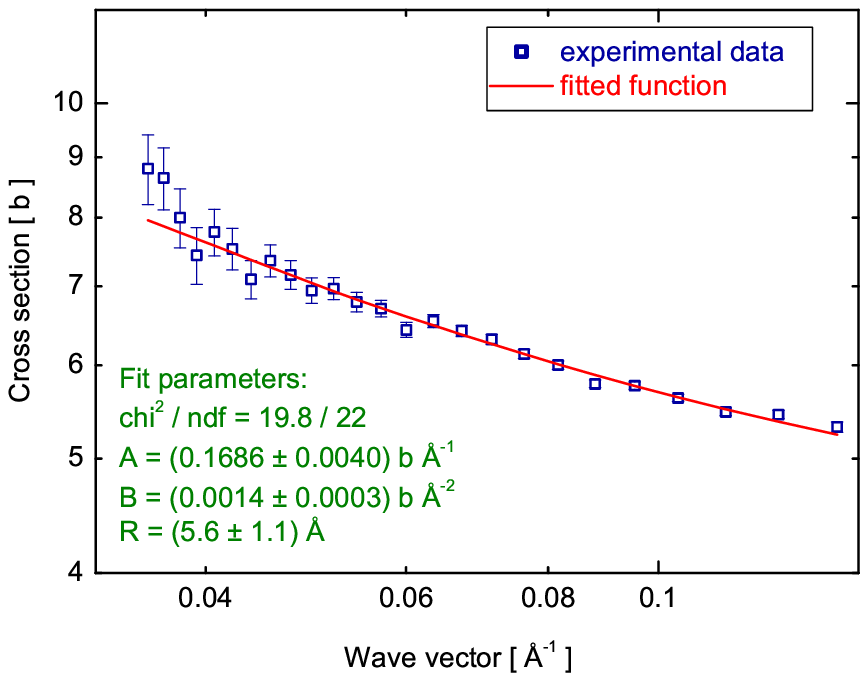}{The total cross section per deuterium molecule for sD$_2$ at 14~\kelvin. The curve through the experimental
points is a fit according to Eq.~\ref{eq:model1}.\label{s14}}{1.0}{0}{!htb}
\epsbild{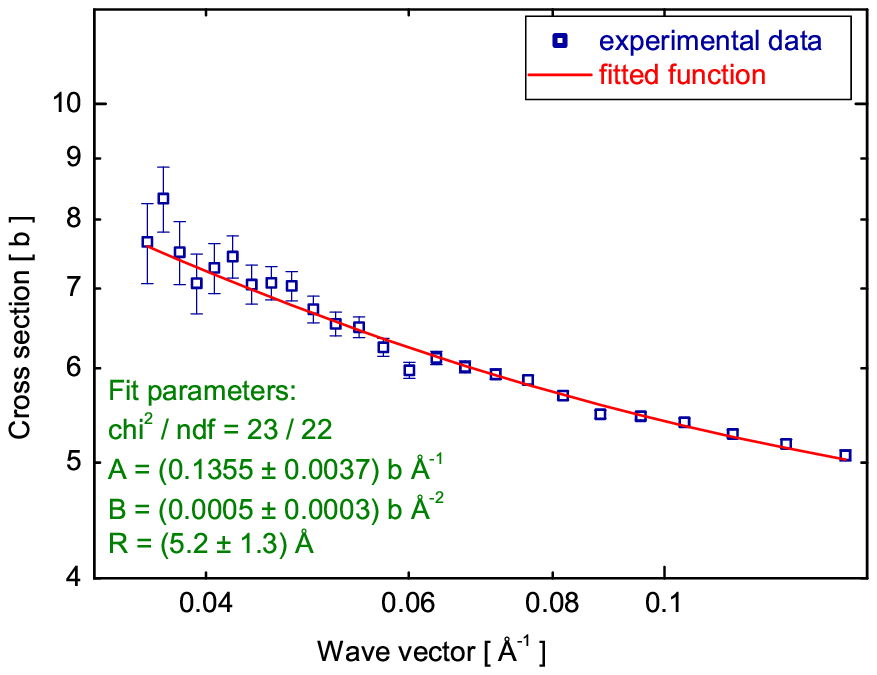}{The total cross section per deuterium molecule for sD$_2$ at 12~\kelvin. The curve through the experimental
points is a fit according to Eq.~\ref{eq:model1}.\label{s12}}{1.0}{0}{!htb}

\clearpage

The interpretation of the total cross section measurement for the deuterium crystal at temperatures below 12~\kelvin~is also based on 
the model which includes both incoherent inelastic scattering (proportional to 1/$k$) and coherent elastic scattering (proportional to 1/$k^2$). 
The function which fits the best the data taken with the crystal at the temperatures below 12~\kelvin~has a form:
\begin{equation}
\sigma = \sigma_0+ \frac{A}{k}+ \frac{B}{k^2}  \label{eq:coherent2}
\end{equation} 
where parameter $A$ is a measure of inelastic scattering and $B$ gives an information about the strength of coherent elastic scattering.  
\epsbild{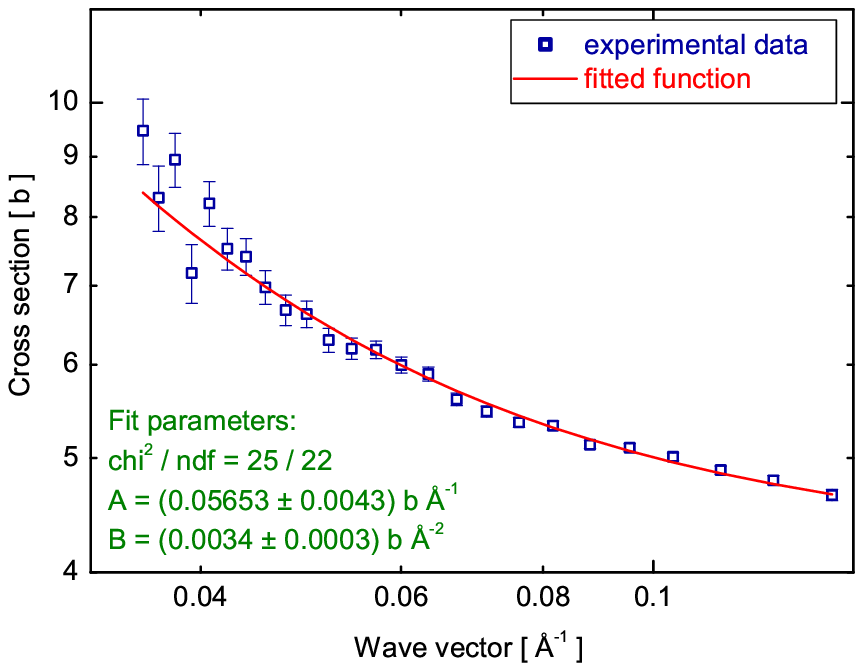}{The total cross section per deuterium molecule for solid deuterium at 5~\kelvin~vs. the neutron wave vector $k$. The curve through the experimental
points is a fit according to Eq.~\ref{eq:coherent2}. \label{s5}}{1.0}{0}{!htb}
Figure~\ref{s5} shows that the experimental point are well represented by the curve of the form~\ref{eq:coherent2}.

The value for A = (0.056 $\pm$ 0.011)~\barn\usk\reciprocal\angstrom~is now about 3 times smaller than that obtained for crystal measurements at 18~\kelvin. 
\epsbild{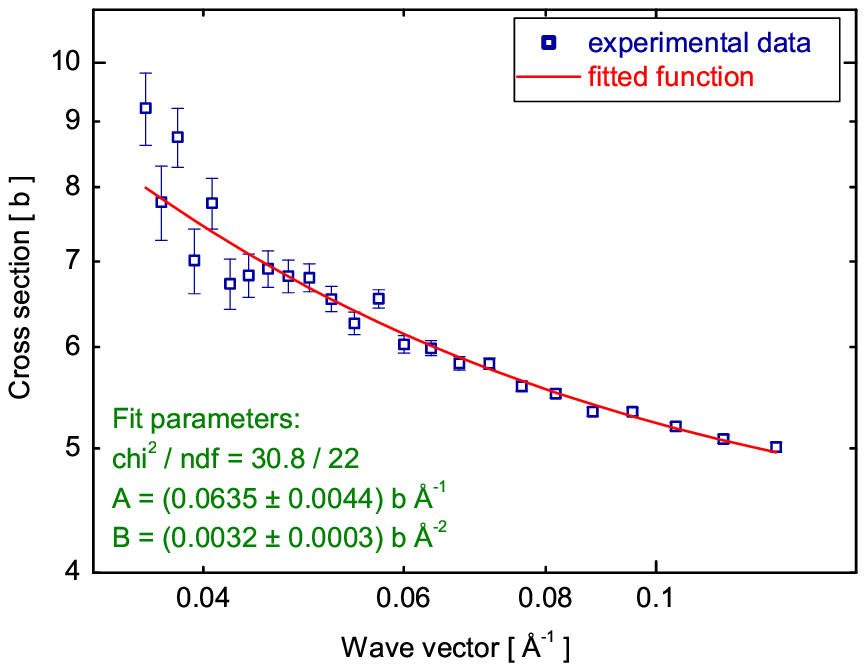}{The total cross section per deuterium molecule for solid deuterium at 10~\kelvin~plotted versus the neutron wave vector. The curve through the experimental
points is a fit according to Eq.~\ref{eq:coherent2}.\label{s10}}{1.0}{0}{!hbt}
\epsbild{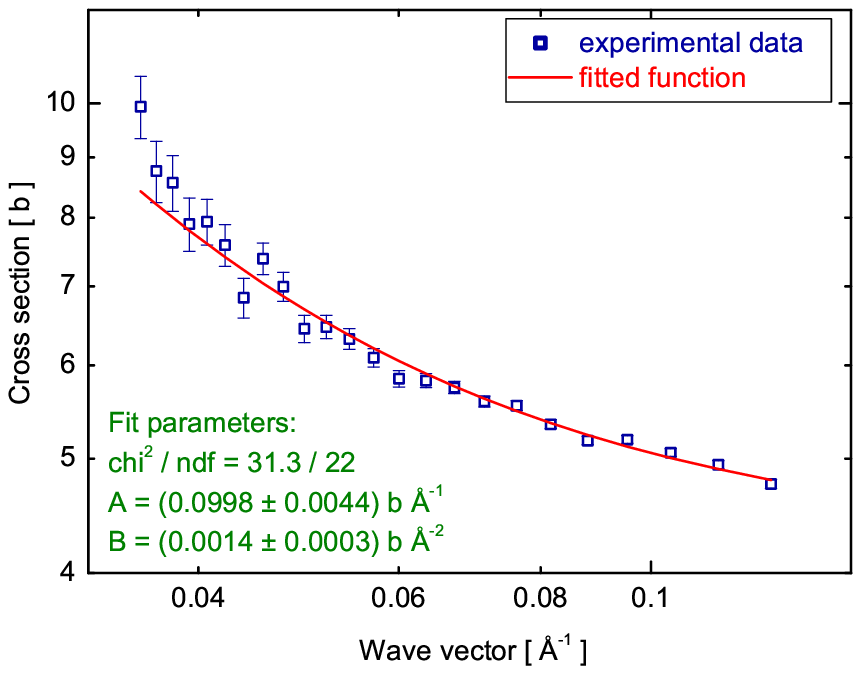}{The total cross section per deuterium molecule for solid deuterium at 6~\kelvin~plotted versus the neutron wave vector. The curve through the experimental
points is a fit according to Eq.~\ref{eq:coherent2}.\label{s6}}{1.0}{0}{!hbt}

The same analysis was applied to the data from the transmission measurements for crystal at temperatures of 10~\kelvin~and 6~\kelvin. 
Figures~\ref{s10} and~\ref{s6} show the experimental data and the obtained fits for 10~\kelvin~and 6~\kelvin, respectively.

We compared the experimental values of parameter $A$ with the one phonon up-scattering cross sections (Eq.~\ref{eq:upscatt}) calculated for the initial
neutron wavelength of 60~\angstrom~(or wave vector $k$ = 0.1~\reciprocal\angstrom).
In Fig.~\ref{upscatt1} the experimental and theoretical values, given in~\barn\usk\metre\usk\reciprocal\second, are plotted versus the temperature of
solid deuterium. It is seen that the experimental results reveal a different trend
then theory. They agree well only for the crystal temperature of 18~\kelvin.
\epsbild{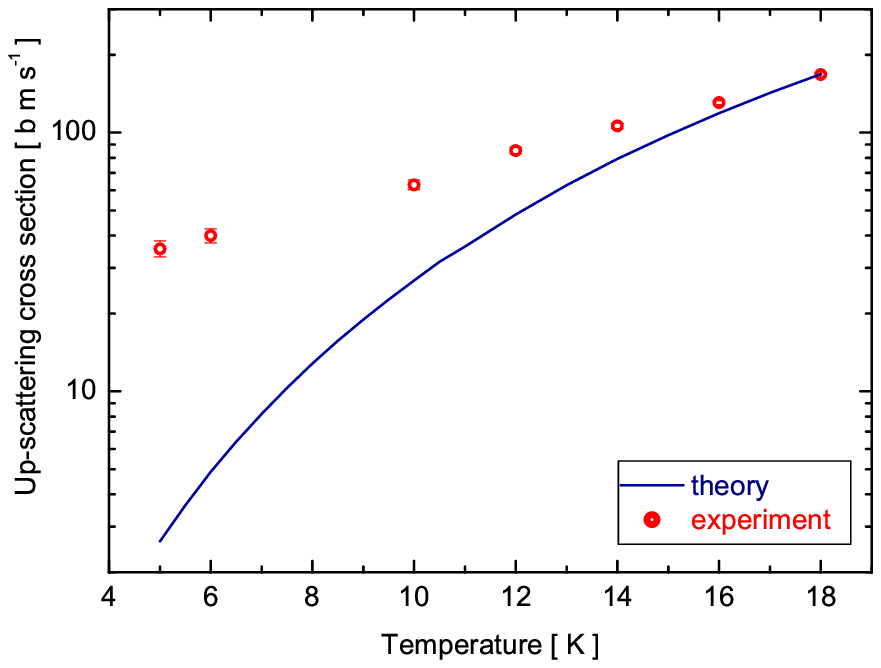}{The up-scattering cross section vs. temperature for solid D$_2$ as measured (red circles) and
as follows from the one phonon up-scattering calculations (blue line).\label{upscatt1}}{1.0}{0}{!htb}

To check the reproducibility of the measurement and sample preparation procedures we repeated the transmission measurements with 
another deuterium crystal with high ortho-deuterium concentration prepared in the same way. 
In Figs.~\ref{comp} and~\ref{stest2} the curves are plotted for both crystal samples at temperatures 18~\kelvin~and 5~\kelvin, respectively.
We conclude that both measurements are consistent. 

\epsbild{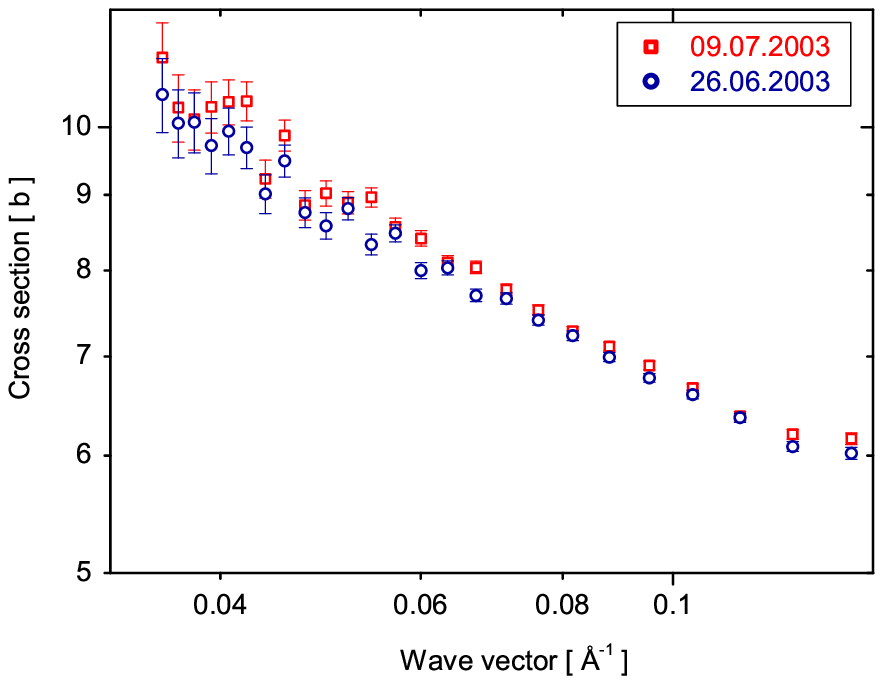}{The total cross section per deuterium molecule for solid deuterium at 18~\kelvin~for two different crystals prepared in the same way. \label{comp}}{1.0}{0}{!hbt}
\epsbild{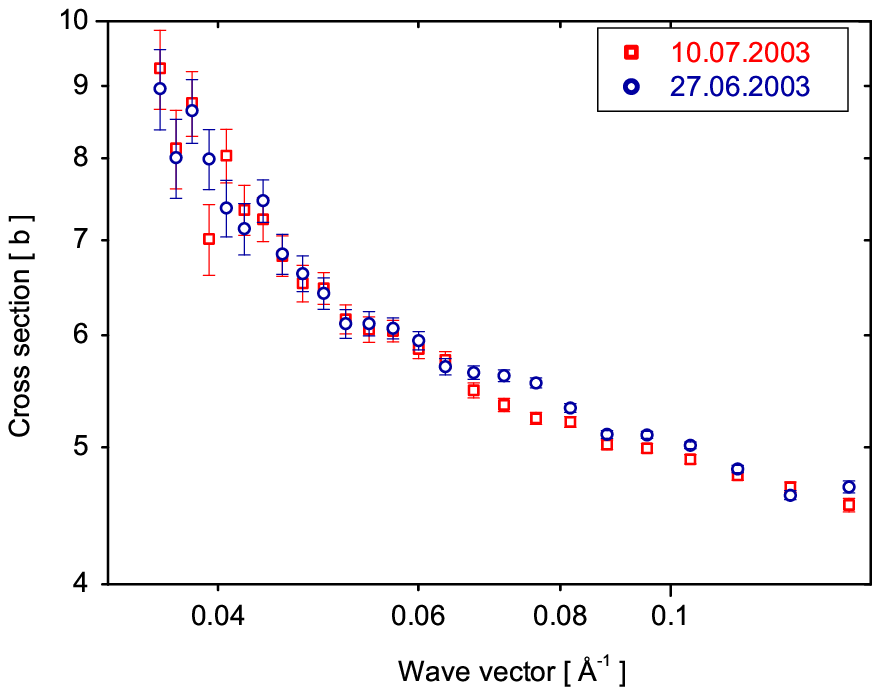}{The total cross section per deuterium molecule for solid deuterium at 5~\kelvin~as obtained for two crystals prepared in the same way. \label{stest2}}{1.0}{0}{!htb}

\clearpage
\subsection{UCN data} \label{ucnresults}
Figure~\ref{ucnld2sd2} shows plots of the neutron transmission loss cross section 
versus neutron wave vector $k$ for the liquid ortho-deuterium ($c_0$ = (98.2 $\pm$ 0.9)$\%$) and solid ortho-deuterium at temperatures of 18~\kelvin~and 5~\kelvin.
The up-scattering cross section for the neutron wavelength of 550~\angstrom~(or an energy of 270~\nano\electronvolt) is 2.7 times smaller for solid deuterium at 18~\kelvin~as compared with that for liquid deuterium and agrees with the value obtained from the VCN analysis. In Fig.~\ref{ucnld2sd2} one also sees that, over
the full energy range covered, at 5~\kelvin~the inelastic cross section is a factor 2.2 smaller than at 18~\kelvin. This is 2 times less than the result obtained from the VCN analysis. 
  
Figure~\ref{ucnsd2all} shows the cross section as a function of the deuterium crystal temperature. One can notice the tendency 
of suppression of the total cross section with decreasing temperature of the crystal. However, the difference in the cross section values between 5~\kelvin~and 10~\kelvin~is
not visible. It is noteworthy that the measurement of the UCN yield from the model of solid-deuterium UCN source performed by Serebrov et al.~\cite{Ser1} 
does not show any difference between the crystal at 10~\kelvin~and 5~\kelvin~either. Figure~\ref{sd210-5} shows the total cross sections for solid deuterium at temperatures of
10~\kelvin~and 5~\kelvin~from both VCN and UCN measurements. It is seen that at the shorter wavelengths (30~\angstrom~- 100~\angstrom) the cross section
values for crystal at 10~\kelvin~are higher as comapred with 5~\kelvin~data, which may indicate that the coherent scattering plays a crucial role 
at the larger wavelengths.
This effect is especially important at temperatures below 10~\kelvin~where the incoherent up-scattering cross section is relatively low. The cross section
curves for solid deuterium at 18~\kelvin~and 5~\kelvin~are plotted in Fig.~\ref{sd218-5}.  
\epsbild{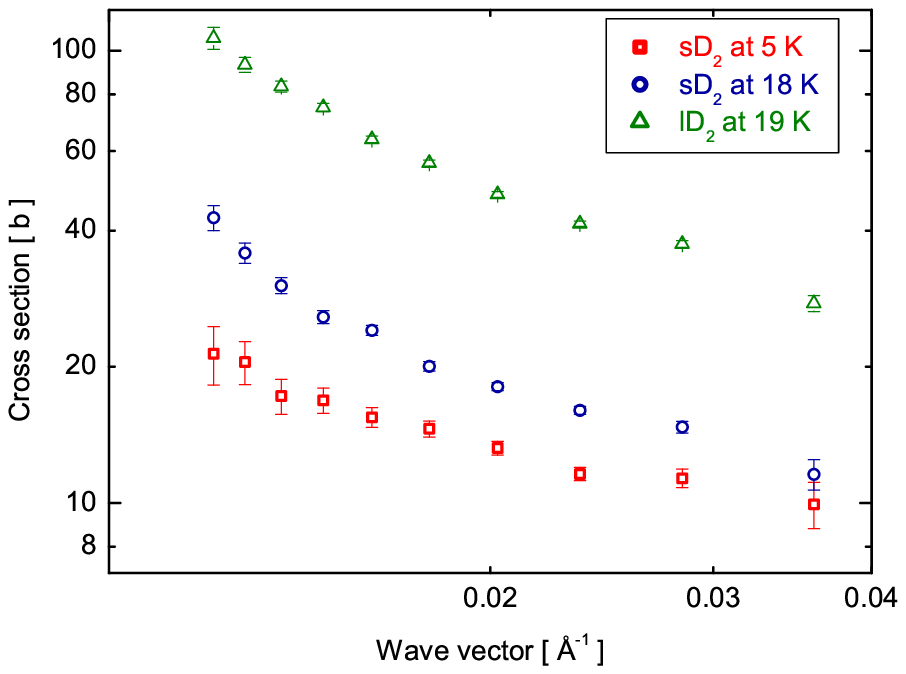}{The total cross section per deuterium molecule for liquid deuterium at a temperature of 19~\kelvin~and solid deuterium at temperatures 18~\kelvin~and 5~\kelvin. \label{ucnld2sd2}}{1.0}{0}{!htb}   
\epsbild{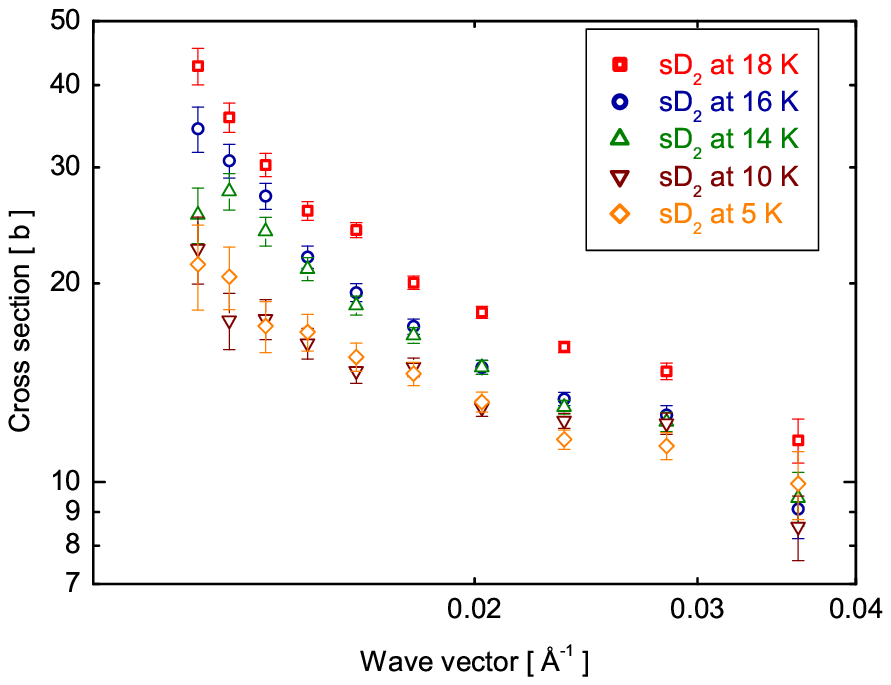}{The total cross section per deuterium molecule for solid deuterium at temperatures 18~\kelvin, 16~\kelvin, 14~\kelvin, 10~\kelvin~and 5~\kelvin~plotted vs. wave vector. \label{ucnsd2all}}{1.0}{0}{!htb}   
\epsbild{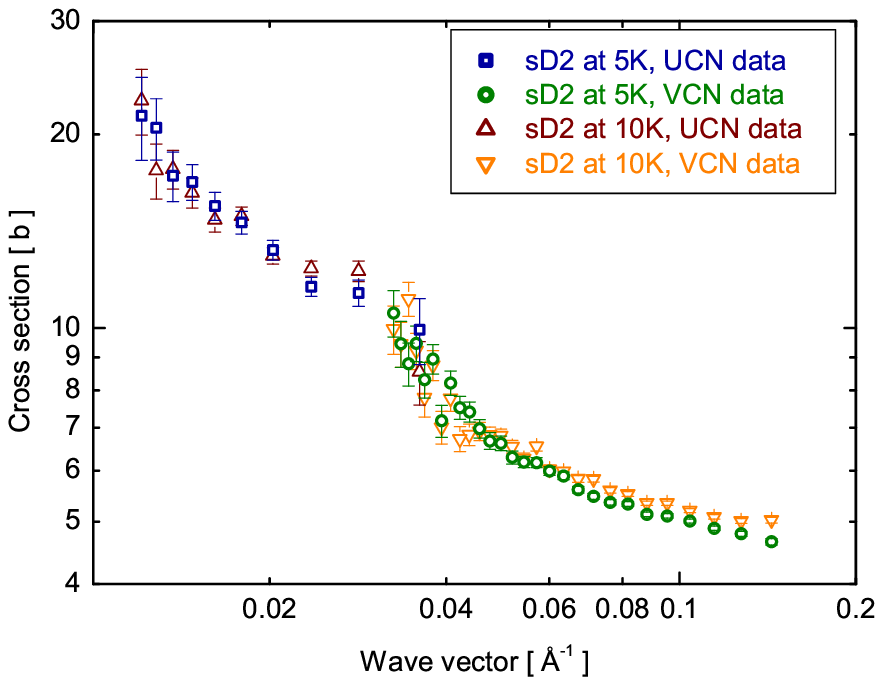}{The total cross section per deuterium molecule for solid deuterium at temperatures 10~\kelvin~and 5~\kelvin as obtained from the measurements with VCN and UCN. In the wave vector range between 0.03 and 0.04~\reciprocal\angstrom~the statistics is low thus the systematics effects dominate. This is the reason of increase of the cross section in this region. \label{sd210-5}}{1.0}{0}{!htb}   
\epsbild{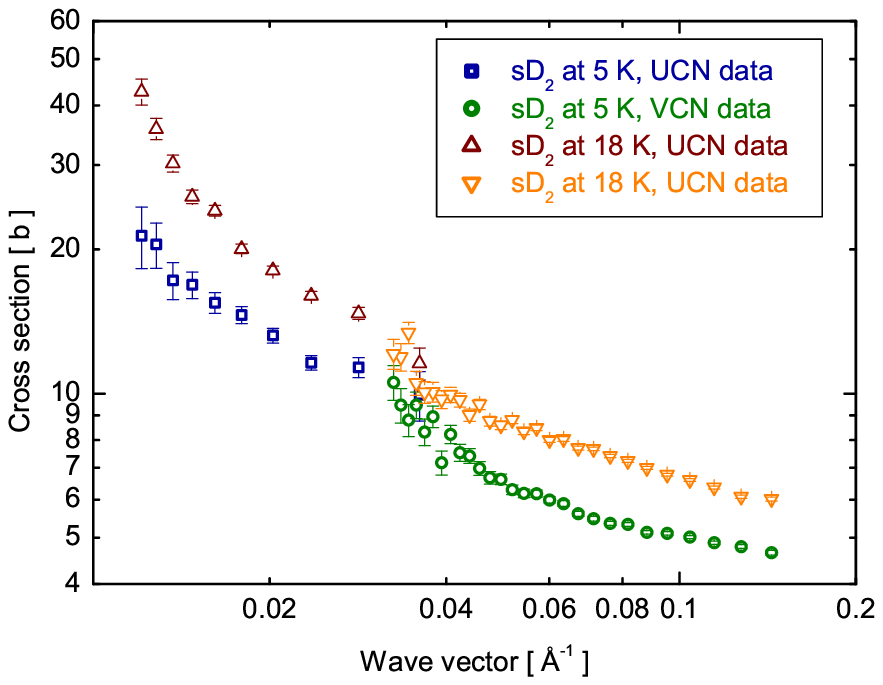}{The total cross section per deuterium molecule for solid deuterium at temperatures 18~\kelvin~and 5~\kelvin as obtained from the measurements with VCN and UCN.  \label{sd218-5}}{1.0}{0}{!htb}

\chapter{Conclusions}
For the construction of a high density superthermal UCN source
several parameters need to be optimized. The production of UCN in
the converter medium and the losses of UCN in this medium are
defining the overall efficiency at which such a source can
operate. Also the extraction efficiency from the source into an experiment itself is of great importance.

To investigate the losses of UCN before they can exit the
solid deuterium converter, neutron transmission measurements through solid deuterium samples have
been performed with VCN and
UCN beams. The storage lifetime related
to these UCN losses in solid deuterium is dominated by four
contributions~\cite{Mor}:
\begin{itemize}
\item
the nuclear absorption of neutrons on deuterons,
\item
temperature dependent phonon - UCN interaction in the deuterium crystal,
\item
UCN up-scattering processes by contaminants such as H$_2$,
\item
the presence of para-D$_2$ which behaves as a contaminant.
\end{itemize}

The work presented in this thesis was devoted to the understanding of
the temperature dependent losses namely losses
due to up-scattering of UCN by picking up energy from phonons present
in the deuterium.

We have shown that the inelastic cross section for neutron
wavelength range between 40~\angstrom~and 500~\angstrom~in
deuterium decreases by a factor 2.8 at the solidification of
liquid deuterium (section~\ref{vcnresults}). This is in agreement
with the calculations done by Keinert~\cite{Kei} for a neutron
wavelength of 640~\angstrom~as well as with the results presented
by Serebrov~\cite{Ser1}. The experimental value of incoherent
cross section obtained from the liquid deuterium data is (4.19
$\pm$ 0.11)~\barn~ and agrees with the theoretical value to within
one standard deviation.

For solid deuterium the one phonon up-scattering is believed to be
the only mechanism which is responsible for temperature dependent
losses of UCN (inelastic cross section). The interpretation method
of the solid deuterium data that we used is based on a model
developed by Steyerl and collaborators. It includes thermal
inelastic scattering and the coherent elastic scattering on
spherical inhomogeneities which are present in the material. Our
measurements carried out with very cold neutrons (in the
wavelength range of 40~\angstrom~ - 190~\angstrom) give results
that are consistent with this model. However, our data as obtained
with ultra-cold neutrons (wavelength range 210~\angstrom~-
500~\angstrom) cannot be properly described by this model. The
most likely explanation for this difference lies in the wavelength
dependence of the neutron scattering on the material structure:
the sensitivity to the scattering on inhomogeneities in the
material is increased significantly at the larger
wavelengths~\cite{Pok3}.

From the fits according to the model mentioned above we were able to distinguish
the inelastic part (parameter $A$) of the total cross section (section~\ref{vcnresults}) for
the temperatures of the crystal between 18 and 5~\kelvin.
The comparison between up-scattering cross sections at different crystal temperatures shows that for wavelength range of 40~\angstrom~ - 190~\angstrom:
\begin{itemize}
\item
At the temperature of 10~\kelvin~the up-scattering cross section decreases 12 times as compared with liquid deuterium
and 4.2 times as compared with solid deuterium at 18~\kelvin.
\item
Further cooling of the crystal to the temperature of 5~\kelvin~ suppress the up-scattering only by 10$\%$.
\end{itemize}
At the wavelengths of 210~\angstrom~- 500~\angstrom~we observed that:
\begin{itemize}
\item
The up-scattering cross section for a crystal temperature of 10~\kelvin~decreases 6 times as compared with that for liquid deuterium and 2.2 times as compared with
solid deuterium at 18~\kelvin. This is about 2 times less than the result obtained from the VCN analysis.
\item
There is no visible reduction of inelastic cross section between measurements with the crystal at 10~\kelvin~and 5~\kelvin.
\end{itemize}
A possible explanation of those effects could be connected with a strongly
enhanced coherent scattering of UCN. For a full understanding of this phenomenon more extensive models are required.

The results from neutron transmission studies are in accordance with optical investigation
of deuterium crystal and with Raman spectroscopy. From the pictures that
were taken with the digital camera we can see that the transparency of
crystal is changing with the temperature and the number of cracks
is increasing with decreasing temperature. The analysis of Raman
spectra shows that the orientation of the crystallites is changing
during cooling which may influence the coherent scattering of
neutrons. Clearly, measurements of this type will prove to be very important in further
optimising the way sD$_2$ is prepared and kept in the UCN source.

\appendix
\chapter{Time of flight spectrum and calibration}
\section{VCN experiment} \label{vcnan}
\subsection{Dead time correction of time of flight spectra}
The time-of-flight (TOF) spectra were measured for VCN packages that passed
the neutron chopper. The raw spectra have to be corrected for the effect of
dead time - specific for detectors and subsequent electronics.
When the detecting system processes an event no additional events can be registered during a time interval of approximately 20~\micro\second.
The influence of the detector dead time on the count rate can be calculated
by analysing the distribution of time interval between subsequent events.

The expected exponential distribution for an ideal (zero dead time) system follows from the Poisson statistics.
The TOF spectra are divided into 2~\milli\second~bins and for each bin we build a distribution of time difference between subsequent events.
Figure \ref{timedistr} shows an example of this distribution for TOF between 20 and 22~\milli\second~taken in one run.

The analysis consist of two steps.
In the first one we determine the average dead time correction for every 2~\milli\second~bin:
\begin{itemize}
\item
An exponential function is fitted to an exponential function to the data between 0.05 and 0.35~\milli\second~(Fig.~\ref{timedistr}).
\item
The integral of the fitted exponential distribution function extrapolated
to zero time difference is calculated.
\item
The ratio of the integral of the fitted function to the observed number of events is determined and it is this number that gives the dead time correction.
\end{itemize}

\epsbild{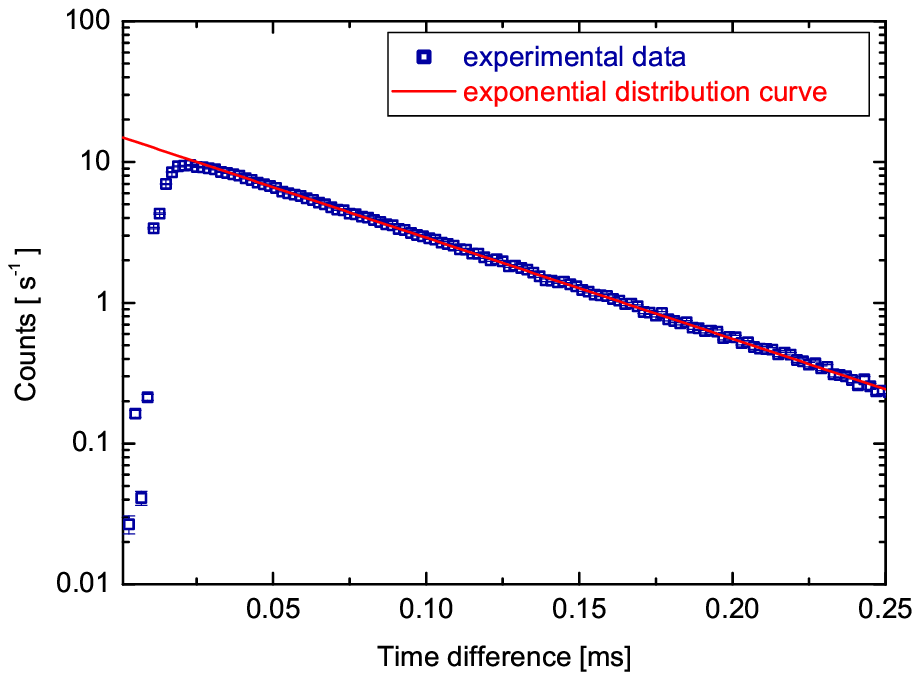}{Time difference distribution of subsequent events for VCN's with time-of-flight between 20 and 22~\milli\second~with an exponential distribution curve $y = P_0\cdot{}\exp{(P_1x)}$. \label{timedistr}}{.8}{0}{!hbt}
\epsbild{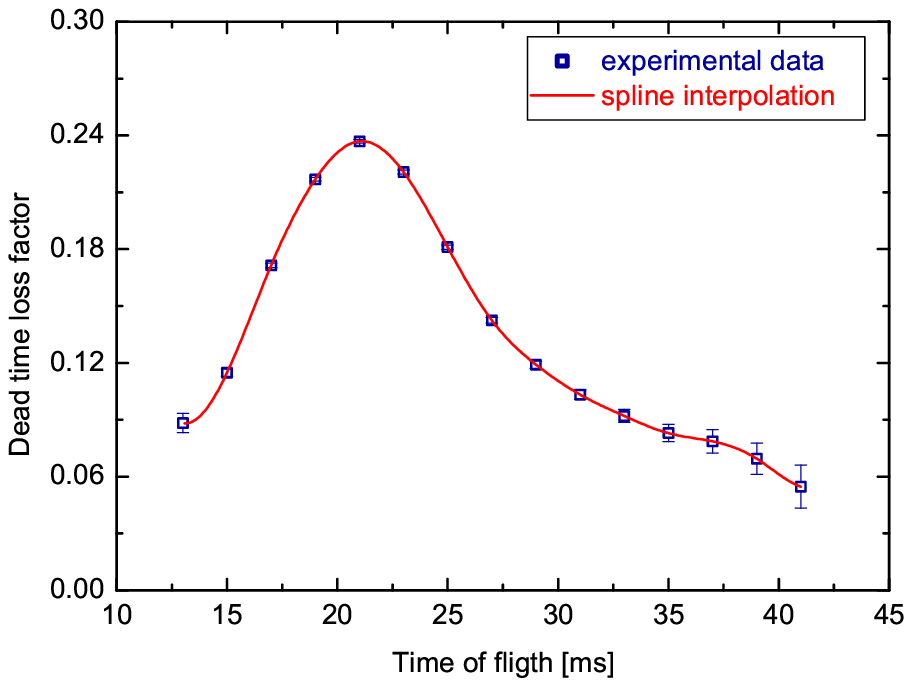}{Dead time loss factor interpolated with cubic spline. \label{spline}}{.8}{0}{!htb}
\epsbild{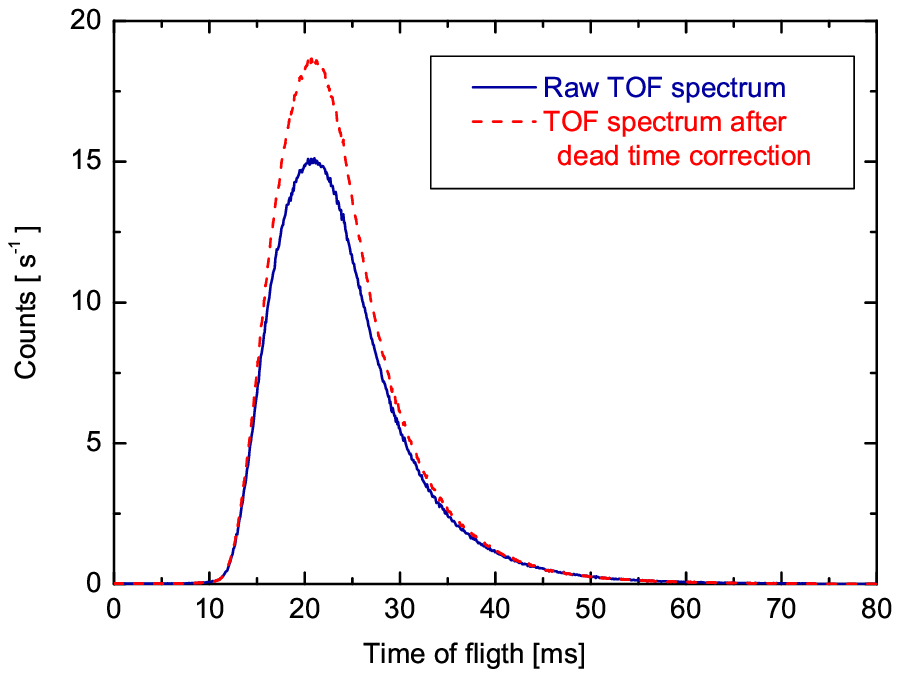}{VCN TOF spectra without (blue solid line) and with (red dashed line) dead time correction for the empty cell. The bin width is 80~\micro\second. \label{tofdead}}{.8}{0}{!htb}
In the second step the dead time loss factor as a function of TOF is interpolated using cubic splines (see Fig.~\ref{spline}).
The interpolation is restricted to the values for which 0.6 $< \chi^2/Ndf <$ 1.7. $\chi^2/Ndf$ values refer to
the fitted exponential functions from the first step. Figure~\ref{tofdead} shows the effect of the dead time correction.
The dead time correction changes the shape of the spectrum and can reach 25$\%$ in the maximum of the TOF distribution.
\clearpage
\subsection{Deconvolution of the time-of-flight spectra} \label{vcndeconv}
The raw TOF spectra are convolved with the chopper opening function
as described in section~\ref{vcnchopper}. To obtain the true TOF spectra
a deconvolution procedure is necessary. We tried two methods (i) direct, based on FFT technique and (ii) iterative.
The stable result in the direct deconvolution is strongly dependent on the input data as FFT techniques are sensitive
to noise. Since the measured TOF spectra carry a non-negligible fraction of noise, the iteration method seems more appropriate.
The iteration method assumes an analytical form of
the unfolded function (i.e. the true spectrum) with free parameter and performs a convolution with the known chopper opening function.
The resulting spectrum should be a good approximation to the raw spectrum as recorded by DAQ system.
To optimize the fit, the following expression for chi-square is evaluated:
\begin{equation}
\chi^2=\left(f(i)-y(i)\right)^2*w(i)
\end{equation}
where $f(i)$ is the folded theoretical function, $y(i)$ is the value experimentally
measured and $w(i)$ is the weight of the data point (inverse square of the statistical error).
The deconvolution factor is given by
\begin{equation}
d(i)=\frac{g(i)}{f(i)}
\end{equation}
where $g(i)$, $f(i)$ are the values before and after the convolution respectively.
The deconvolved spectra (with their natural statistical fluctuations) are then obtained by multiplying the raw data by the deconvolution factor.

The analytic function used to parameterise the VCN TOF spectra is a sum of two modified Maxwellian time distributions and a constant background.
\begin{equation}
f(x) = p_1 + \frac{p_3}{(x-p_2)^7}\exp{\left(-\frac{p_4}{(x-p_2)^2}\right)} + \frac{p_5}{(x-p_2)^{11}}\exp{\left(-\frac{p_6}{(x-p_2)^2}\right)}
\end{equation}
Figure~\ref{deconv_1} shows the experimental data and the fitted function $f$. The effect of deconvolution is presented in Fig.~\ref{deconv_2}. The deconvolution factor $d$ is shown in Fig.~\ref{deconv_4} and the deconvolved experimental spectrum is presented in Fig.~\ref{deconv_3}.
\epsbild{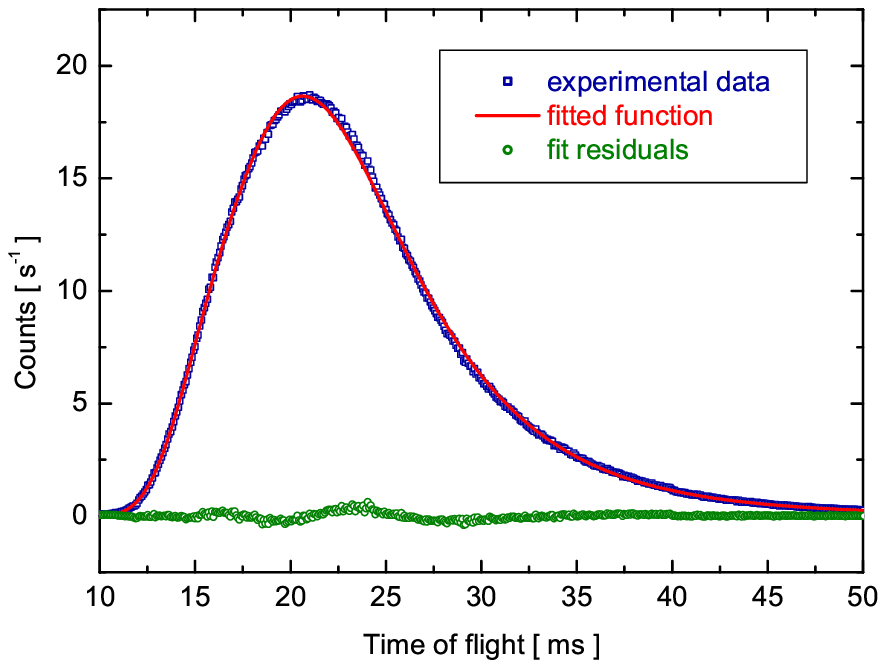}{An experimental time of flight spectrum fitted with the analytical function $f$(red line).
The green circles are the fit residuals. \label{deconv_1}}{.7}{0}{!htb}
\epsbild{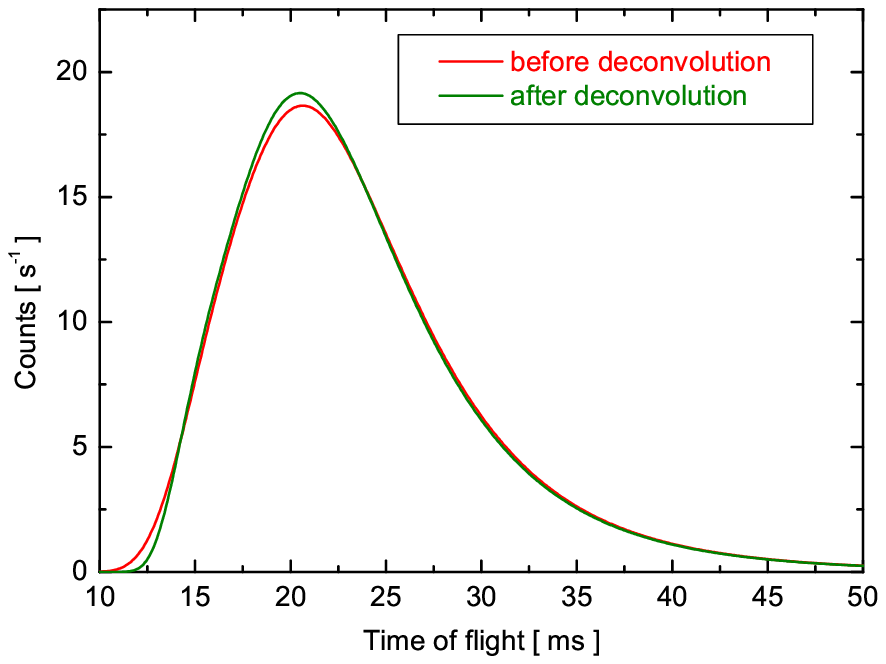}{Function $f$ before deconvolution (red line) and after deconvolution $g$ (green line).\label{deconv_2}}{.7}{0}{!htb}
\epsbild{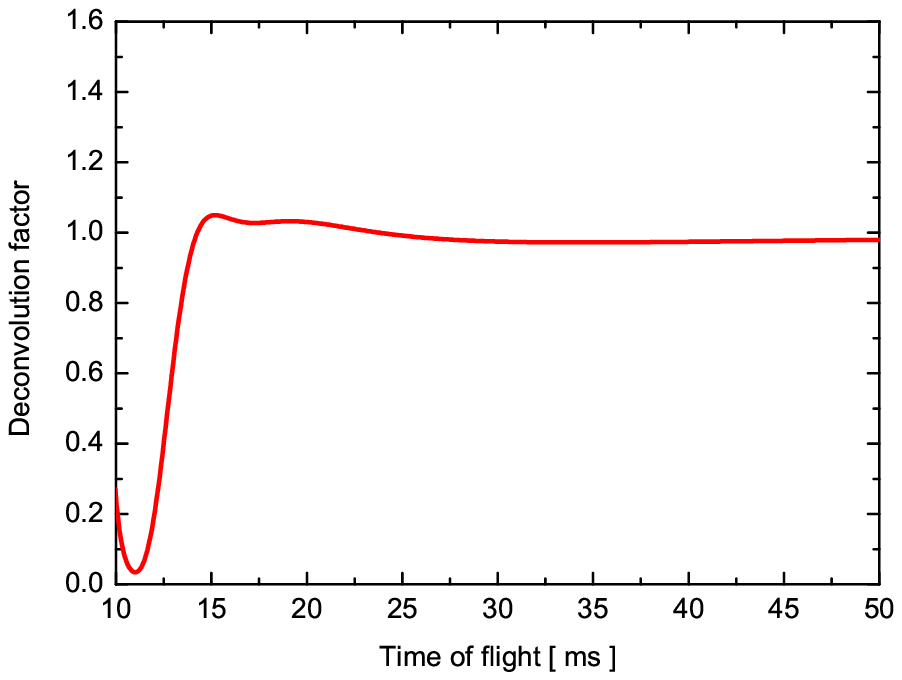}{A plot of the deconvolution factor $g/f$. \label{deconv_4}}{.7}{0}{!htb}
\epsbild{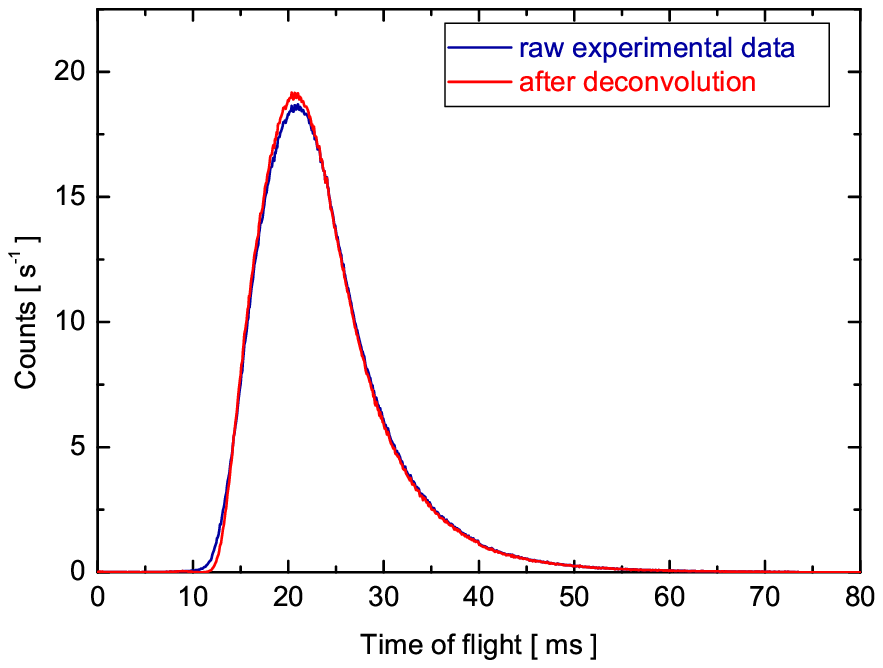}{Time of flight spectra before and after deconvolution. \label{deconv_3}}{.7}{0}{!htb}
\clearpage
\subsection{Zero time calibration method} \label{vcncal}
The TOF spectrum needs to be corrected for an effect that arises because of a difference in the time at which the chopper start signal is generated and the actual time at which the neutrons start passing through the chopper. This effect will appear as a shift in the TOF spectrum. This section describes the calibration method to determine the zero point on the time scale.

For calibration, data was collected with the detector placed at two positions: ``front'' (F) and ``extended'' (E), with
the same collimation (section~\ref{setup}).
The difference in the flight path between these two configurations is $s_{EF}$ = 511~\milli\metre.
Figure \ref{calib1} shows the time-of-flight spectra recorded at these two positions.
\epsbild{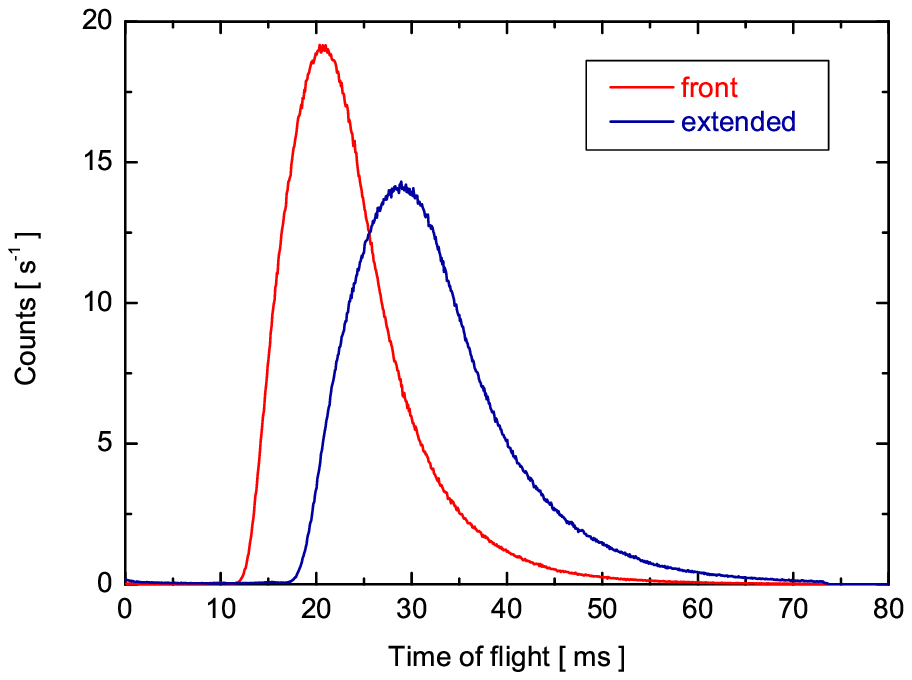}{Time of flight spectra for ``front'' and ``extended'' detector positions. \label{calib1}}{.8}{0}{!htb}
\epsbild{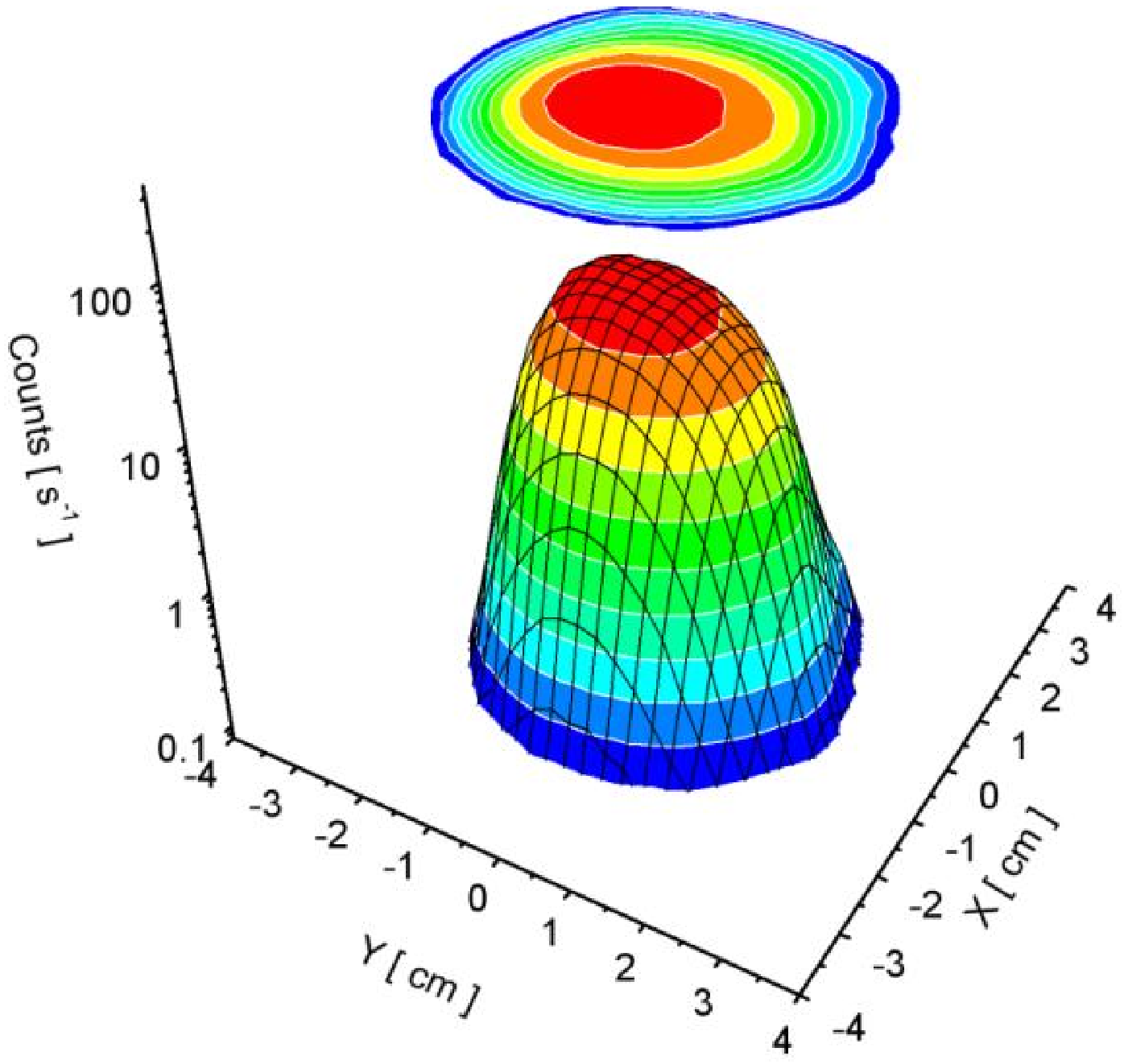}{A beam profile as measured for the ``front'' detector position. \label{beamclose}}{0.8}{0}{!htb}
\epsbild{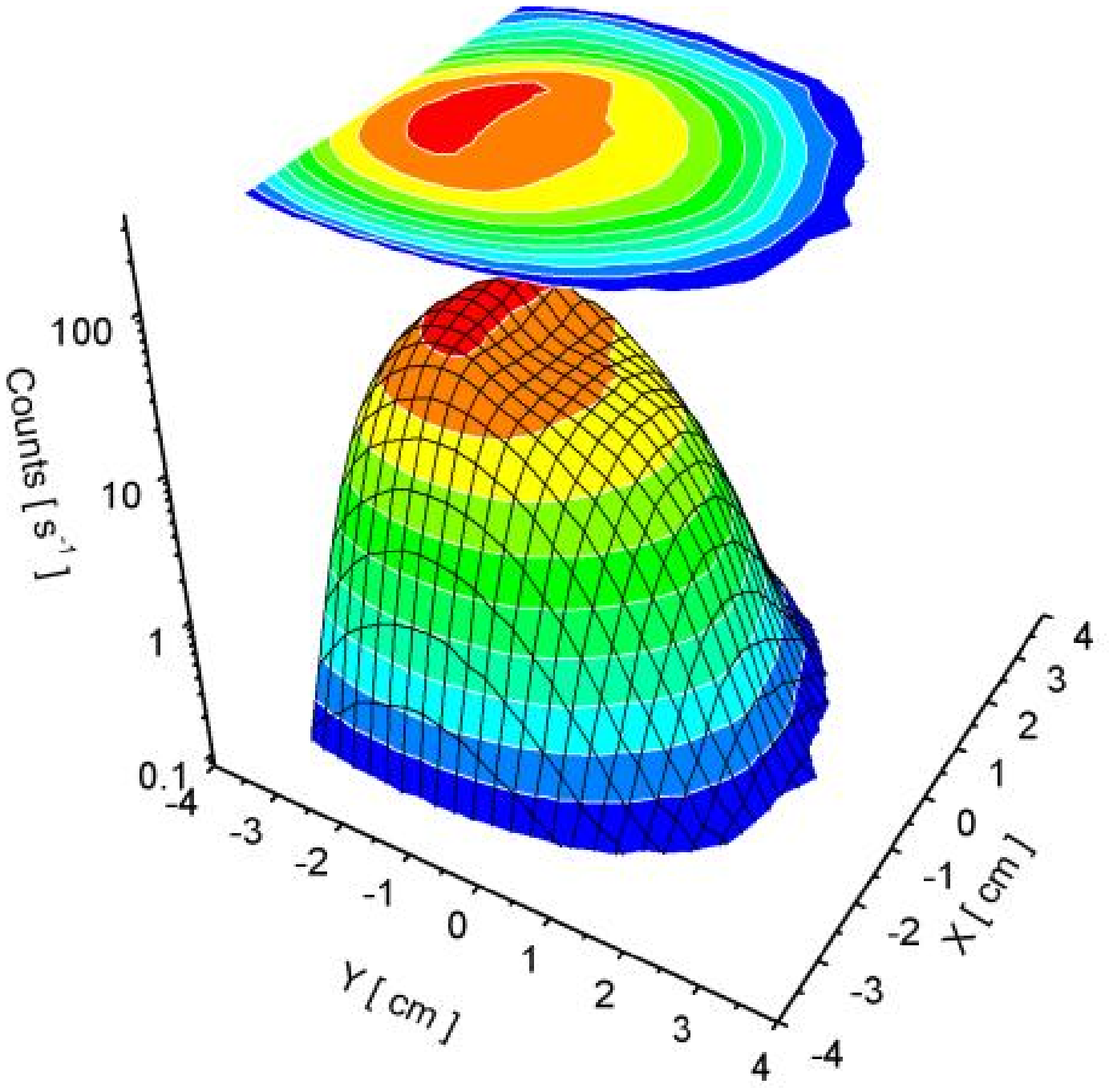}{A beam profile as measured for the ``extended'' beam position. \label{beamext}}{0.8}{0}{!htb}
\clearpage
Assuming that the mean arrival time does not change with spreading of the TOF spectrum, the mean velocity
can be calculated by dividing the distance $s_{EF}$ by the difference
in the mean arrival time at both positions:
\begin{equation}
v = \frac{s_{EF}}{t_E-t_F}
\end{equation}
A true mean arrival time at the ``front'' position is then determined by dividing the length of the flight path ($s_F$ = 1436~\milli\metre~) by the mean velocity:
\begin{equation}
t_T=\frac{s_F}{v}
\end{equation}
The correction to the arrival time was than calculated by subtracting the measured time from the true time.

There are two contributions to the systematic error connected with this calibration method:
\begin{itemize}
\item
gravitational and beam divergence effects,
\item
 determination of the distance with the assumption that neutrons react with the $^3$He in the middle of the detector.
\end{itemize}
\epsbild{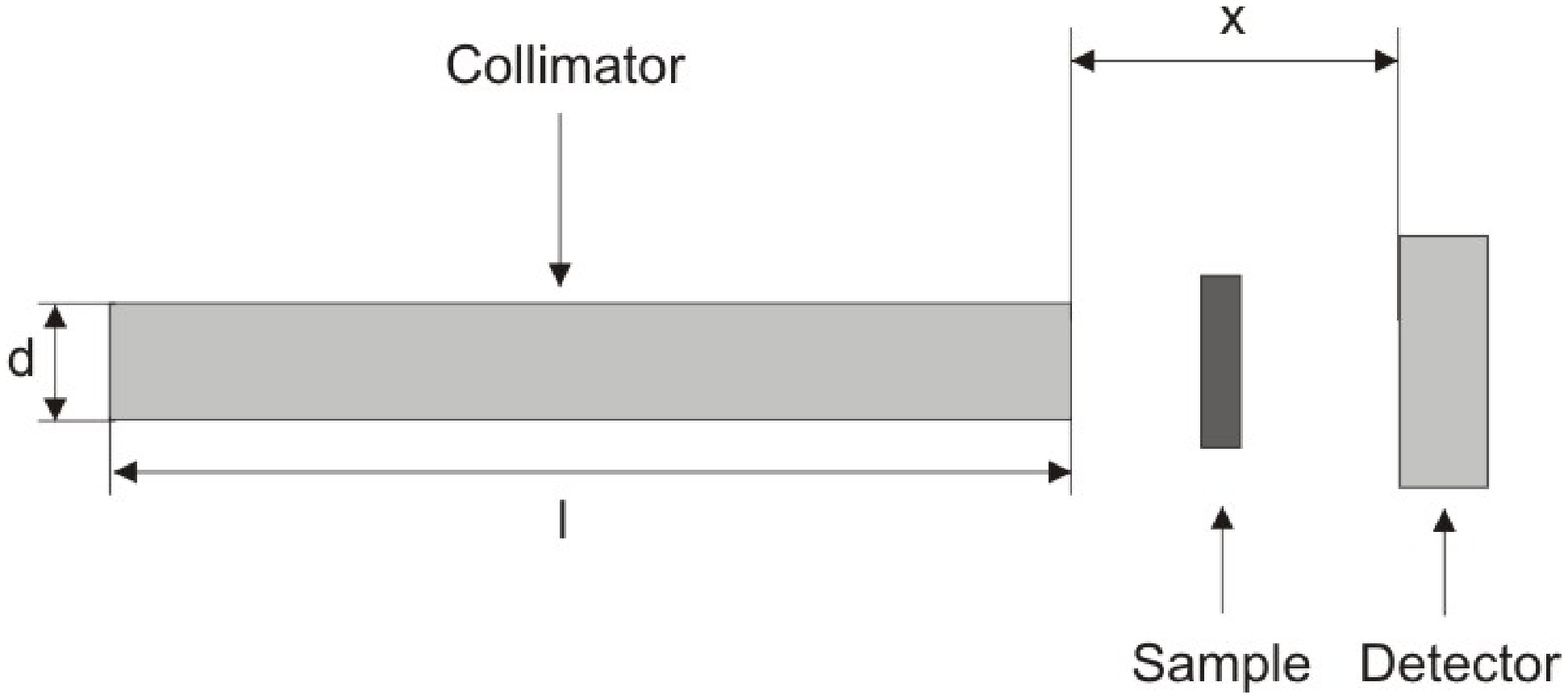}{A schema showing the dimensions needed for calculations of the beam spot. \label{beamspot}}{0.7}{0}{!htb}
To minimise this uncertainty, the TOF spectra have been built for the events falling into the collimator projected spot
area defined by the spot size parameter $b$:
\begin{equation}
b = d*\left(\frac{2x}{l}+1\right)
\end{equation}
where $d = 20~\milli\metre$ is the diameter of the collimator, $l = 1260~\milli\metre$ is the collimation length and $x$ is the distance
from the collimator to the detector (see Fig.~\ref{beamspot}). Thus the beam spot size is 26~\milli\meter~for the ``front'' detector localization ($x$ = 174~\milli\metre) and
42~\milli\metre~for the ``extended'' detector position ($x$ = 685~\milli\metre). Figures \ref{beamclose} and \ref{beamext}
show the beam profiles for the ``front'' and ``extended'' detector position.
Divergence of the beam and gravitation cause missing of events at the ``extended'' detector position, and consequently, changing of the mean arrival time.
The contribution from this effect to the calibration error was estimated to be of 0.04~\milli\second.
The uncertainty in determining the exact flight path length gives a systematic error of 0.28 ms.

The obtained correction to the arrival time (shift in the TOF spectrum) is (0.80 $\pm$ 0.28) ms. The effect of calibration is shown in Fig.~\ref{calib3}: one curve shows a TOF spectrum before and the other one after calibration.\\
\epsbild{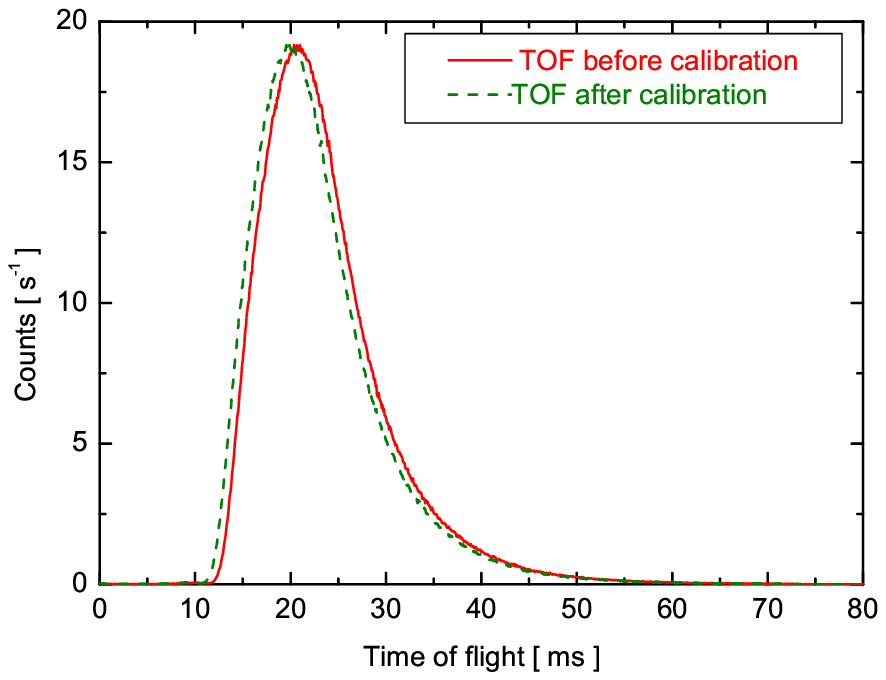}{Time of flight spectra before and after calibration. \label{calib3}}{.8}{0}{!htb}
In order to check the reliability of this method we calculate the time-of-flight distribution at the
close position from known distribution at the ``extended'' position. The comparison is shown in Fig. \ref{calib2}.
\epsbild{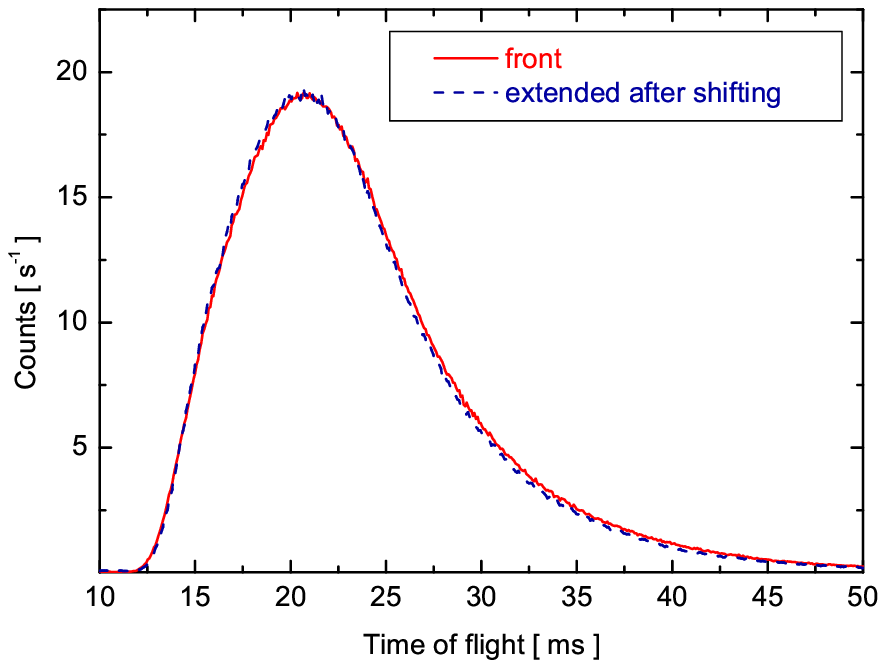}{Comparison between the measured spectrum at the ``front'' position (solid line) and transformed to that position from a measurement at the ``extended'' position (dashed line). \label{calib2}}{.8}{0}{!htb}
\subsection{Background measurements}
The background rate for the VCN experiment was measured to be about 2 events per second. The measurement was performed with the main shutter closed when the reactor was running.
The main contribution comes directly from the turbine.
\clearpage
\section{UCN experiment}
\subsection{Dead time consideration}
The neutron flux at the UCN test position is considerably lower than that at the VCN beam: the UCN flux of the order of 20~\centi\rpsquare\metre\usk\reciprocal\second~for v = 13~\metre\per\second~and about 4000~\centi\rpsquare\metre\usk\reciprocal\second~at the velocity of 70~\metre\per\second.
Consequently, for the experiments carried out with the UCN beam, no dead time corrections are required.
\subsection{Calibration of the timing in the TOF spectra} \label{ucncal}
As is the case of the VCN experimental setup, the neutron chopper
operating in the UCN setup generates a start signal that
does not exactly coincide with the opening of the neutron window
itself (section~\ref{vcnchopper}).
This causes shift in the TOF spectra and needs to
be corrected by a calibration procedure.
Two independent methods have been developed to calibrate UCN
time-of-flight spectra.

The first one is based on a direct experimental determination
of the number of steps the stepping motor moves between the moment the
TTL start signal appears and the moment the neutron start passing through the chopper.
The chopper is driven slowly, step by step, by a control program running in the LabView environment.
This is why the exact angular position of the
stepping motor at which the TTL signal appears can be determined. Likewise by
driving the chopper further, step by step, the exact angular
position of the stepping motor at which the UCN pass through the
chopper and are detected by the detector is determined, too. One thus
knows the exact number of steps $N_{SW}$ between the TTL start signal and
the chopper opening. Knowing also the frequency $\nu$ at which the
stepping motor is driven one can then calculate the time offset $t_0$ to be applied to raw TOF spectra. In a similar procedure, one can also experimentally
determine the number of steps $N_{OC}$ of the motor between opening
and closing the chopper window as well as the number of steps of a full chopper cycle.
This method was applied for the measurements with the beam collimator and without it.
The obtained results are collected in Tab.~\ref{tab1}.
\begin{table}[!htb]
\begin{center}
\begin{tabular}{|l|c|c|c|} \hline
Setup & $N_{SW}$ & $N_{OC}$ & Full chopper turn \\
\hline
With collimator & 591 & 60 & 864 \\
\hline
Without collimator & 595 & 52 & 864 \\
\hline
\end{tabular}
\end{center}
\caption{Results from the calibration measurements \label{tab1}}
\end{table}

In the calculations we have
assumed that $t$ = 0 in the centre of the chopper opening.
For the measurements with the chopper speed of 2500 steps/s (3~\hertz)
the calculated offset is:
\begin{equation}
t_0 = \frac{N_{SW}+0.5N_{OC}}{2500} = 248~\milli\second
\end{equation}
The precision of this results, limited by the uncertainty in counting
the number of steps performed, is 1~\milli\second.

The other method of calibrating
the TOF spectra is based on measuring the arrival time of the fastest (slowest) neutrons at
various chopper speeds and fitting a linear function to the data. Examples for the measurements without collimator
are plotted in Figs \ref{slow} and \ref{fast}.
The time offset between TTL signal and chopper window opening is
then given by
\begin{equation}
t_0 = \frac{P_1}{\nu}
\end{equation}
where $P_1$ is the fit parameter and $\nu$ is the chopper frequency.
The calculated offset (average from both measurements) is $t_0$ = (250 $\pm$ 2)~\milli\second.

Both these calibration methods rely on a precise knowledge of the
chopper speed. The chopper has been tested and has shown to be
operating stable over long periods. The chopper speed scales
linearly with the true disks speed. Both calibration methods gave consistent results.
Figures~\ref{ucncal_1} and \ref{ucncal_2} show a time-of-flight spectrum with no calibration applied
and a plot of the same spectrum with proper timing calibration.
\epsbild{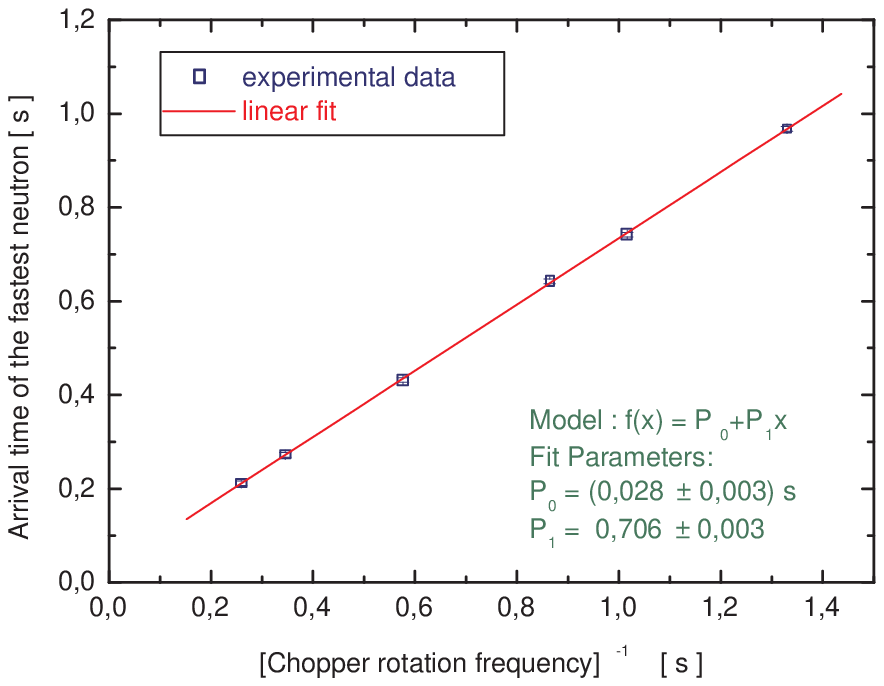}{The calibration of the UCN TOF spectrum based on the arrival time of the slowest neutron. \label{slow}}{.8}{0}{H}
\epsbild{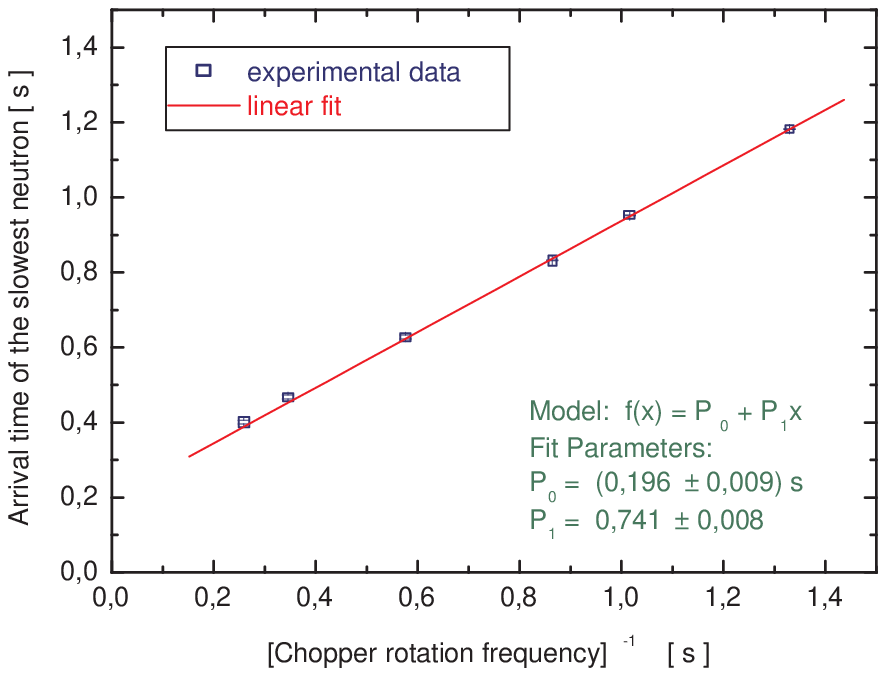}{The calibration of the UCN TOF spectrum based on the arrival time of the fastest neutron. \label{fast} }{.8}{0}{H}
\subsection{Background measurements}
The background for the UCN experiment appears to be of the order of 0.1 counts per second.
It was measured with the main shutter closed and the reactor running.
\epsbild{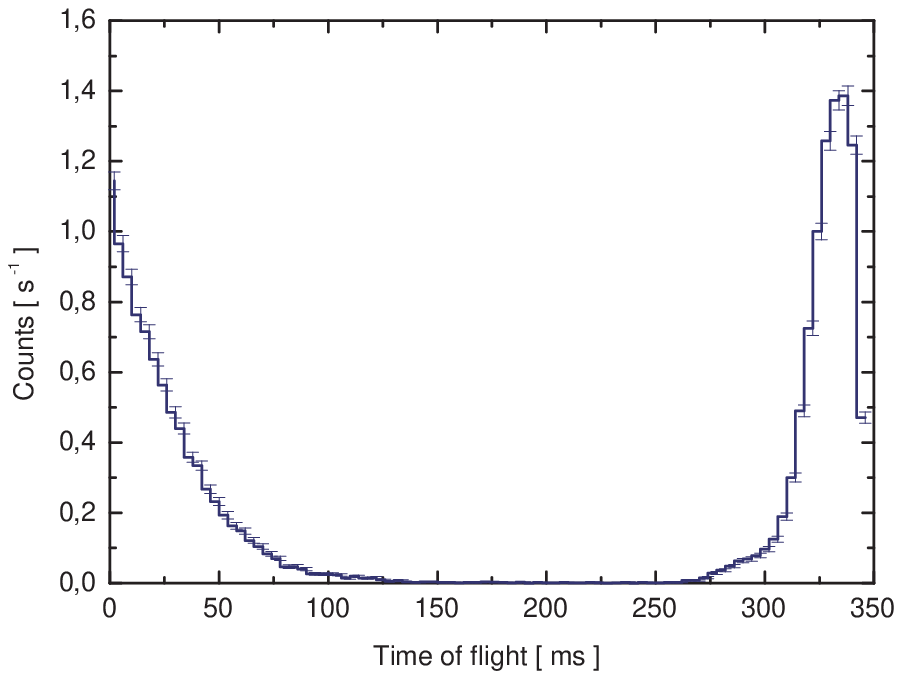}{A time-of-flight spectrum before calibration. The spectrum is ``split'' into two parts by the start signal which
is generated during the chopper window is open. The bin width is 4~\milli\second.\label{ucncal_1}}{1.0}{0}{!hbt}
\epsbild{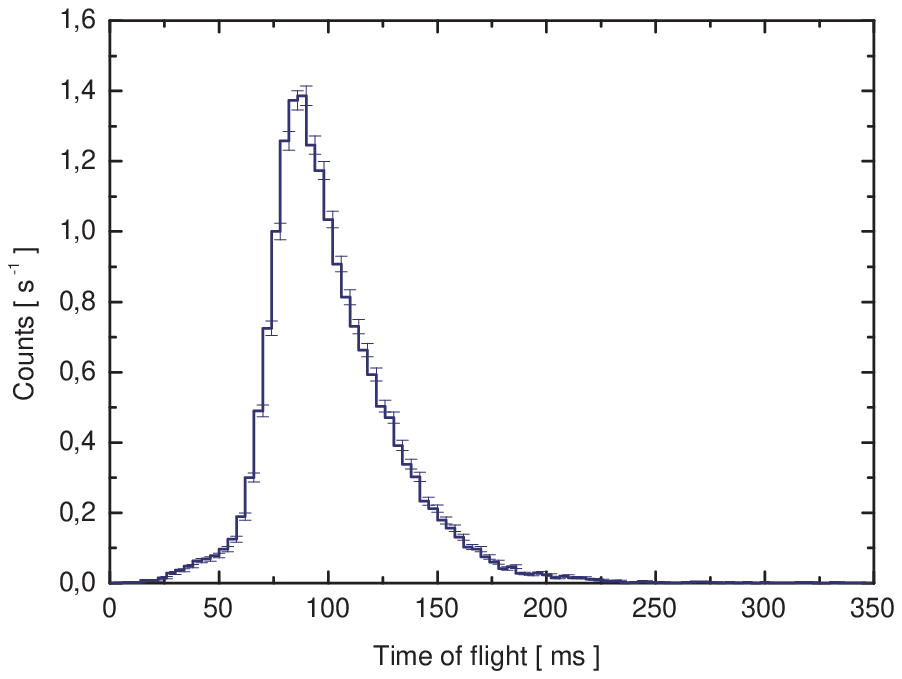}{A time-of-flight spectrum with time calibration. The bin width is 4~\milli\second.  \label{ucncal_2}}{1.0}{0}{!htb}


\begin{thebibliography}{99}
\bibitem{Huf} P.R. Huffman \emph{et al.},
\emph{Progress towards magnetic trapping of Ultra-Cold Neutrons},
Nucl. Instr. Meth. \textbf{A440}, 522-527 (2000).
\bibitem{Doy} J.M. Doyle, S.K. Lamoreaux,
Europhys. Lett. \textbf{26}, 253 (1994).
\bibitem{Gol2} R. Golub, D. Richardson, S.K. Lamoreaux,
\emph{Ultra-Cold Neutrons},
(Adam Hilger, Bristol, 1991).
\bibitem{Abe} H. Abele, S. Bae\ss{}ler and A. Westphal,
Lect. Notes Phys. \textbf{631}, 355-366 (2003).
\bibitem{Nes} V. Nesvizhevsky \emph{et al.},
Nature (London) \textbf{415}, 297 (2002).
\bibitem{Gol} R. Golub and J.M. Pendlenbury,
Phys. Lett. \textbf{53A}, 133 (1975).
\bibitem{Pok1} Y.Pokotilovski, 
\emph{ESS Special Expert Meeting - UCN Factory Workshop}, 
Febr. 22-23, 2002, Wien, http://www.ati.ac.at/~neutrweb/ess/ess.html
\bibitem{Liu1} C.-Y. Liu,
\emph{A Superthermal Ultra-Cold Neutron Source},
Dissertation, Princeton University, 2002. 
\bibitem{Liu2} C.-Y. Liu,
\emph{Physics of superthermal UCN production in SD$_2$ and other materials},
Proceedings of the 3rd UCN Workshop, Pushkin, 2001.
http://nrd.pnpi.spb.ru/SEREBROV/3rd/talks/20/chen.pdf 
\bibitem{Doy1} http://www.doylegroup.harvard.edu/neutron/neutron.html
\bibitem{Lanl} http://p25ext.lanl.gov/edm/edm.html
\bibitem{Suss} http://www.sussex.ac.uk/physics/research/particle/ppghmp.htm
\bibitem{Mas} Y. Masuda \emph{et al.}, 
\emph{Spallation ulracold-neutron production in superfluid helium},
Phys. Rev. Lett. \textbf{89}, 284801 (2003).  
\bibitem{Gol1} R. Golub and K. Boning,
\emph{New Type of Low Temperature Source of Ultra-Cold Neutrons and Production of Continous Beams of UCN},
Z. Phys. \textbf{B51}, 95 (1983).
\bibitem{Yu} Z.-Ch. Yu, S.S. Malik, R. Golub, 
\emph{A thin film source of Ultra-Cold Neutrons},
Z. Phys. \textbf{B62}, 137 (1986).
\bibitem{Pok2} Y. Pokotilovski, 
Nucl. Instr. Meth. \textbf{A356}, 412 (1995).
\bibitem{Psi} http://ucn.web.psi.ch/
\bibitem{Hill} R.E. Hill \emph{et al.}, 
\emph{Performance of the prototype LANL solid deuterium ultra-cold neutron source},
Nucl. Instr. Meth. \textbf{A440}, 674 (2000).
\bibitem{Kir} K. Kirch \emph{et al.}, 
\emph{Status of the New Los Alamos UCN Source},
16th Intern. Conf. on the Application of Accelerators in Research and Industry, CAARI 2000, Denton, 2000, AIP Conf. Proc. \textbf{576}, 289 (2001).
\bibitem{Sau} A. Saunders \emph{et al.},
\emph{Demonstration of a solid deuterium source of ultra-cold neutrons},
nucl-ex/0312021, Phys. Lett. B \textbf{593}, 55 (2004), in press. 
\bibitem{Mor} C.L. Morris \emph{et al.}, 
Phys. Rev. Lett. \textbf{89}, 272501 (2002).
\bibitem{Ser} A. Serebrov \emph{et al.},
Nucl. Instr. Meth. \textbf{A440}, 658 (2000).
\bibitem{Alt} I.S. Altarev \emph{et al.}
\emph{A liquid hydrogen source of of ultra-cold neutrons},
Phys. Lett. \textbf{B51}, 95 (1983).
\bibitem{Lov} S.W. Lovesey,
\emph{Theory of neutron scattering from condensed matter},
(Clarendon press, Oxford, 1984), Vol I.
\bibitem{Sch} L.I. Schiff,
\emph{Quantum Mechanics},
3rd ed (New York: McGraw-Hill, 1968).
\bibitem{Eng} G. Engelmann \emph{et al.},
\emph{Comparison Between Very-Slow-Neutron Transmission and Small-Angle Neutron-Scattering Experiments},
Z. Phys. \textbf{B35}, 345-349 (1979).
\bibitem{Bod} K. Bodek \emph{et al.},
\emph{An apparatus for the investigation of solid D$_2$ with respect to ultra-cold neutron sources},
accepted for publication in Nucl. Instr. Meth.
\bibitem{Liu3} C.-Y. Liu, A.R. Young, S.K. Lamoreaux,
\emph{UCN upscattering rates in a molecular deuterium crystal},
Phys. Rev. \textbf{B62}, R3581 (2000). 
\bibitem{Sil} I.F. Silvera,
\emph{Solid molecular hydrogens in the condensed phase},
Rev. Mod. Phys. \textbf{52}, No. 2, Part I, 393-452 (1980).
\bibitem{Nie2} M. Nielsen, H. Bjerrum Moller,
\emph{Lattice Dynamics of Solid Deuterium by Inelastic Neutron Scattering},
Phys. Rev. \textbf{B3}, 4383 (1971).
\bibitem{mpa} MPA-3 Manual version 1.47, October 10, 2002.
\bibitem{Kra} J. Van Krankendonk,
\emph{Solid Hydrogen},
(Plenum, New York, 1983).
\bibitem{Sou} P.C. Souers,
\emph{Hydrogen Properties for Fusion Energy},
(University of California, Berkeley, 1986).
\bibitem{Ned} J.A. Nelder, R. Mead, 
Comput. J. \textbf{13}, 317 (1965).
\bibitem{Fle} J. Fletcher, 
Comput. J. \textbf{13}, 317 (1970).
\bibitem{Kei} J. Keinert, unpublished.
\bibitem{Ser1} A. Serebrov \emph{et al.},
\emph{Studies of Solid-Deuterium Source of Ultracold Neutrons},
unpublished.
\bibitem{Pok3} Y. Pokotilovski \emph{et al.},
\emph{Differential Neutron Spectrometry in the Very Low Neutron Energy Range. 
Neutron Cross Sections for Zr, Al, Polyethylene and Liquid Fluoropolymers},
Communication of the Joint Institute for Nuclear Research, E3-2003-138, Dubna, 2003. 
\end{thebibliography}
\end{document}